\documentclass[12pt]{article}

\usepackage{scicite}
\usepackage{graphicx}

\usepackage{times}

\usepackage{comment}

\title{Measuring Social Media Network Effects}

\author
{Sinan Aral,$^{1\ast}$ Seth G. Benzell,$^{2}$ Avinash Collis,$^{3}$ Christos Nicolaides$^{4}$ \\
\\
\normalsize{$^{1}$Massachusetts Institute of Technology,}
\normalsize{$^{2}$Chapman University,}\\
\normalsize{$^{3}$Carnegie Mellon University,}
\normalsize{$^{4}$University of Cyprus}\\
\normalsize{$^\ast$To whom correspondence should be addressed; E-mail:  sinan@mit.edu.}\\
\normalsize{Authors listed in alphabetical order.}
}

\begin{document} 


\baselineskip24pt


\maketitle 
\textbf{
Network effects, which dictate that the utility from consuming a good increases with the total number of other consumers consuming it, are widely regarded as critical to the digital economy. But recent theory and evidence suggest that \emph{local} network effects---the economic value created by specific social network connections---are what drive value in networked online platforms. Here, we use representative, incentive-compatible online choice experiments involving 19,923 Facebook, Instagram, LinkedIn, and X users in the United States to provide the first large-scale, empirical measurement of local network effects in the digital economy and to measure heterogeneity in the value of network connections across social media platforms. Our analysis reveals that social media platform value ranges from \$78 to \$101 per consumer, per month, on average, and that 8.1-23.7\% of that value is explained by local network effects. We also find that 1) stronger ties are more valuable on Facebook and Instagram, while weaker ties are more valuable on LinkedIn and X; 2) connections known through work are most valuable on LinkedIn and least valuable on Facebook, and people who are looking for work value LinkedIn significantly more and Facebook significantly less than people who are not looking for work; 3) men value connections to women on social media significantly more than they value connections to other men, particularly on Instagram, Facebook and X, while women value connections to men and women equally across platforms; 4) consumers value network connections on any platform more if they are also connected with them on other platforms, suggesting that social media platforms are complements, not substitutes; 5) white consumers value relationships with other white consumers significantly more than they value relationships with non-white consumers on Facebook while, on Instagram, connections to alters eighteen years old or younger are valued significantly more than any other age group---two patterns not seen on any other platforms. The social media platforms we studied each individually generate between \$53B and \$215B in consumer surplus per year in the US alone. These results suggest that social media generates significant value, that local network effects drive a substantial fraction of that value and that the sources and contours of these effects vary across platforms, consumers, and connections.
}

Numerous economic models explain how network effects operate~\cite{rohlfs1974theory, katz1985network, rochet2003platform, parker2005two} and affect market competition~\cite{katz1994systems, shapiro1999information, parker2016platform}, product design~\cite{parker2005two}, platform strategy~\cite{belleflamme2018platforms}, systems compatibility~\cite{wang2010survival, liu2011compatibility}, antitrust regulation~\cite{benzell2022multi}, market failure~\cite{liebowitz1994network}, prices~\cite{weyl2010price}, and lock-in to potentially adverse or sub-optimal technologies~\cite{farrell2007coordination}. Some empirical work measures economic network effects in the aggregate, estimating the effect of the total size of a product or platform's user base on economic value creation, using observational data~\cite{brynjolfsson1996network, bramoulle2009identification}, field experiments~\cite{rajkumar2022causal, boudreau2021promoting}, and natural experiments~\cite{farronato2024dog}. But, for decades, most scientific literature only considered network effects in the aggregate, modeling and measuring the effect of the \emph{total} number of consumers of a product or platform on that product or platform's economic value and treating each consumer, and therefore each connection between consumers in the network, interchangeably. 

Recently, however, as the economy became more networked, the concept of economic network effects expanded to consider how social networks affect economic value. Tucker (2008)  is an early example of an empirical paper which allowed a network effect to depend on the identity of technology adopters \cite{tucker2008identifying}. A new theory of ``local network effects," which evolved from consideration of networks of geographically proximate consumers~\cite{goolsbee2002evidence, corrocher2009me, kim2022local}, now considers how social network connections and structures create network effects through socially proximate consumers~\cite{sundararajan2008local}. Rather than asking how the total number of consumers of a product affects a product's value, these new local network effects theories posit that a product's value is created, at least in part, through subsets of consumers connected by the product in social networks. These new theories also seek to explain how different types of network connections and structures determine a product's value.

Sociological studies of the value of network connections have examined self reports of connection value~\cite{marsden1984measuring, fischer1982we, wellman1990different, reagans2005preferences}, measured the strength of ties~\cite{granovetter1973strength, granovetter1985economic, lin1981social, bruggeman2016strength, rajkumar2022causal, aral2016future}, examined the social capital and social support implications of network ties~\cite{putnam2000bowling, lin2002social, burt2000network, burt2005brokerage, hall1985social, coleman1988social}, measured the importance of digital network connections~\cite{wellman2001does, onnela2007structure, petroczi2007measuring, gilbert2009predicting, jones2013inferring, burke2011social, wellman1999living, ellison2007benefits, ellison2010little, ellison2014cultivating} and valued ties based on strategic outcomes related to information advantage and power~\cite{burt2003social, burt2004structural, kim2017strength, aral2011diversity, rost2011strength}. But sociologists have not examined economic valuations of network ties or measured tie value in the context of economic network effects. On the other hand, economic studies have estimated total platform valuations in the context of network effects in the aggregate ~\cite{brynjolfsson1996network, bramoulle2009identification, boudreau2021promoting, tucker2008identifying, farronato2024dog, brynjolfsson2019using, allcott2020welfare, mosquera2020economic}, investigated the role of heterogeneity in network formation ~\cite{Mele2017, Badev2021}, and studied the implications of specific types of network ties for economic outcomes such as income mobility~\cite{chetty2022social1, chetty2022social2}. But economists have not measured consumers’ economic valuations of network connections or measured tie value in the context of economic network effects. As a result, a large and important gap exists, between the sociology of networks and the economics of network effects, in our understanding of local network effects.\footnote{See SI subsection 2.1 for a more complete discussion.}

In today's digital economy, which is increasingly characterized by products consumed by networks of consumers together, this lack of understanding of local network effects is critical for several reasons. First, local network effects are essential to understanding the welfare implications of social media platforms and their regulation. As social media platforms expand internationally and as governments attempt to regulate them across geographies, precise measurement of the value created by social network connections is essential to guiding policies that maximize welfare. Effective antitrust regulation, merger oversight, national bans on foreign social media platforms, and interoperability and network portability legislation all depend on understanding the value created by social network connections and how that value is shaped by the nature of the connections and the social network structures in which they are embedded. For example, when governments seek to break up networked monopolies, unwind anti-competitive acquisitions, or ban foreign social media, understanding the welfare implications of breaking the networked ties enabled by these platforms is a first-order consideration. Similarly, the welfare implications of legislation that forces networked platforms to be interoperable or that allows consumers to take their networks with them when they leave one platform for another, can only be understood in the context of precise measurement of the network value created by enabling such interoperability or network portability. 

Our study addresses this challenge by being the first to obtain valuations, at the connection and platform level, for multiple major social media platforms simultaneously using a common methodology. This enables direct cross-platform comparisons that reveal how platforms differ in the nature of the value they create. Our results show, for example, that job seekers value LinkedIn significantly more and Facebook significantly less than non-job-seekers, providing direct evidence that these platforms serve distinct economic functions. The finding that connections to people consumers see frequently in person are valued far more than connections to people they never meet offline raises important questions about whether social media primarily complements or substitutes for offline social life. An additional policy-relevant finding is that consumers value connections to the same person more when they are connected on multiple platforms. This is a result with direct implications for competition policy and one that can only be observed in a study that measures valuations across platforms simultaneously.

Second, local network effects are critical to platform strategy. How network connections create economic value directly implicates the most important strategic decisions faced by networked platforms. The first ever model of network effects theorized that how different network connections create value, for example if consumers have ``communities of interest" or ``a few principal contacts," dramatically changes platforms' optimal go-to-market strategies and pricing~\cite{rohlfs1974theory}. If consumers are socially clustered, as we know they are in human social networks~\cite{watts1998collective}, recruiting communities of connected consumers will create greater consumer welfare and total platform value than recruiting disjoint sets of unconnected consumers. Local network effects also implicate platform monetization, user acquisition, and algorithmic connection recommendation strategies for the same reason~\cite{rajkumar2022causal}. As different connections and network structures may create dramatically different value for consumers and platforms, how platforms recruit users, recommend connections, monetize their services, and how these decisions affect welfare all depend on the contours of the value created by local network effects. 

Finally, local network effects determine, in part, why some platforms succeed while others fail. Analysis of the early platform battle between Facebook and MySpace suggests that Facebook's go-to-market strategy, which focused on cultivating connections among college friends, created stronger local network effects than MySpace's recruitment strategy, which yielded negative assortativity, loose connections and, thus, weak local network effects~\cite{ahn2007analysis}. The evidence suggests that Facebook overcame MySpace's significant installed base advantage by creating stronger local network effects~\cite{aral2021hype}.

Despite the importance of local network effects for regulation, competition, platform strategy, and the measurement of economic welfare, we currently lack methods for measuring local network effects in the digital economy. To address this challenge we adapt the framework of massive, incentive compatible online choice experiments to measure local network effects among 19,923 representative users of Facebook, Instagram, LinkedIn and X across the United States. The incentive compatible choice experiment framework is well understood and has previously been used to measure total welfare gains from the consumption of a variety of digital goods and services~\cite{brynjolfsson2019using, allcott2020welfare, mosquera2020economic}. Choice experiments illicit truthful and accurate valuations of products by offering consumers a series of choices in exchange for monetary compensation equal to their willingness to accept deactivation of a product if and only if they cease using the product for specified periods. The experiments are made consequential by informing participants that some fraction of the study pool will be randomly selected to receive payments based on their valuations after the survey concludes. In incentive compatible online choice experiments, compliance with deactivation is monitored by requiring subjects to provide access to their online accounts so researchers can verify that they deactivated the products before payments are made.

While previous incentive compatible choice experiments have measured the total value of social media platforms by offering consumers payments equal to their expressed platform valuations upon platform deactivation, we offer consumers payments for deactivating the platforms and, separately, solicit incentive compatible estimates of specific connections. By randomly varying payment offers and the types of connections selected for valuation, our approach allows us to precisely measure total platform value, platform value arising from local network effects and the heterogeneity of local network effects across consumers, connections and platforms.

We conducted two waves of surveys containing incentive compatible online choice experiments including Becker-DeGroot-Marschak (BDM) lotteries~\cite{becker1964measuring}, multiple-price lists~\cite{andersen2006elicitation}, and a ranked choice method~\cite{krosnick1988test} adapted to measure the value of networked connections and local network effects, which we call Connection Rank Choice (CRC), across four social media platforms over three years. The first wave, conducted between February 2021 and May 2022, recruited 10,528 respondents representative of the U.S. internet population from Lucid---a high quality online survey platform ~\cite{coppock2019validating} widely used in academic research ~\cite{pennycook2021shifting, benzell2020rationing, solis2021covid}. The second wave, conducted between November and December 2023, recruited 9,395 respondents representative of the U.S. internet population, again from Lucid. Every survey respondent was verified by Lucid to be a monthly active user of the platforms for which we were conducting incentive-compatible choice experiments.

During the first wave, after respondents were recruited, they were immediately asked to connect with our research accounts on the four platforms. When connecting to those accounts, respondents were informed that they would provide account information to the research team including their connection counts, connection lists and demographic profile information. This gave us access to rich data about respondents' profiles and social networks, which we describe below. After connecting with our research accounts, respondents were asked a series of survey questions about their age, gender, ethnicity, household income, political leaning and time spent using each of the platforms. Following the survey, respondents were asked to list their top four closest connections and ``any four connections excluding the four closest connections'', which created a list of eight connections for every respondent totaling 84,224 network connections. Then, following Brynjolffson et al~\cite{brynjolfsson2019using} and Allcott et al~\cite{allcott2020welfare}, we provided respondents a Becker-DeGroot-Marschak (BDM) lottery to illicit their total platform valuation for each platform---asking them to provide a valuation they were willing to accept to deactivate the platform for a month. After the platform valuation question, we asked respondents to rank how valuable they find their eight connections using Connection Rank Choice (CRC), a rank order question adapted to measure the value of networked connections and local network effects.

Once the connections, the BDM platform valuations and connection rankings were provided, the research team manually verified all of the responses by navigating to the respondents' accounts and checking each response against each platform's data to ensure survey accuracy. For each connection named in the survey to whom the respondent is connected on the platform, we checked whether that connection appears in the list of the respondent's connections on the platform. We also collected information about the respondents, their eight ranked connections and their relationship to each connection. On Facebook, for example, we collected respondents' connection lists and matched the names with U.S. Census data from the U.S. Social Security Administration to estimate the race and gender of each of the respondents' connections.\footnote{https://www.ssa.gov/oact/babynames/limits.html} For LinkedIn, we collected respondents' profile information, including their educational attainment and career history. For X, we collected respondents' post history and lists of followers and followees using the X API. Instagram was the only platform from which we were unable to gather additional information due to restrictions imposed by the platform.

During the second wave survey, in addition to the demographic and platform usage questions listed above, we also asked respondents if they were currently looking for work and validated our total platform valuation measurement using a multiple price list valuation elicitation instrument equivalent to a series of single binary discrete choice experiments across a range of valuations~\cite{andersen2006elicitation,brynjolfsson2019using}. This method is also well understood and has been validated across a number of different product and platform valuation studies~\cite{brynjolfsson2019using}. We randomly varied the upper limit of the price range to be \$100, \$150, or \$200 to account for anchoring effects. One out of every hundred respondents were randomly selected and offered to follow up on their choices with incentive compatible payments (see SI more details about incentive-compatibility in both waves). The research protocol in each Wave was separately evaluated and approved by the Massachusetts Institute of Technology’s (MIT’s) Institutional Review Board and the analysis in Wave two was preregistered with the Open Science Foundation (OSF).\footnote{https://osf.io/qs4kn/?view\_only=5cee408999b5484dafcc9626e88c953b}

To evaluate the results of these surveys, we developed a model of local network effects based on a latent topics/redundant signal framework. The model allows for rich patterns of complementarity and substitution across types of connections and information transmitted. The model is described in detail in SI section 3.2.3. It motivates our two main sets of analyses: First, an analysis of the total value created by the platform and the share of the value created by local network effects; Second, an analysis of the relative value of different types of connection, which are interpreted as the product of the value of the information transmitted and likelihood the connection yields that information.

\begin{figure}[h!]
\centering
    \includegraphics[width=\linewidth]{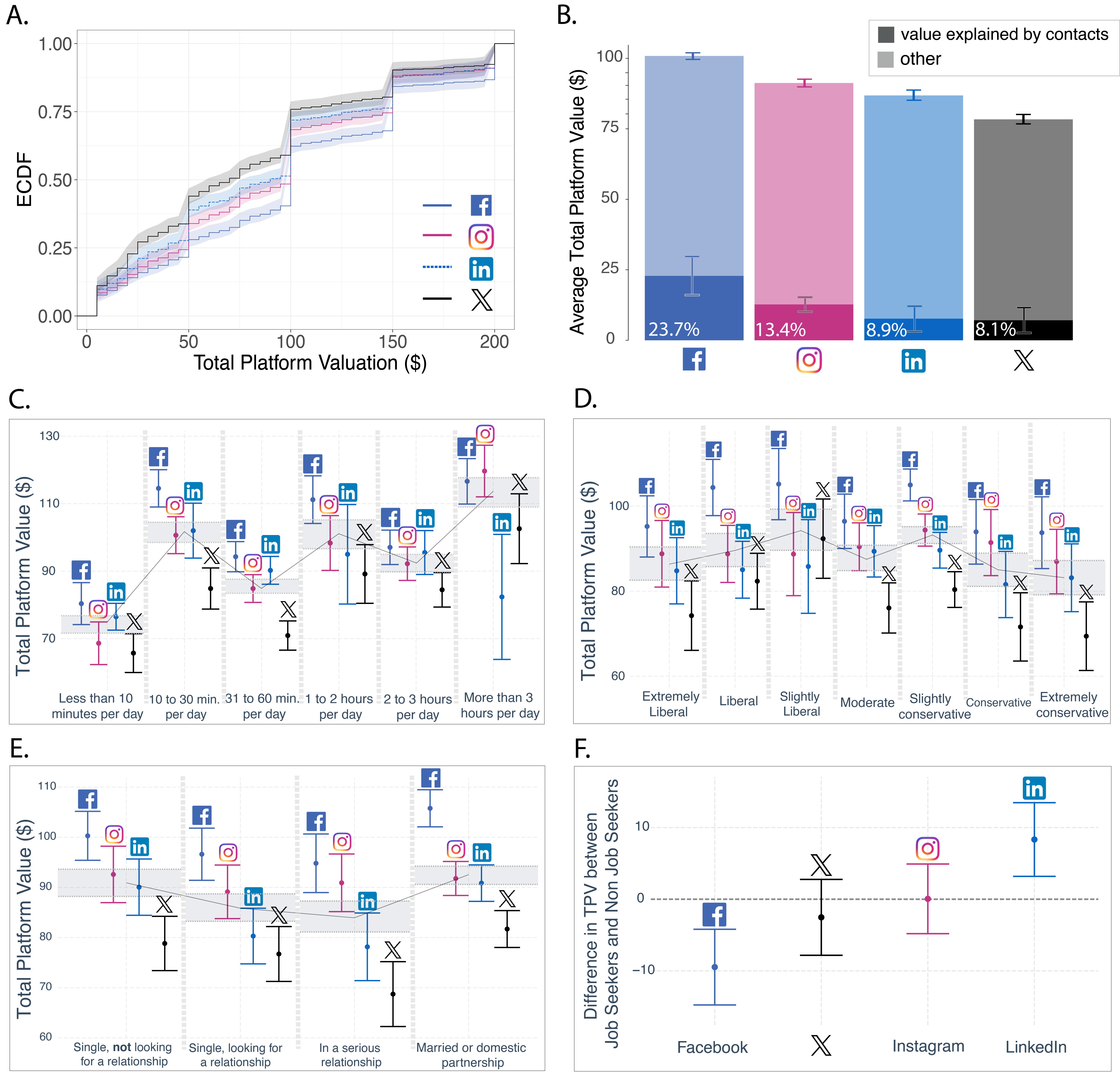}
    \caption{Total Platform Value and Local Network Effects. \small{This figure summarizes results on total monthly platform valuations from study 2. Panel A plots the empirical CDF of monthly valuations of retaining access to the platform for one month in contemporaneous US Dollars. Panel B plots the average of these values by platform, as well as the estimated share of that value created by local network effects. Panels C reports average valuations and 95\% confidence intervals conditional on platform and time spent on that platform. Panel D reports average valuations and 95\% confidence intervals conditional on platform and political ideology. Panel E reports average valuations and 95\% confidence intervals conditional on platform and relationship status. Panel F reports differences between job seekers and non-job seekers of average valuations and 95\% confidence intervals conditional on platform. }}
    \label{fig1}
\end{figure}

Figure~1 presents estimates of total platform value, the share of platform value explained by direct connections, and how platform value varies with time spent on the platforms, political leaning, relationship status and job seeking behavior.\footnote{For full statistical reporting of estimated differences between conditional averages and associated p-values for Figures 1 through 4, see SI section 6.} Results of our choice experiment measurements show that social media platform value ranges from \$78 to \$101 per consumer, per month, on average with Facebook having the highest estimated platform value of \$101 per month, Instagram \$91 per month, LinkedIn \$87 per month and X \$78 per month (Fig.~1A). Using equation [1], we estimate that between 8.1-23.7\% of total platform value is created by local network effects or the value consumers receive from their network connections (Fig.~1B). Facebook displayed the strongest local network effects as it had the highest share of platform value associated with direct connections. X displayed the weakest local network effects as it had the lowest share of platform value associated with direct connections. The $R^2$ of the regressions informing these estimates are low, consistent with strong individual heterogeneity in the share of platform value derived from local network effects. However, all local network effect estimates were significantly different from zero, indicating that a substantial fraction of platform value is derived from direct connections above and beyond aggregate network effects derived from the total installed base of the platforms for the average user. 

There is a positive relationship between time spent on the platforms and platform valuations (Fig.~1C), but no clear relationship between a platform's valuation and consumers' political ideology (Fig.~1D). Married and partnered individuals have significantly higher valuations for Facebook and LinkedIn than individuals who are currently in or are looking for serious relationships (Fig.~1E), perhaps because married or partnered individuals are more focused on their careers and socializing with their families online. Along similar lines, people who are actively looking for work value LinkedIn significantly more, and Facebook significantly less, than people who are not looking for work (Fig.~1F). 

\begin{figure}[h!]
\centering
    \includegraphics[width=\linewidth]{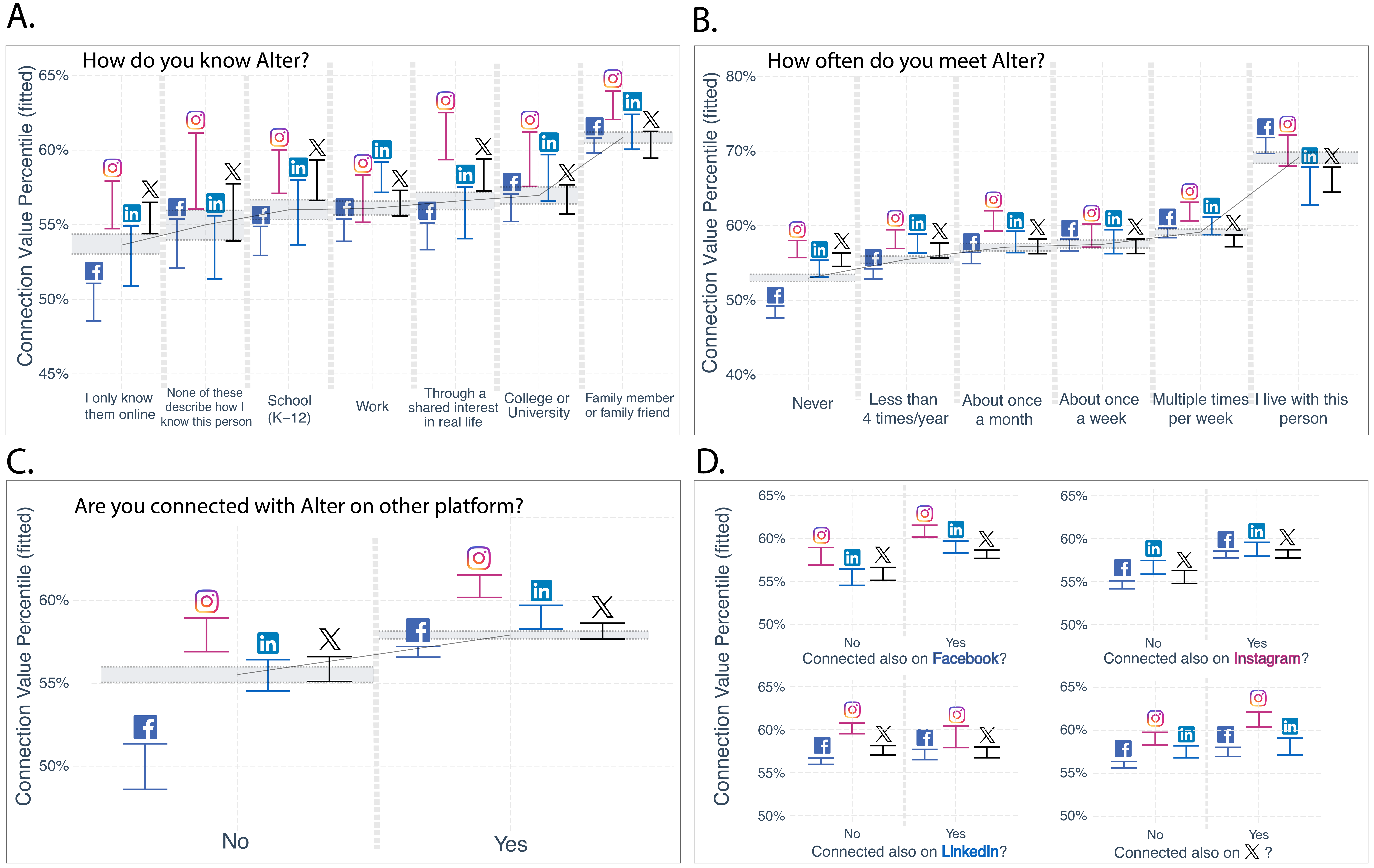}
    \caption{Tie Strength and Connection Value. \small {This figure summarizes results on the relative value of social media connections from study 1. Outcome is Connection Value Percentile (CVP) as defined in equation [2], with higher values indicating a stronger preference. Panel A reports average CVP and 95\% confidence interval conditional on platform and how the ego (survey taker) reports knowing the alter. Panel B reports average CVP and 95\% confidence interval conditional on platform and how often the ego reports seeing the alter. Panel C reports average CVP and 95\% confidence interval conditional on platform and whether the ego and alter are connected on at least one other platform. Panel D reports average CVP and 95\% confidence interval by platform of the ego-alter pair being connected on one of three other studied platforms. }}
    \label{fig2}

\end{figure}

We refer to the value of a single connection, as distinct from the total value of local network effects, as connection value. Consumers value closer, stronger ties more than weaker, arms length ties across all platforms  (Fig 2.). Specifically, digital connections to family members and family friends are more valuable than contacts known only online (Fig. 2A). Value from network connections is increasing in interaction intensity, with a sharp increase in value from digital connections to people we live with and a sharp drop in value for people who we never meet in person, especially for connections made through Facebook (Fig. 2B). For every platform, connections made on more than one platform are valued significantly more than connections made on only a single platform (Fig. 2C). This strongly suggests that connections to the same people across platforms are either imperfect substitutes or even complements, perhaps due to a desire for redundancy (if one platform were to be interrupted) or because different information is more usefully transmitted on different channels. There is heterogeneity in the extent to which connection on an alternate platform reduces CVP on the focal platform, suggesting varying degrees of complementarity across pairs of platforms (Fig. 2D).\footnote{Of course, purely correlational evidence is not conclusive. The ideal test of this hypothesis would use exogenous changes to network connections to identify their elasticity of substitution. For a complete discussion of measurement of complementarity and substitution, see SI section S.2.6} Being connected with someone on Facebook, Instagram or LinkedIn is associated with higher valuations for that same person on every other platform. However, being connected with someone on X is associated with higher valuations for that person  on LinkedIn, but not on Instagram and Facebook, suggesting lower relative complementarity.

Men value connections to women on social media significantly more than they value connections to other men, particularly on Instagram, Facebook and X, while women value connections to men and women equally across all platforms (Fig. 3A). The relationship valuations we estimated also reveal preferences for race homophily, as same race relationships are valued more highly than mixed race relationships, on average, across social media (Fig. 3B). Interestingly, the higher average valuations for race homophily are driven by a single platform. White consumers value relationships with other white consumers significantly more than they value relationships with non-white consumers on Facebook, while same race relationships are not valued significantly more than mixed race relationships on any other platform. The data also reveal value disassortativity by age on one platform---Instagram. Valuations for connections younger than oneself are significantly higher on Instagram than valuations for connections that are older or of the same age (Fig. 3C). Furthermore, on Instagram, connections to alters eighteen years old or younger are valued significantly more than any other age group, which are patterns not seen on any other platform (Fig. 3D).

These individual characteristic gradients are jointly confirmed in a rank-ordered, exploded, logit estimated on the pooled Wave 1 rankings. Nearly every positive correlate of CVP retains an independent, statistically significant coefficient when entered together in a pooled regression. This suggests that contact frequency, relational source, cross-platform connection, and demographic homophily each capture distinct sources of value. In a pooled regression, contact frequency is the most important predictor, then relationship-type and cross-platform connection status, with relative and absolute demographics being the least important. By anchoring the estimated marginal utilities from connection characteristics to average connection value, we also estimate monthly monetary valuations for different characteristics. The largest value is for `I live with this person' which raises connection value by 4.6 cents a month on Facebook. From a policy perspective, the most important takeaway is that being connected on another platform remains a positive predictor of connection value after controlling for other dyadic characteristics, further supporting the theory that connections on different platforms are complementary. See SI section 3.1.2 for full results and details.

\begin{figure}[h!]
\centering
    \includegraphics[width=\linewidth]{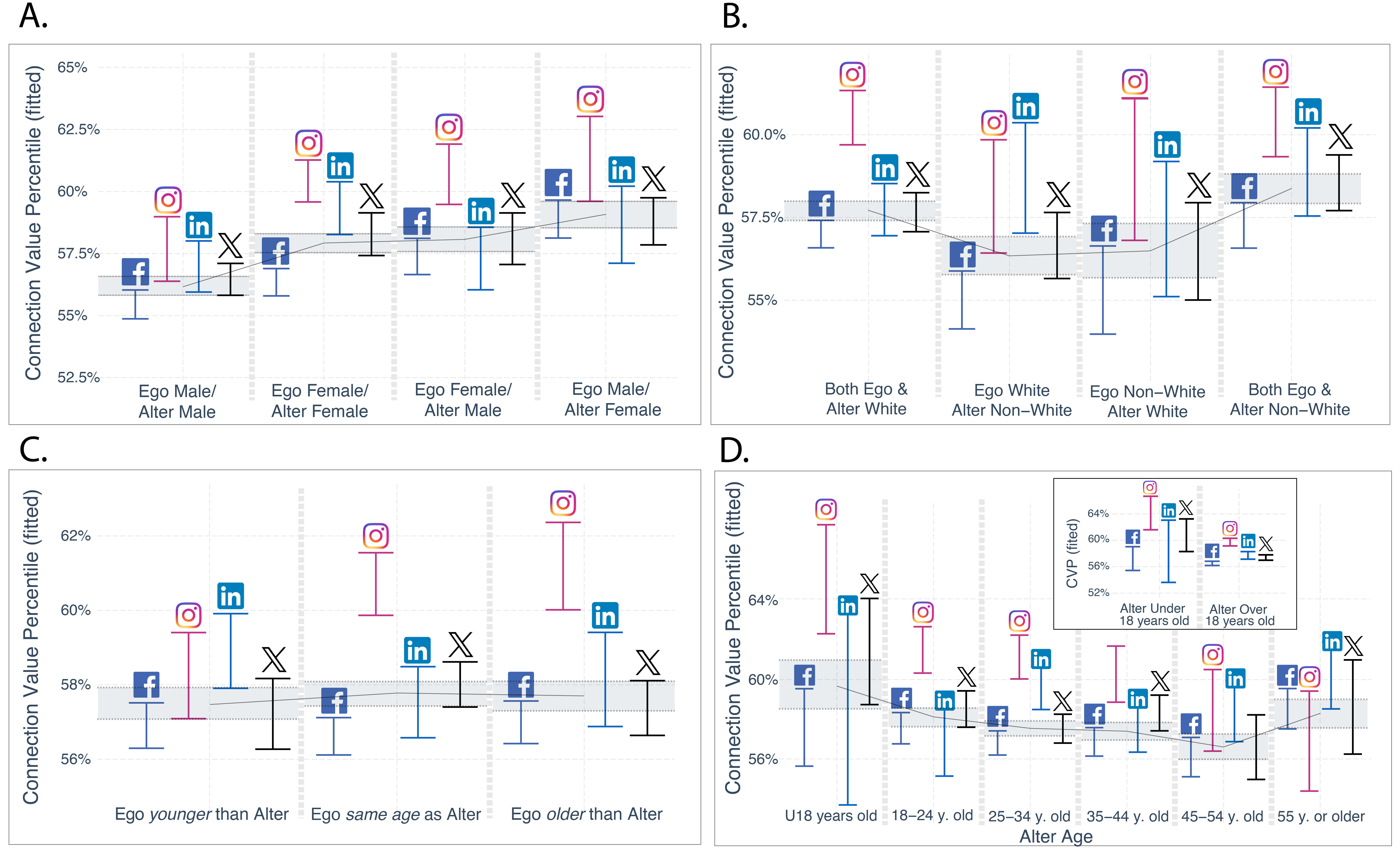}
    \caption{Relational Demographics and Connection Value. \small {This figure summarizes results on the relative value of social media connections from study 1. Outcome is Connection Value Percentile (CVP) as defined in the text, with higher values indicating a stronger preference. Panel A reports CVP and 95\% CIs conditional on platform and ego (survey taker) and alter gender (an `other' category, with wide confidence intervals, is omitted). Panel B reports CVP and 95\% CIs conditional on platform, and ego and alter race. 
    Panel C reports CVP and 95\% CIs conditional on platform, and ego and alter's relative age, where same age indicates being in the same age bin. Panel D reports CVP and 95\% CIs conditional on platform and alter's age bin, and includes an insert where age-bins are split alternatively: under age 18 and otherwise.}}
    \label{fig3}

\end{figure}


\begin{figure}[h!]
\centering
    \includegraphics[width=\linewidth]{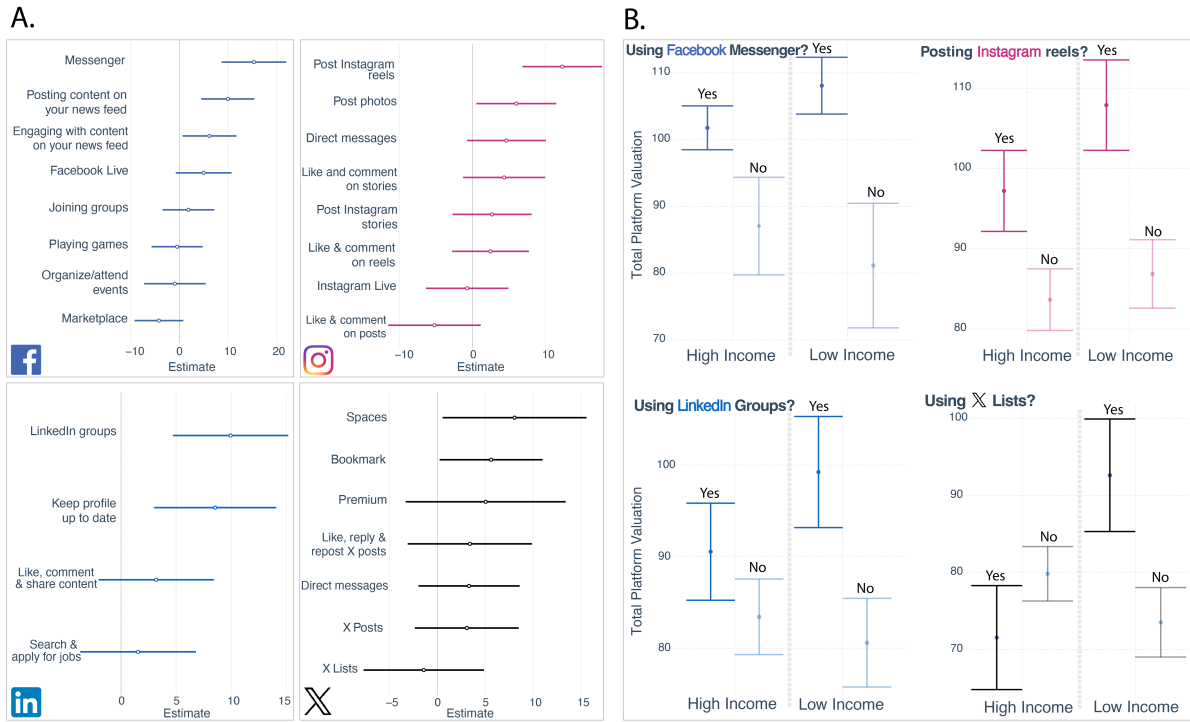}
    \caption{Relational Demographics and Connection Value. \small {This figure summarizes results on how TPV varies with platform feature use from study 2. Panel A reports, for each platform, the increase in TPV associated with feature use. Coefficients and 95\% confidence intervals are derived from a regression of total platform value on egos' usage of each platform feature. Panel B reports how platform value varies with feature use and ego income. Plotted are conditional average platform values for users and non-users of the feature, for egos of different income groups. The split for high vs. low income is at one hundred thousand dollars. SI Table S11 reports the point estimate and p-values for two differences: The difference between conditional average total platform value for high and low income egos who use the feature, and the difference in differences between conditional average total platform value for high income feature users and non-users less the difference between low-income feature users and non-users.  }}
    \label{fig3}

\end{figure}

Analysis of the value of platform features is also instructive (Fig. 4). First, posting content on one's Facebook feed, posting new photos and reels to Instagram and posting profile updates, like new educational credentials or achievements, job transitions or work promotions to LinkedIn, all add significantly to higher platform valuations (Fig. 4A). These results are in line with the higher valuations for followers than for followees on Instagram and X reported earlier. The ability to express ourselves to our networks and to update the world about ourselves and our thoughts provides significant value to consumers. This is in line with research like Filippas et. al. \cite{production_of_social_media} who argue that we desire attention online and that follower-followee relationships in part reflect the bartering of attention.  Second, group interactions add significant value beyond dyadic network connections. For example, the use of LinkedIn Groups and X Spaces both significantly increase consumers' valuations of LinkedIn and X on average (Fig. 4A). Results on the value of group interactions lend support to recent work documenting the importance of higher order group relationships, characterized by simplicial complexes in networks, expanding our theoretical conceptualization of networked value beyond dyadic interactions~\cite{sarker2024higher, sarkersinan2024working}. Third, the use of messaging features that enable direct, often synchronous, real-time communication with others increase consumers' valuations. For example, use of Facebook Messenger, on average, increases a consumer's valuation of Facebook by approximately \$12 per month (Fig. 4A). Results on direct communications features, which add value beyond our estimates of the value of networked connections, suggest that our estimates of the importance of local network effects may be conservative as these features capture the value of not just having network connections, but being able to communicate with them directly in real-time. Fourth, features that allow users to organize and curate content add significantly to platform valuations. For example, the ability to bookmark X posts adds approximately \$5 per month, on average, to consumers' valuations of X. These results are in line with prior evidence of the value of content curation and organization features demonstrating, for example, that the ability to curate playlists on music streaming services increases consumers' willingness to pay for those services~\cite{oestreicher2013content}.


Perhaps even more instructive, however, is how different consumers value the same features differently. Figure 4B reports conditional average platform values for four features — Facebook Messenger, Instagram Reels, LinkedIn Groups, and X Lists — stratified by income, with the high/low cutoff drawn at \$100{,}000 annually. Two patterns emerge. First, low-income users who adopt these features value them significantly more than high-income users who adopt them. Second, the gap between feature users and non-users is significantly larger for low-income than for high-income consumers (SI Table S11). Together, these results indicate that social media networks deliver value to lower-income consumers in fundamentally different ways than to higher-income consumers, and that this divergence is concentrated precisely in features designed to extend reach beyond one's existing network.

When the purposes of these features are considered, we believe the results support the use of social media networks by lower income consumers to achieve social mobility, echoing research demonstrating that social media network connections to higher socio-economic status (SES) peers is ``among the strongest predictors of upward income mobility identified to date"~\cite{chetty2022social1, chetty2022social2}. For example, LinkedIn's official description of the Groups feature notes that it is intended to enable consumers to ``build relationships... that can enhance your skills and knowledge... seek feedback... [and to] learn from others' experiences through group discussions." Similarly, a key element of Facebook Messenger is that it allows consumers to speak to people they are not connected to on Facebook, which potentially enables them to break through to new social networks with higher SES, value and resources that don't exist in their own social networks. X Lists enable consumers to ``manage and guide conversations... monitor competitors... highlight local or small businesses... [and] track leads passing through their sales funnels." These three features are designed to enable consumers to reach beyond their own social networks, to enhance skills and knowledge through interactions with those not yet in their own social networks and to seek feedback, collaboration and professional development opportunities in networks not accessible through direct connections. Such use cases are the precise enablers of social mobility described by Chetty et al~\cite{chetty2022social1, chetty2022social2} in their large-scale analysis of Facebook.

Access to the creator economy may be another driver of value for low income consumers. Instagram describes Reels as enabling consumers to ``create multi-clip videos... select from a variety of effects created by Instagram and creators around the world... share your reels with your followers on Feed, or make them available to the wider Instagram community... [and] recommend your Reel on Facebook." Thus, Reels not only enables participation in the creator economy, but also content sharing beyond one's direct network. As the creator economy attracts aspiring, amateur creators seeking to participate in influencer marketing, platform payouts and the monetization of short-form video, it is perhaps not surprising that lower income consumers value Instagram Reels more than higher income consumers.

Combining our total platform valuation estimates with up-to-date counts of monthly active users (MAUs) reported by each platform, we can estimate the total consumer surplus generated by each platform in the US, monthly and annually. Using similar methods, Allcott et. al. ~\cite{allcott2020welfare} estimated Facebook generates \$31 Billion of consumer surplus per month and \$372 Billion annually. We estimate that Facebook generates \$17.9 Billion per month and \$214.9 Billion annually, Instagram generates \$12.6 Billion per month and \$151.7 Billion annually, LinkedIn generates \$8.7 Billion per month and \$104.9 Billion annually, and X generates \$4.4 Billion per month and \$52.6 Billion annually. Given that 97\% of total welfare gains from technological innovations are captured by consumers, these estimates seem reasonable~\cite{nordhaus2004schumpeterian}. The similarity and consistency of our estimates compared to Allcott et. al. ~\cite{allcott2020welfare} also give us confidence in our shared approach to estimating the total consumer welfare gains generated by these four social media platforms in the United States. It is not surprising that our estimates are slightly lower than Allcott et al.~\cite{allcott2020welfare} given they only focus on highly active Facebook users (consumers who use Facebook more than 15 minutes per day).

Although our work is the first to quantify the value and structure of local network effects in the digital economy and to describe a systematic and rigorous method for doing so, it is not without its limitations. First, in our setting, we do not have access to exogenous variation in the number or types of respondents' social network connections. As such, causal estimates of the marginal value of adding or losing specific types of connections are elusive and omitted variables could bias our estimates of connection value. We took several steps to estimate and counteract our vulnerability to the lack of experimental variation in connections in our analysis. Our pre-registered ordinary least squares (OLS) and LASSO specifications, which include numerous controls and interactions, go a long was toward controlling for omitted variables. These results show that, while controlling for many other factors, the number of connections (or alternatively, followers and followees) are highly significant predictors of platform valuations. Our pre-registered analyses demonstrate that between 14\%-25\% of the variation in platform value is explained by our predictors lending support to our post hoc control of omitted variables. Furthermore, interactions between connection counts and platform engagement are also predictive of platform value, suggesting that, as respondents engage more with the platforms, increases in their platform valuations increase with their connection counts. Taken together, these results suggest that local network effects are a consistent, significant and economically meaningful driver of platform value and consumer surplus. 

We encourage future work to measure local network effects using exogenous variation in the number and types of respondents' connections to causally identify the contributions of direct connections to platform value. We know such analyses are possible given prior large scale experiments that randomly varied the number and types of connections on social platforms to measure, for example, the strength of weak ties~\cite{rajkumar2022causal}. Experiments varying the connection network could also provide insight into the value of alters' connections to an ego's valuation of the connection. Understanding the value of these second and higher-order connections to user value is another promising direction for future work. 

Second, all incentive compatible choice experiments are potentially vulnerable to anchoring effects in the solicitation of valuations. When presenting potential valuations to respondents, the maximum values presented could anchor their perceptions of typical or acceptable valuations. We directly addressed potential anchoring effects using a number of preregistered designs. First, in Wave 1, we provided respondents a slider of valuations and an open-ended free response value solicitation that exceeded the slider range. Second, we randomly assigned different maximum offer values in Wave 2 to experimentally vary distinct anchors enabling a systematic examination of our vulnerability to anchoring effects. Controlling for random assignment to anchors has a minimal impact on point estimates of connection valuations, which remain positive and significant on all platforms, suggesting that our results are robust to anchoring effects. 

Third, platform and connection value solicitation in our context is self reported and inherently subjective, making efforts to generalize findings such as ours more challenging. Our approach to this problem goes to the core of our research design. We present a rigorous attempt to quantify inherently subjective valuations by constructing an incentive compatible value solicitation mechanism that provides real monetary incentives for the revelation of truthful and accurate valuations of platforms and connections. Yet, incentive compatibility itself has weaknesses applicable to this entire literature. Most importantly, it depends on participants having a reasonable expectation of payments upon deactivation, which may be especially problematic for people with higher true valuations. Prior incentive compatible valuations have approached this problem in different ways. Allcott et. al. ~\cite{allcott2020welfare} allow for a free response in the BDM value solicitation, but winsorize maximum replies at \$170, as this is the maximum payment they allow in their randomized BDM offer, though this cutoff is not made known to participants. Burztyn et. al. ~\cite{bursztyn2023product} provide a multiple price list with a \$200 maximum and fit a thin tailed curve through their results, allowing for imputted values above \$200, and conclude this does not change their findings. 

We analyze different winsorization cutoffs for maximum value and learn several things. First, pooling all platforms, less than 1\% of respondents have valuations greater than \$1000. This seems reasonable given some creators earn tens of millions of dollars a year from their social media presences and could not earn this revenue without social media. Second, this interpretation is supported by the fact that X and Instragram are the platforms most impacted by winsorization, suggesting the long tail of high valuers is most important on those platforms where creator economies are most prevalent, compared to Facebook and LinkedIn. These facts suggest a need for some caution when interpreting the aggregate welfare implications of our estimates if a substantial fraction of value is driven by a small number of high valuers who are imprecisely measured. However, the fact that less than 1\% of our sample are high valuers makes us more confident in our total platform valuations. Furthermore, winsorization and maximum values do not effect our relative connection value estimates or platform or connection valuations for typical platform users, giving us confidence in our results. In addition, our approach is at or beyond the current state of the art, which has extensively solicited valuations with incentive compatible designs even when such solicitations promise to enforce BDM contracts under circumstances known to experimenters ex-ante to be impossible (9). It has been shown, even in these situations, that the promise of payments themselves incentivize participants to respond to value solicitation questions truthfully and accurately. Furthermore, our extensive real platform data collection mitigates vulnerabilities associated with self reporting of platform network and use variables.

Fourth, network sampling and tie valuation are inherently challenging. Since we cannot ask respondents to value all connections, we must sample their networks in a way that is both representative and likely to choose connections respondents know well. Soliciting connection value may also be unfamiliar for and inconsistent across respondents. We therefore sampled the top four connections and any other four connections and asked respondents for relative rankings rather than numeric valuations, which allows us to ensure consistency across comparisons while estimating numeric valuations by combining total connection values with relative rankings to understand how valuations vary by tie type and network structure. While prior connection value research uses name generators, typically soliciting between three and fifteen connections, and solicits tie closeness using self reported Likert scales, binary indicators of whether someone is a ``connection," or observations of time spent together, no prior work applies incentive compatibility to connection value solicitations. 

Our solicitation of top and other connections ensures sampling from the entire connection distribution and those whom respondents know well. But it may be that respondents are better able to value close connections than other connections. To understand the reliability of our connection value estimates, we reanalyzed our CVP results using only the four closest connections given these are the connections over which consumers are likely to have the strongest and most consistent preferences. If these results differed from results based on all eight connections, we might be concerned that respondents are not able to value the four other connections as accurately or consistently as top connections. Our analysis shows, however, that the relative rankings are robust to the acquaintance generation process and return similar and consistent results (see S4.1 in the SI).

Fifth, we may be concerned that an online sample is biased or not representative. However, Lucid is a highly reliable online survey platform ~\cite{coppock2019validating} widely used in academic research ~\cite{pennycook2021shifting, benzell2020rationing, solis2021covid} to solicit representative samples of US respondents, and the size of our study population, which approaches 20K respondents, makes us less vulnerable to the idiosyncrasies of a small online sample. By comparison, our sample is over six times larger than that of Allcott et. al. ~\cite{allcott2020welfare}, which offered the ``the largest-scale experimental evidence available to date" when it was published in 2020. Moreover, we required Lucid to verify that every survey respondent was a monthly active user of the platforms for which we were conducting incentive compatible choice experiments, performed multiple attention checks during the experiments filtering out inattentive respondents, required respondents to connect with our research accounts to verify their reporting of platform usage, verified each of their 84K network connections by examining connection lists on their profiles, and collected substantial additional information, by survey and by downloading observational data, to control for observable differences between respondents and to ensure the reliability and representativeness of our analysis. We then used the additional information we collected to analyze pre-registered robustness checks that showed the strong and consistent contributions of connection value to total platform value holding observable factors constant (see SI Tables S12-S21). We believe these measures support the robustness and validity of our data collection from our large representative sample and mitigate against inattentive responses and missing variables.

Finally, asking respondents to value platforms and connections on a monthly basis may either over- or under-estimate the long term value of connections. For example, it could be that giving up a social platform and your connections on it for a month imposes a low cost because you know you will be back online and in touch in a month. If respondents had been asked to give up platforms and connections for six months or a year, they might have assigned higher values, making our one-month-based value solicitation undervalue platforms and connections. On the other hand, short term connection value may be high if respondents want to communicate with their connections immediately. When confronted with a more permanent, year long obstacle to connection, they might seek alternative modes of communication, substituting away from social media based connections entirely, lessening their value. We follow the literature and solicit values based on deactivation for four weeks. However, we encourage future work in this domain to examine different time horizons, like six months and a year, to understand how valuations depend on elasticities of substitution.

Despite these limitations, our work presents the first large-scale, empirical measurement of local network effects in the digital economy and the largest-scale experimental evidence on the consumer surplus generated by social media to date. Our incentive compatible online choice experiments revealed that between 8.1-23.7\% of social media's total value emanates from the value consumers receive from their direct connections and that the social media platforms we studied each generate between \$53 billion and \$215 billion a year in consumer surplus in the United States alone. Our analysis also showed that stronger ties are more valuable on Facebook and Instagram, while weaker ties are more valuable on LinkedIn and X, that men value connections to women on social media significantly more than they value connections to other men, while women value connections to men and women equally across all platforms, that white consumers value relationships with other white consumers significantly more than they value relationships with non-white consumers on Facebook while, on Instagram, connections to alters eighteen years old or younger are valued significantly more than any other age group (patterns not seen on any other platform) and, as Chetty et al~\cite{chetty2022social1, chetty2022social2} predicted, low income consumers value social media features that enable them to reach across networks to higher socio-economic status connections more than high income consumers. These results are essential to understanding the welfare implications of social media platforms and inform policymakers seeking to regulate social media and managers setting platform monetization, user acquisition, and algorithmic connection recommendation strategies. We hope the methods developed here provide a foundation for future analyses of local network effects and the welfare effects of social media.

\subsection*{Materials and Methods} 
We analyzed three main measurements, derived from our choice experiments, to quantify network effects in the social media economy: total platform valuation (TPV), the share of platform value explained by connections (SPV) and connection value percentile (CVP). Total platform valuation is the average total monthly value of a platform implied by our choice experiments measured in U.S. dollars. The value of social media platforms derives, in part, from direct network connections and, in part, from other platform features like news feeds, job search features, marketplaces, groups, reels, lists and short form videos. The share of platform value explained by connections (SPV) is the fraction of total platform value (TPV) derived from consumers' direct connections. 
 We estimate SPV by first evaluating the following regression:

\begin{equation}
TPV_{p,i} \;=\; \delta_p \;+\; \lambda_p \left(1 - e^{-z_p\,C_{p,i}}\right) + \varepsilon_i ,
\end{equation}

\[
z_p \;\equiv\; -\log(1-\pi_p), 
\]

This functional form is motivated by a redundant signal model where receiving the benefit of connection on a platform requires successfully transmitting a message at least once, but transmitting the message additional times is redundant. For example, the message may represent an important social event or job opportunity. For a longer discussion, see section S3.2.3 in the SI. $\pi_p$ is the probability that a single connection provides the signal, $\lambda_p$ is the value of receiving the signal, and $\delta_p$ is the value of platform participation in the absence of receiving the signal. $TPV_{p,i}$ is consumer $i$'s total monthly valuation for platform $p$, and $C_{p,i}$ is the number of users consumer $i$ is directly connected to on platform $p$. On Facebook and LinkedIn these are friends and direct connections. On Twitter and Instagram, two separate terms for $\pi$ and $\lambda$, the probability of receiving the benefit of connection and the value of that signal being delivered, for followers and followees were estimated. The results of these regressions are reported in the SI (Table S22). If we define $TPV_p$ as the average of $TPV$ for all users ($\sum_iTPV_{p,i}/I$, where $I$ is the total number of respondents), then the share of platform value not explained by connections is the platform's intercept term ($\frac{\delta_p}{TPV_p}$) and the share of platform value explained by connections $SPV_p$ is $1-\frac{\delta_p}{TPV_p}$.




Finally, Connection Value Percentile ($CVP$) measures the relative value of a particular network connection. Egos in our study named and described up to eight alters they are connected to on the focal platform. On a subsequent survey page, we solicited the relative importance of these contacts using a rank-order question (connection Ranked Choice (CRC)).\footnote{The text of the relevant question is ``Please rank the following connections from most valued on [PLATFORM] (top) by you to least valued on [PLATFORM] (bottom) by you.'' Note that we do not give detailed instructions to survey takers about what is meant by “most valued.” Interpreting these results as informative about the nature of connection complementarity across platforms requires interpreting importance as marginal value and the evidence suggests such an interpretation was employed by participants (see Section S2.6 in the SI for more discussion).} Using the ranking provided, we calculated $CVP$ for each ego-alter pair as \begin{equation}
    CVP_{i,j,p} = 100(1 - \frac{CR_{i,j,p} -1}{TCR_{i,p}})
\end{equation}

where $CR_{i,j,p}$ is the ranking given to connection $j$ by ego $i$ on platform $p$ on our relative ranking scale and $CFR_{i,p}$ is the total number of connections ranked by consumer $i$ on platform $p$. We then estimated the relative value of different types of connections by analyzing conditional average $CVPs$ for egos and connections of different types.\footnote{Our results are very similar when restricting attention to only top four connections, indicating the correlates of CVP we find are not solely driven by small cardinal differences in value between e.g. 5th and 8th ranked connections, but rather likely reflect substantive cardinal differences in value between these connections.}

\subsection*{Acknowledgments}We thank D. Eckles, E. Brynjolfsson and participants of the MIT Conference on
Digital Experimentation, the Networks, Markets and the State conference at NYU Abu Dhabi, the Workshop on Information Systems
Economics, the MIT Social Analytics Lab and the IDE Research Seminar for their
helpful comments on earlier drafts of this manuscript. We also
thank members of the MIT Initiative on the Digital Economy for their intellectual support. The study
was approved by the MIT Institutional Review Board (IRB Protocols
E-3229 and E-5633).


\bibliographystyle{unsrt} 
\bibliography{pnas-sample}
\end{document}


\date{}

\baselineskip24pt

\maketitle 

\vspace{2cm}
{\raggedright
{\bf This PDF file includes:}\\[-1.65ex]
Description of Data Collection\\[-1.65ex]
Results\\[-1.65ex]
Robustness Analyses\\[-1.65ex]

References}
\pagebreak
\tableofcontents
\renewcommand{\thetable}{S\arabic{table}}
\renewcommand{\thesection}{S\arabic{section}}
\renewcommand{\thesubsection}{\thesection.\arabic{subsection}}
\renewcommand{\thefigure}{S\arabic{figure}} 
\renewcommand{\theequation}{\arabic{equation}} 
\makeatletter 
\def\tagform@#1{\maketag@@@{(S\ignorespaces#1\unskip\@@italiccorr)}}
\makeatother


\pagebreak

\section{Supplemental Materials and IRB Approval}

This document contains supplemental materials for "Measuring Social Media Network Effects.'' Our research was approved by the Massachusetts Institute of Technology’s (MIT’s) Institutional Review Board\footnote{IRB protocol ID for Wave 1: E-3229, IRB protocol ID for Wave 2: E-5633} and analyses were pre-registered at the Open Science Foundation (OSF).\footnote{https://osf.io/qs4kn/?view\_only=5cee408999b5484dafcc9626e88c953b} For Wave 1, we ask respondents to follow our research accounts on social media platforms. For Wave 2, we collect email addresses of respondents for following up with them for additional rewards. For both waves of the surveys, we permanently deleted all personally identifying information (social media handles, email addresses, and names of their connections) as soon as the data collection was completed. As noted, the MIT IRB reviewed and approved all of these protocols. The survey instruments for Instagram for both waves are appended at the end of this document. The survey instruments for other platforms are identical.

\section{Research Design and Data Collection}

\subsection{A Gap Between the Sociology of Networks and the Economics of Network Effects}\label{sociology review}

Our research speaks to literature from sociology on the nature and role of networks, and from economics on the value of network effects. While both fields have established the importance of relationships, including online relationships, ours is the first paper to measure local network effects, to assign economic value to relationships in an incentive compatible way and to distinguish heterogeneity in the value of network effects across social media platforms.

\subsubsection{Measuring Connections and Their Value}

While we are unaware of other papers valuing connections on social media, there have been sociological studies of the value of ties and the predictors of high-value relationships. Many studies have measured the strength of ties~\cite{aral2016future, marsden1984measuring, fischer1982what}. For example, in Marsden and Campbell~\cite{marsden1984measuring}, which measured predictors of tie importance, dependent variables included self-reported closeness on a scale of 1-3\footnote{Namely: `acquaintance (coded 1), a good friend (2), or a very close friend (3)'} and a composite set of objective measures of closeness (e.g. time spent together). The study conducted surveys of individuals in three cities (1,013 native-born white males in the Detroit metropolitan area, 496 non-institutionalized adults in Aurora, Illinois, and 820 drawn from the eligible voters in an unnamed West German city). The study began with each participant self-reporting the names of their three closest friends. Then, many features of their relationship were coded, including both measures that were thought to predict closeness (such as whether the two were related) and measures of closeness. The study found that kinship was a relatively strong predictor of closeness, but that the explanatory variables had low predictive power overall. 

A second classic study, with elements similar to our own by Fischer~\cite{fischer1982what}, asked ``What do we mean by friend?'' In this study the dependent closeness variable was binary---whether a person was classified by the individual as a friend. The study was conducted on 1050 adults living in northern California and began with the interviewer asking open-ended questions about the subject's relationships, to generate a list of 2-26 acquaintances per subject. The subject then reported several facts about the individual and their relationship. 59\% of acquaintances qualified as friends, while 41\% did not.  The study found no strong patterns in which types of contacts were considered friends, as an overwhelming majority of non-relative acquaintances were classified this way.

Other studies have looked into how different friendships provide different types of value, without trying to construct a measure of overall value (e.g. Wellman and Wortley~\cite{wellman1990different}) as well as which relationships tend to be emotionally intense over the long term (e.g. Roberts et al.~\cite{roberts2011costs}). In the first study, different measures of social support (e.g. home health visits, financial assistance) are associated with different types of acquaintances. In the latter study, 25 students were asked, over an 18-month period, who they felt emotionally intimate with. Features of these relationships were solicited and the study found kinship ties positively predicted enduring, emotionally close relationships. 

Our dependent variable (CVP) has advantages over these and other approaches. First, and perhaps most importantly, it quantifies value in an incentive compatible choice framework. Such measurement enables economic analyses relevant to social, industrial and judicial policy by standardizing value measurement in a quantitative framework that is comparable across consumers and likely to solicit high fidelity measures of economic value due to incentive compatibility. Second, soliciting CVP over eight options also provides more information than a binary classification or a 1-3 scale. Rigorous quantification of subjective value is important because objective measures, like time spent, may not be sufficient to capture the subjective value provided by social media platforms as we may communicate less frequently with some very high value relationships. Third, a rank-ordered list may be easier for subjects to reason over than a binary classification or scale, when these use vaguely defined terms~\cite{krosnick1988test}. Finally, in this setting, the rank-ordering task is conceptually easier than soliciting cardinal connection preferences through BDM or a similar mechanism. While BDM may work well for soliciting the overall value of a platform, it would be cognitively hard to estimate the value of a particular connection in monetary amounts.

Our approach shares some weaknesses with previous approaches. We rely on self-reported acquaintance names and evaluate relative value over this set of names. However, in section \ref{FVP_robust} we provide robustness analyses related to the concern that our results are driven by our acquaintance generation process by analyzing data restricted to just the four closest contacts (omitting the four other top-of-mind contacts). This restricts attention to the subset of names that the subject is likely to have the strongest and most reliable feelings about. Our analyses indicate our relative ranking is robust to the acquaintance generation process. Moreover, if two connections are similarly ranked, forcing respondents to select one over the other may increase cognitive load and result in random responses~\cite{descioli2009alliance}. However, relative ranking is still a much easier question framing compared to separately soliciting connection valuations~\cite{descioli2009alliance}.

\subsubsection{Networks as Social Capital and Social Support}

Seminal work by Granovetter~\cite{granovetter1973strength} proposed the importance of weak ties for accessing information and opportunities, including job opportunities. Rajkumar et al.~\cite{rajkumar2022causal} conducted a large-scale randomized experiment on LinkedIn to causally test the impact of weak ties on employment outcomes. They found evidence supporting Granovetter’s thesis that weak ties have the greatest impact on job mobility while strong ties have minimal impact, with the caveat that the relationship is non-linear, meaning moderately weak ties provide greater employment opportunities than strong ties or very weak ties.

Coleman~\cite{coleman1988social} posits that social capital—comprising relationships, trust, and norms within social structures—plays a crucial role in developing human capital, such as skills and knowledge. He identifies three primary forms of social capital: obligations and expectations within relationships, information channels that facilitate knowledge exchange, and social norms that guide behavior. Coleman emphasizes the significance of “closure” in social networks, where tightly-knit communities effectively enforce norms and trust. He illustrates this with examples like family dynamics and community involvement, highlighting that the presence of social capital can enhance educational outcomes and individual development. Unlike physical or human capital, social capital is embedded in social relationships and is often a collective asset, benefiting not just individuals but the broader community. Coleman’s work underscores the importance of social structures in shaping individual actions and outcomes.

Recently, Chetty et al.~\cite{chetty2022social1, chetty2022social2} conducted one of the largest studies examining the impact of social capital on economic mobility in the United States. Utilizing data from 21 billion Facebook friendships, they introduced the concept of “economic connectedness,” which measures the degree of interaction between individuals from different socioeconomic backgrounds. Their research revealed that communities with higher economic connectedness exhibit greater upward income mobility for children from low-income families. The studies identified two primary factors influencing economic connectedness: “exposure,” referring to the presence of individuals from diverse socioeconomic backgrounds within shared environments, and “friending bias,” the tendency to form friendships within one’s own socioeconomic group. Both limited exposure and high friending bias were found to equally impede cross-class friendships. These findings suggest that policies aimed at increasing socioeconomic diversity in communal settings and encouraging cross-class interactions can enhance economic mobility.

\subsubsection{Strategic Implications of Network Ties}

There is also a sociological literature on the strategic value of networks. Like this manuscript, research of this type emphasizes that not all ties are created equal. In particular, the value and impact of a tie may depend on overall network structure. 

One prominent theory of how networked social structure drives value was proposed by Burt \cite{burt2000network}. Burt distinguishes between connections that broker across ‘structural holes’ and connections that simply increase the density of connections between nodes that are already closely connected. He argues that the sociological evidence is consistent with the former being most valuable, but with both playing important roles. Burt finds evidence consistent with this theory in data on the networks of managers employed in a large corporation \cite{burt2004structural}. Managers who bridged structural holes of disparate business entities were paid more and were noted to have better ideas. Building on the theory of structural holes, \cite{podolny2001networks} argues that those that bridge holes should have disproportionate value in settings with high practical uncertainty but low social uncertainty. Similarly, \cite{white2001markets} emphasizes that real economies are not pure markets conducted by a Walrassian auctioneer, but rather embedded in social networks that can inspire managerial strategies or retard innovation. 

Another theory of how network connection topology relates to value was developed in \cite{jackson2003strategic}. This model, relevant for political and social as well as economic settings, suggests that when individuals self-interestedly form or disconnect links, there are typically multiple equilibria, and none are guaranteed to be efficient. Papers have considered strategic action in the context of political networks as well. One foundational paper in this domain is \cite{padgett1993robust}. This paper evaluates the rise of the Medici banking family to dominate renaissance Florence. The authors, Padgett and Ansell, document how Cosimo de Medici made a series of marriages and business deals that placed him at the center of the political, social and economic world. They theorize his foresight set up his dynasty for centuries of success. Relatedly, \cite{benzell2021network} found royal marriage networks play a major causal role in preventing wars in Early Modern Europe.

\subsubsection{Digital Network Connection Research}

Several papers have also looked at the importance of connections on digital platforms. Wellman et al.~\cite{wellman2001does} analyzed data from one of the largest-scale web surveys in 1998 to study whether internet use affects social capital. The key finding of their work is that online interactions supplement face-to-face communication. Online network connections maintain and enhance offline face-to-face connections, particularly connections which are socially and geographically dispersed. Moreover, internet use increases involvement in offline organizational and political activity. Similarly, Burke et al.~\cite{burke2011social} studied the impact of using Facebook on social capital. They find that receiving messages from friends on Facebook is linked with increases in bridging social capital. Wellman and Hampton~\cite{wellman1999living} also find that the internet strengthens real-life local links within neighborhoods and households. 

From a network science perspective, Onnela et al.~\cite{onnela2007structure} analyze communication patterns of millions of mobile phone users to construct a population-level social network and explore the relationship between network topology and tie strength. They find that there is a correlation between tie strength and the local network structure around the tie. They further explore how removing ties impacts network integrity. Removing weak ties results in network collapse, while removing strong ties has little impact on network integrity while still playing an important role in maintaining local communities.

Another literature investigates the role of social media connections in creating and maintaining self-reported social capital. In this literature, social capital may be “bridging” i.e. increasing a user’s access to novel or socially distant information, or “bonding” i.e. deepening existing ties \cite{putnam2000bowling, williams2006social}. The concept of bridging ties is connected to Burt’s structural holes \cite{burt2000network}, as these holes are why these bridges are valuable. It is also connected to the concept of ‘weak ties’ or the more distant relations who in aggregate provide many important opportunities \cite{granovetter1973strength}. One early survey of 800 MSU undergraduates users found more intense Facebook users had more perceived social capital \cite{ellison2007benefits}, especially “bridging” social capital as measured through questions like “Interacting with people at MSU makes me feel like a part of a larger community”. A 2011 survey of 614 Midwestern US university nonfaculty staff also found that those with more intense Facebook friends and who performed more Facebook relationship maintenance activities had stronger perceived bridging capital both online and offline \cite{ellison2014cultivating}. Research suggests that this is due to Facebook lowering the cost of maintaining social ties \cite{ellison2010little}.

Several studies have also measured tie strength quantitatively using surveys and digital trace data. Petróczi et al.~\cite{petroczi2007measuring} developed an 11-question survey scale to provide a continuous measure of tie strength in online communities. Gilbert and Karahalios~\cite{gilbert2009predicting} instead measured tie strength using a prediction model based on social media data. Jones et al.~\cite{jones2013inferring} focused on inferring real-world tie strength using Facebook data. Facebook users were asked to name their closest friends in real life. In this study, interaction frequency on Facebook successfully predicted real-world strong ties. This provides confidence in our methods, as we also ask people to name their close ties.

\subsubsection{Economic Measurement of Network Effects and the Research Gap}

In addition to the previous literature, several economic studies have estimated total platform valuations in the context of network effects in the aggregate. Brynjolfsson and Kemerer~\cite{brynjolfsson1996network} studied the spreadsheet market in the 80s and 90s to measure aggregate network effects in this market and its impact on pricing. Using a hedonic model to estimate the impact of network externalities and other software features on pricing, they find that a 1\% increase in the installed base is associated with a 0.75\% increase in price.

Tucker~\cite{tucker2008identifying} explores how formal and informal social influences affect technology adoption in the presence of network externalities. Analyzing the rollout of a video-messaging tool within a firm, she finds that adoption by managers and employees in key network positions significantly boosts uptake among others, whereas adoption by ordinary workers has minimal impact. Boudreau~\cite{boudreau2021promoting} investigates how subjective expectations influence the early adoption of platforms with network effects. Using a field experiment, he finds that optimistic statements about future user base size significantly boost adoption rates more than disclosures of the current user base. The findings underscore the importance of managing market expectations to overcome the initial “chicken-and-egg” problem in platform adoption.

On the methodological side, Bramoulle et al.~\cite{bramoulle2009identification} introduce a method to identify peer effects within social networks. By leveraging the structure of these networks, they distinguish between endogenous peer influences and contextual factors, providing conditions under which peer effects can be empirically identified.

Recent literature has also looked at quantifying the total platform value of various social media platforms. Brynjolfsson et al.~\cite{brynjolfsson2019using} introduced the method of incentivized choice experiments, similar to the ones we use in this paper, to quantify the total value of various popular digital apps. Allcott et al.~\cite{allcott2020welfare} build upon this work and estimate the causal impact of Facebook on subjective well-being in addition to quantifying Facebook’s total platform value. We compare our total platform valuations in this paper with those of these papers in the SOM. What is missing from this literature on economic measurement of network effects is that these studies have not measured consumers’ economic valuations of network connections, measured tie value in the context of economic network effects in an incentive-compatible way or distinguished heterogeneity in the value of network effects across social media platforms—all elements of what we do in this paper.

To summarize, sociological research has extensively explored the value of social network connections, focusing on aspects like tie strength, social capital, and digital interactions. However, sociologists have not directly measured how individuals economically value these ties, particularly within the framework of economic network effects. Conversely, economic studies have assessed the overall value of platforms and the impact of specific network connections on economic outcomes such as income mobility, but they have not measured consumers’ economic valuations of network connections or measured tie value in the context of economic network effects. As a result, a large and important gap exists, between the sociology of networks and the economics of network effects, in our understanding of local network effects. This highlights a significant gap between sociology and economics in measuring local network effects, something we aim to fill in this paper. In this paper, we assign economic value to relationships in an incentive-compatible way and distinguish heterogeneity in the value of network effects across social media platforms. These specific questions and measurement frameworks are essential to the primary policy and strategy questions facing lawmakers and platform managers today.

\subsection{Data Collection Through Lucid Survey Platform}

We conducted online surveys of US users regarding their use of four major social media platforms: Facebook, X (formerly known as Twitter), Instagram, and LinkedIn. Subjects were recruited through Lucid, a market research firm. Lucid provides high-quality responses for online surveys and is one of the largest marketplaces for online sampling in the US \cite{coppock2019validating}. Approximately 375,000 respondents take part in studies every day on Lucid and over 30 million respondents pass through the exchange every year~\cite{coppock2019validating}. Coppock and McClellan~\cite{coppock2019validating} ran several classic experiments and compared results on Lucid to other traditionally well-regarded polling and survey platforms like the American National Election Studies (ANES). While these platforms differ in that Lucid participation is `opt-in' and standard polling uses probability sampling, they found that the demographics and experimental results of the Lucid samples were comparable to those obtained from standard US probability samples.

Many academic studies have used Lucid. For example, Pennycook et al.~\cite{pennycook2021shifting} conducted survey experiments on Lucid to test interventions that can reduce misinformation sharing online. Solís Arce et al.~\cite{solis2021covid} studied factors that explain COVID-19 vaccine acceptance by conducting a large-scale survey on Lucid. Benzell et al.~\cite{benzell2020rationing} conducted a survey on Lucid to understand people's preferences for keeping businesses and other locations open during the COVID-19 pandemic.

Lucid provides us with users of the four social media platforms that we study---Facebook, X, Instagram, and LinkedIn---and verifies that participants are monthly active users of the platform under consideration. Subjects are paid directly by Lucid (the exact amount is determined based on their demographic profile), and we pay Lucid for the participant fee and survey administration costs. Lucid provides additional responses at no cost for responses that are rejected for failing to pass an attention check. 

We conduct two Waves of surveys. Wave 1 was conducted from 2/24/2021 to 5/10/2022 and collected 10,528 valid responses. Wave 2 was conducted from 11/1/2023 to 12/19/2023 and collected 9,395 valid responses.

\subsection{Wave 1}

In the Wave 1 survey, our goal was to understand how the characteristics of consumers' connections relate to how they value those connections on social media. We refer to the subjects of these surveys as “egos,” and the network connections they are asked to characterize and evaluate as “alters.” For this wave, consumers on each of the four platforms that we study were recruited and asked to report on their use of one platform.

\subsubsection{Wave 1 Instrument Summary}

Before launching Wave 1, we created a research account on each of the four platforms studied. In the first step of the Wave 1 study, we asked respondents to connect with our research account. This step ensures data quality and allows collection of additional data of interest. At this stage, respondents could choose to exit the survey.  After connecting to our research account, ego respondents were asked about their demographics (age, gender, ethnicity, income and political ideology) and usage of social media platforms, including the amount of time they spend on these platforms. These questions were asked with responses provided as dropdown lists with options.

In the next part of the survey, egos were asked to identify eight individuals (alters). The first four were their four closest connections (who they may or may not be connected to on the platform).\footnote{The question text is: ``Name your most important contacts (first name and last initial) ranked in order from most important to 4th most important. These contacts could include friends, family, and colleagues or any other type of relationships.''} The second four were ``any four connections excluding the four closest connections" that they were ``connected to on [platform].'' We then asked the egos to characterize their relationships with these alters. More specifically, for each of these connections, we asked for their age, gender, ethnicity, how the ego knows them, how often the ego sees this person face to face, and if they are connected to them on other social media platforms.

In the next section of the survey, (following Brynjolfsson et al.~\cite{brynjolfsson2019using} and Allcott et al.~\cite{allcott2020welfare}), we solicited egos’ total overall valuations of social media platforms using the Becker DeGroot Marschak (BDM) method~\cite{becker1964measuring}, by asking them for the minimum dollar amount they were willing to accept (WTA) to give up access to the platform for one month. Consumers provided answers by moving a slider to enter a value between \$0 and \$100, with an option to enter values greater than \$100 in an open-ended text box. 

We informed the respondents that we would select 1 in 200 respondents and follow up with them to receive rewards of at least \$200. Once we verified that they entered all the details of their connections accurately, they would be eligible to receive \$200. In addition to this, they would be eligible to receive additional rewards based on the outcome of their BDM lottery. The procedure for the additional rewards was as follows. We drew a random amount of money from a distribution. If the random offer was greater than the valuation the respondent entered, they would receive the random offer as an additional reward if and only if they deactivated their account for one month. If the random offer was less than their valuation, they would not have to do anything, and they would not receive any additional reward. This mechanism is incentive-compatible and is widely used in experimental economics to solicit valuations~\cite{burchardi2021testing}.

We followed up with respondents selected for additional rewards through Lucid. As per our agreement with Lucid, we did not have direct contact with the respondents in Wave 1 and Lucid reached out to the selected respondents through their portal, but no respondents replied. To increase the chances of response, in Wave 2 we acquired respondents’ emails and reached out to them directly. Still, respondents answered their valuation solicitation question under the impression that responses were incentive-compatible. This is in line with other recent incentive compatible implementations of BDM in the social media context, such as Bursztyn et al.~\cite{bursztyn2023product} who, in the context of TikTok and Instagram, promised to enforce BDM contracts only under circumstances known to the experimenters ex-ante to be impossible, preserving incentive compatibility without enforcement.

After the platform valuation question, we asked respondents to rank how valuable they find their eight connections using a rank order question adapted to measure the value of networked connections and local network effects, which we call Connection Ranked Choice (CRC).

Consumers of all four platforms received the same survey instrument with a slight modification for Facebook and LinkedIn, where we only asked egos if they are connected to alters, unlike Instagram and X, where we asked egos if they follow alters and if alters follow them. A key aspect of Wave 1 is that egos are required to follow/friend our research account on the platform to qualify for the study. When egos follow/friend our research account, we can see their entire set of public profile information for that platform, as well as their list of contacts, if it is public. We reject observations which do not choose to follow our account. We collected valid responses from 4,149 Facebook users, 2,899 X users, 1,981 Instagram users, and 1,499 LinkedIn users.

\subsubsection{Dependent Variable: Connection Value Percentile (CVP)}

To understand what characteristics lead to highly valued connections on digital platforms, we constructed a measure of the relative value of connections, called ``Connection Value Percentile'' (CVP). This measure is derived from our Wave 1 questionnaire about individual connections. The egos in our study name and describe up to eight alters they are connected to on the focal platform. On a subsequent survey page, we solicit the relative importance of these contacts using a rank-order question (Connection Ranked Choice (CRC)). The exact text of the relevant question is ``Please rank the following connections from most valued on [PLATFORM] (top) by you to least valued on [PLATFORM] (bottom) by you.'' 

Using the provided ranking we calculate CVP as \begin{equation}
    CVP_{i,j,p} = 100(1 - \frac{CR_{i,j,p} -1}{TCR_{i,p}})
\end{equation}

where $CR_{i,j,p}$ is the ranking given to connection $j$ by ego $i$ on platform $p$ on our relative ranking scale and $TCR_{i,p}$ is the total number of connections ranked by consumer $i$ on platform $p$. This measure evaluates which relationships are more valued on the  platform. To summarize these relationships, we calculate average CVPs conditional on a given dyadic characteristic. Larger CVPs for one dyadic characteristic versus another indicate that those types of relationships are valued more highly, on average.

\subsubsection{Explanatory Variables: Ego Demographics, Dyadic Characteristics, and Platform Usage}

For our Wave 1 analysis, the first set of data collected is the information from the ego's public social media profiles. For Facebook, we downloaded respondents' entire connection list and matched these names with census data from the US Social Security Administration to estimate the race and gender of alters.\footnote{https://www.ssa.gov/oact/babynames/limits.html} For LinkedIn, we downloaded respondents’ profiles and their education and career history. For X, we used the X API to download rich information about respondents’ profile information and posting activity, as well as the number of followers and followees. Instagram was the only platform from which we were unable to gather complementary information due to restrictions imposed by the platform. Table ~\ref{table:1} summarizes the data we were able to collect by platforms.

\begin{table}[h!]
\centering
\begin{tabular}{c| c} 
 Platform & Additional Information Collected \\ [0.5ex] 
 \hline
 Facebook & Ego’s public friend list  \\ 
 LinkedIn & Ego’s profile (e.g. resume information)  \\
 X & Ego’s Timeline history, list of followers and followees  \\
 Instagram & N/A  \\ [1ex] 
\end{tabular}
\caption{Summary of additional information collected by survey platform. Section \ref{Additional_data_collect_tables} provides additional information on data collected from LinkedIn and X participants. }
\label{table:1}
\end{table}

Our main set of explanatory variables are alter characteristics as reported in the survey. The alter characteristics solicited were: age, gender, ethnicity, how  ego knows them (school/ college/ shared interest/ family member/ only know them online), how often ego sees this person face to face (or lives with them), and if they are connected to them on any of the other social media platforms. Other explanatory variables are the ego characteristics (gender, age, ethnicity, income, political affiliation) and platform usage characteristics (time spent on each of the four platforms) self-reported in the survey.

\subsection{Wave 2}

Wave 2 is focused on determining the predictors of ego's overall platform valuation and the role of social network connections as a predictor. Wave 2 follows the design of Wave 1. However, there is a focus on ascertaining total platform value in an incentive-compatible manner as well as directly soliciting the count of the ego's connections. 

\subsubsection{Wave 2 Instrument Summary}

We first recruited users of one of the four social media platforms from Lucid to participate in our study. Respondents were asked for their email addresses at the beginning of the study so that we could follow up with them if they were selected for additional rewards. If they choose not to share their email address, they exited the survey. At this stage 1195 or 30\% of Facebook recruited respondents exited the survey; 644 or 19\% of Instagram recruited respondents exited the survey; 648 or 17.7\% of LinkedIn recruited respondents exited the survey; and 691 or 18.8\% of X recruited respondents exited the survey. Given that 70-82\% of recruited users were retained after being asked to give their email address and agreeing to the terms of service, we feel retention is not a significant issue for our study although we cannot compare respondents who opted out of the experiments to those who were retained since those who exited did not answer survey questions or provide permission for other data collection. Users (egos) were then asked about their demographics (age, gender, ethnicity, income, and political ideology), usage of the platform, and their total number of connections on the platform. For users of Facebook and LinkedIn, we collected their total number of connections on these platforms directly. For users of Instagram and X, we collected both the number of followers and followees. 

Unlike Wave 1, we also collected information about which features consumers used on each of the platforms. On Facebook, respondents were given the option to select one or more features from posting on the news feed, to engaging with content from connections, following pages, playing games, joining groups, organizing and attending events, buying and selling on Marketplace, messaging on Messenger, and using Facebook Live. On Instagram, respondents were asked to select one or more options from posting photos/ reels/ stories, liking and commenting on photos/ reels/ stories, streaming on Instagram Live, and direct messaging other users. On LinkedIn, respondents were asked to select one or more options from updating their profile, using groups, liking, commenting, and sharing content, and searching and applying for jobs. On X, respondents were asked to select one or more options from posting posts, liking/ replying to/ reposting posts, direct messaging other users, bookmarking posts, using X Lists, using X Spaces, and subscribing to X Premium. 

In the next section of the survey, we solicited egos’ valuations of social media platforms by asking them for the minimum dollar value they were willing to accept (WTA) to give up access to a platform for one month using a variation of incentive-compatible BDM, following Brynjolfsson et al.~\cite{brynjolfsson2019using}. Valuations were solicited using a multiple price list~\cite{andersen2006elicitation}, and the upper limit of the price range was randomly selected to be either \$100, \$150, or \$200 to account for anchoring effects. The multiple price list is analogous to a discrete choice experiment as described in Brynjolfsson et al.~\cite{brynjolfsson2019using} and de Bekker-Grob et al.~\cite{de2012discrete}. 

All platforms used the same survey instrument except for the question related to the usage of platform-specific features, which had different feature options. Respondents who selected a fake social media platform (``Mystro'') as one of the platforms that they use were also filtered out of the survey for failing an attention check. The platform valuation experiments were incentive compatible and provided consumers an opportunity to be paid when the incentive compatibility constraint became binding---when they were verified to have given up the platform and were paid to do so. For participants who received the upper limit of \$100 in the multiple price list question, 1 in 100 respondents were randomly selected and offered to follow up on their choices (1 in 150 for those with upper limit of \$150 and 1 in 200 for those with upper limit of \$200). We made cash offers of a randomly selected value if and only if they deactivated the platform. 70\% of those who were made cash offers to deactivate a platform complied with their offers. We received valid responses from 2470 Facebook users, 2237 X users, 2488 Instagram users, and 2200 LinkedIn users. Summary statistics for the second Wave data are presented in Table~\ref{table:summary_stat_unfiltered2}. All pre-registered results for Wave 2 of the study not included in the main text are included in the sections below. 



\subsubsection{Dependent Variable: Total Platform Value (TPV)}
As mentioned previously, we solicited the minimum dollar value respondents' were willing to accept (WTA) to give up a platform for one month using a BDM and a multiple price list. Respondents made a series of choices between keeping access to the platform or giving it up for one month in exchange for a certain monetary compensation. The price list was offered in increments of \$5, ranging from a minimum of \$5 to a maximum of either \$100, \$150, or \$200, assigned randomly, to ensure our results are robust to anchoring effects~\cite{jack2022multiple}.

Prices were listed in multiple rows on a single page, ranging from the lowest price to the highest price. For each row, respondents made a choice between keeping access to the platform or giving it up for one month in exchange for the price listed in that row. We followed up with randomly selected respondents and selected one of their rows from the multiple price list question. Whatever choice was made by the respondent was offered to them. For example, if they chose to give up access to a platform for \$45 in a particular row, we offered them the chance to actually give up that platform and get paid \$45 at the end of the month after we verified that they complied with their choice. Therefore, our valuations are incentive-compatible. 

Since prices were listed in increasing order, respondents made a switch at some point from keeping access to the platform to giving it up in exchange for a certain price. Our dependent variable, total platform value, is measured by looking at the lowest price at which respondents made this switch. For example, if a particular respondent didn't accept prices until \$80 but switched and started accepting prices from \$85 and above to give up access to a platform for one month, their total platform value for that platform was assessed as \$85.

\subsubsection{Explanatory Variables: Number of Connections, Ego Demographics, and Platform Usage}
For our Wave 2 analysis, our explanatory variables included the ego demographics and platform usage characteristics previously listed. In addition, a key explanatory variable that we measured in this study was the total number of connections a respondent has on a given platform. Previous literature has documented the impact of network effects on platform value~\cite{parker2005two}. Based on this, we hypothesized that platform value is increasing in the number of connections. While we focused on the characteristics of connections in Wave 1, we analyzed the total number of connections and their association with total platform value in Wave 2.

\subsection{Ordinal Versus Cardinal Connection Value Solicitation Designs}\label{ordinal_vs_cardinal}

When soliciting connection value through incentive compatible designs, a natural choice arises in whether to solicit connection values ordinally or cardinally. For example, one option is to solicit connection values by randomly selecting some connections from an ego’s network and soliciting their willingness to accept payments for disconnecting from that connection for a specified period of time. We considered this research design in depth and even conducted exploratory studies to evaluate its reliability, robustness and appropriateness. In the end, several factors made clear that this design is likely unreliable in soliciting connection value.

First, implementing plausible,  connection specific, incentive compatible designs at scale in a fully remote, online setting is technically challenging. The study participant’s full contact list needs to be collected, parsed, and then a random subset selected, all within the initial survey time-frame. In a lab setting, for Instagram, we achieved this by assisting participants in carefully copy-pasting their contacts into a specially designed desktop application. 

Putting aside these concerns, we conducted a laboratory study to test the practicality of the cardinal, incentive compatible connection value solicitation design.\footnote{Experiment approved by Chapman University IRB: IRB-22-304 Measuring Digital Network Effects Using Incentivized Choice Experiments} We recruited 316 students to participate in a one-hour session with optional online follow up. Study participants were asked to report on their entire list of Instagram contacts, directly copied and pasted from their main Instagram account. Using the standard BDM mechanism, we asked subjects their willingness to accept (WTA) offers to disconnect, on Instagram, from 1, 5, or 10 randomly selected followers or followees for one semester (about 3.5 months), as well as for giving up Instagram altogether. The randomly selected contacts’ handles were displayed for the subjects to see as they made their decisions. Of the 316 study participants, 119 were made follow up offers larger than their stated willingness to accept. Of these, 80 remained in compliance with the restrictions for the entire semester and were made compliance payments averaging \$87.17. Figure \ref{INSTAGRAM_STUDY} below summarizes our findings on the WTA of subjects to give up randomly selected connections. There are a few surprising things about the results that led us to reject using this design to measure connection value.

\begin{figure}[htb]
        \centering
        \includegraphics[width=\linewidth]{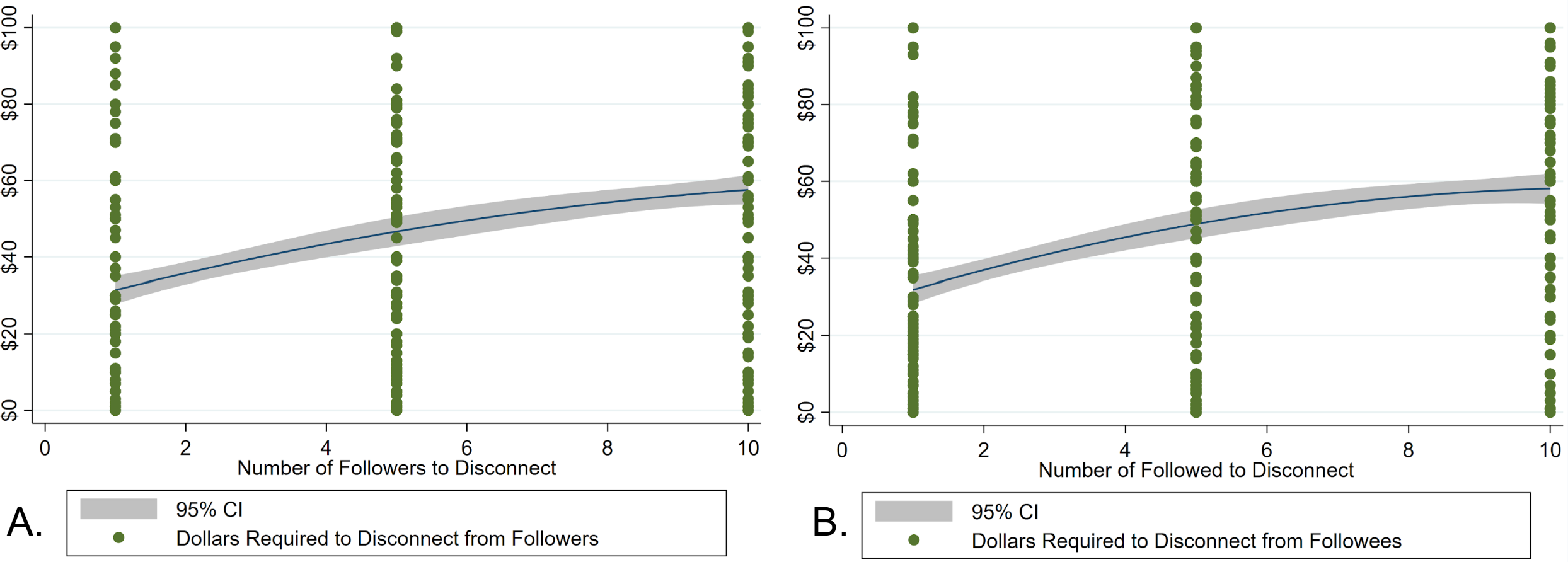} 

    \caption{Scatterplot and quadratic line of best fit of study participants’ minimum required payments (i.e. WTA) to block 1, 5, or 10 randomly selected Instagram contacts. In the left panel, A, the random names were selected from those followed by the study participant (followees). In the right panel, B, the random names were selected from those following the study participant (followers). }
    \label{INSTAGRAM_STUDY}
\end{figure}

First, subjects requested an average of almost \$30 to disconnect from a single randomly selected follower or followee—an implausibly high valuation for a single connection for a period of 3.5 months. Exit interviews suggested that bias in connection valuation may have arisen  from a) a fixed cost of monitoring, b) loss aversion or c) a high cost of the social signal of disconnection. The fixed cost of monitoring arises from subjects’ distaste of being digitally monitored. The problem with this fixed cost of monitoring is that it biases connection value estimates disproportionally and extremely compared to platform value estimates. To understand why, consider this toy example: Suppose for a specific person the actual value of a random connection is 1 penny, and the actual total platform value is \$20. Suppose the cost of monitoring is \$10. The WTA an offer to disconnect from that connection would be  \$10.01, a 1000x bias. In estimating the platform value, we'd estimate \$30, a 1.5x bias. Loss aversion would arise from the psychological cost from losing a monetarily low value resource (a single connection), that is not matched symmetrically by as large a gain from acquiring the resource (or connection). Finally, the negative social signal of disconnecting from a connection may be an additional cost that biases connection value estimates in this design upward, which is itself unrelated to the actual value of the connection. If we perceive a significant social cost to slighting our connection, the incentive compatible disconnection design would not produce a reliable connection value solicitation because it would bias connection value estimates upward significantly and unnecessarily as the concept of disconnection is merely an artifact of the value solicitation method rather than an intrinsic element of the value of the connection.

\begin{figure}
        \centering
        \includegraphics[width=\linewidth]{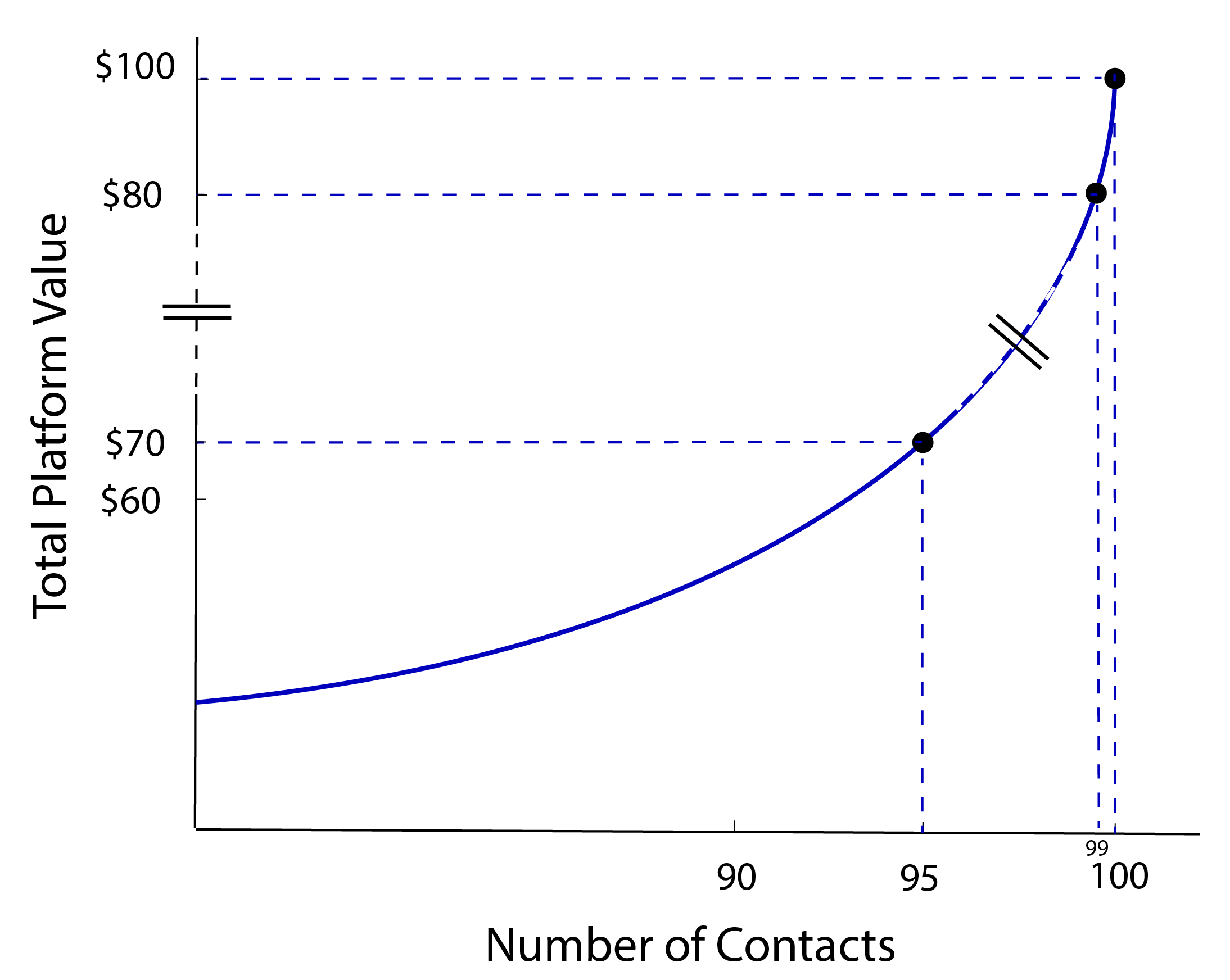} 

    \caption{This figure is a stylized illustration of a utility function with increasing returns to the number of connections. In this example, the platform user would require at least \$20 to disconnect from a single connection, and would require \$30 to disconnect from 5 users - a cost of only \$6 per user. This figure shows how increasing returns to connections generates a decreasing cost of disconnection per-connection as in our lab experiment. The dashed lines on the y-axis and curve indicate the figure is not to scale.} 
    \label{increasing_returns}
\end{figure}

Second, the relationship between the number of connections to be removed and the WTA is non-linear (see Figure \ref{INSTAGRAM_STUDY}), suggesting a strongly increasing marginal value of connections, which is also itself implausible. To understand why, consider Figure \ref{increasing_returns} below. The figure depicts the number of social media contacts on the X-axis, and platform value on the Y-axis. The utility function is upward sloping, indicating an increasing marginal value of contacts or that, for example, the marginal value of 100th connection is significantly higher than the first connection. As the figure illustrates, the immediate corollary of this is that the WTA disconnections has decreasing marginal costs. 

It is not surprising that total platform value, across all users of a platform, would be increasing in the number of connections. If all connections have the same, constant value, then the total value of the entire platform to all users grows with the square of the number of users. This observation is known as `Metcalfe’s Law.’ However, mature networks in long-term equilibria likely have constant or decreasing network effects in the number of users. Otherwise, a very small increase or decrease in the number of participants would have dramatic effects on the value from, and future participation in, the network. We think it implausible from a network stability perspective that losing a single connection is so impactful, and so much more impactful, per-connection, than disconnecting from 5 or 10 contacts. 

We also feel this is an important area of future work. If we can devise reliable strategies for measuring local network effects, an important next step in the work is to empirically validate the shape of the utility function that captures network effects and local network effects, and to understand whether network effects are non-linear, marginally increasing or decreasing in the number of contacts, and how the shape of the network or the type of contacts considered affects the shape of the utility function.

Ultimately we eschewed using the BDM mechanism to solicit connection value for three reasons: (1) the difficulty in having plausible IC at scale in a fully remote online setting including the cognitive difficulty in evaluating the ‘dollar value of a connection’, a much more unusual, alien or distasteful task relative to simply ranking their importance, (2) the theoretically small value of individual connections would be dwarfed by the variance of such estimates making estimation extremely challenging, (3) experimental biases arising from loss aversion, the fixed costs of monitoring and the costs of negative social signals from disconnection, all of which were plausibly displayed in our laboratory study. 

For these reasons, we believe the way forward for connection value solicitation and local network effects estimation is the combination of incentive compatible platform value solicitation designs and connection rank choice, which describe and implement in this study.

\subsection{Measuring Complementarity in Connections}

The elasticity of substitution between a pair of inputs refers to how much the ratio of marginal values of the two inputs changes as a result of an increase in the use of one of the inputs. In the context of social media connections, one might be interested in how the value of a connection on a focal platform changes as a result of a connection being formed on an alternate platform. This elasticity has practical implications for policymakers. For antitrust enforcers, for example, complementarity and substitution patterns across different platforms are relevant to judging the consumer welfare effects of mergers. When communications across a pair of services are substitutes, combinations of social media firms will necessarily have more market power -- a “market” being defined as a set of products which are close substitutes \cite{hovenkamp2020antitrust}. 

In the context of discrete inputs, complementarity is defined as when the value of both inputs together is greater than the sum of the value of each input alone, i.e.:
$$
V(1,1) – V(0,0) \geq V(1,0) – V(0,0) + V(0,1) – V(0,0). 
$$

Where V(X,Y) total outcome value as a function of two binary input choices.\cite{brynjolfsson2013complementarity} 

We can use this definition to determine whether connections to the same person across different platforms are complements in our data. To do so, we interpret CVP as a measure of the marginal value of a connection.  Rearranging the complementarity criteria yields:
$$
V(1,1) – V(0,0) \geq V(1,0) – V(0,0) + V(0,1) – V(0,0)
$$
$$
V(1,1) - V(1,0) - (V(1,0) +V(0,0)) \geq V(1,0) – V(0,0) + V(0,1) – V(0,0)
$$
$$CVP(1,1) - CVP(1,0) \geq CVP (1,0) - CVP(0,1)
$$
$$
CVP(1,1) \geq  CVP(0,1)
$$
This condition implies that the criterion for connections on different platforms to be complementary in our data is for CVP for the combination of connections to be higher than for a single connection alone. As is seen in Figure 2C (and Table \ref{complementarity_table}) this is the case for the connections of each platform and being connected on at least one other platform (all differences significant at $\alpha =.00001$ level). 

However, this analysis faces a major limitation. Connections are not randomly assigned but rather endogenously chosen by users. Therefore, omitted variables may bias our findings. Correlational analysis can only be suggestive, although controlling for observable covariates may help \cite{brynjolfsson2013complementarity}. 

Another barrier to this interpretation arises from ambiguity in how we solicited connection valuations. Note that we do not give detailed instructions to survey takers about what is meant by “most valued.” Interpreting these results as informative about the nature of connection complementarity across platforms requires interpreting importance as marginal value. However, some respondents may have taken a portfolio approach to ranking connections, with their top 4 ranks, for example, corresponding to their preferred team of four connections rather than indicating individual preferences. 

To investigate this point further, we examined the extent to which certain connection types co-occur in rankings. If participants tend to list many friends of the same type at the top of their rankings, this is in tension with the portfolio hypothesis. On the other hand, if participants always have a diverse mix of connections in their top 4, this would suggest a portfolio strategy.

The table below lists 4 ranking co-occurrence measures for 12 focal characteristics. The first row reports the probability of having a different category 2nd ranked connection conditional on a given first ranked connection. The remaining rows give the probability of all top N friends being in the same category.\footnote{The top 8 ranking restricts attention to only participants who ranked all 8 possible friends, creating a slightly different denominator.}

\begin{table}
\centering
\footnotesize
\setlength{\tabcolsep}{4pt}
\caption{Rank concentration by relationship type.}
\label{tab:rank_concentration}
\resizebox{\textwidth}{!}{%
\begin{tabular}{l ccccccc ccc cc}
\toprule
 & \multicolumn{7}{c}{How ego knows alter} & \multicolumn{3}{c}{Relative age} & \multicolumn{2}{c}{Other platform}\\
\cmidrule(lr){2-8}\cmidrule(lr){9-11}\cmidrule(lr){12-13}
 & Online & None & K--12 & Work & Interest & College & Family & Same & Younger & Older & Yes & No\\
\midrule
$P(\text{R2}\notin c \mid \text{R1}\in c)$ & 0.580 & 0.793 & 0.667 & 0.644 & 0.716 & 0.678 & 0.450 & 0.457 & 0.586 & 0.528 & 0.120 & 0.487\\
$P(\text{top }2\subseteq c)$ & 0.0357 & 0.0064 & 0.0357 & 0.0668 & 0.0380 & 0.0423 & 0.2277 & 0.2514 & 0.0986 & 0.1449 & 0.7480 & 0.0963\\
$P(\text{top }4\subseteq c)$ & 0.0204 & 0.0020 & 0.0122 & 0.0191 & 0.0093 & 0.0130 & 0.1189 & 0.1065 & 0.0291 & 0.0510 & 0.6253 & 0.0597\\
$P(\text{top }8\subseteq c)$ & 0.0018 & 0.0002 & 0.0020 & 0.0061 & 0.0012 & 0.0033 & 0.0369 & 0.0292 & 0.0045 & 0.0145 & 0.7019 & 0.0049\\
\bottomrule
\end{tabular}%
}
\vspace{2pt}
\end{table}

On balance, the table should move us substantially away from the view that a portfolio motive governs the rankings. A community-following logic implies that respondents diversify across groups after only a slot or two, so that concentration at the top of the ranking should decay quickly and, crucially, should do so without regard to how valuable a category is. Instead, attributes that earn the highest Connection Value Percentile—cross-platform ties, family, and same-age peers—accumulate at the very top (e.g., over half of people who rank a family connection as their top connection list another family connection as their second most valued, and 88\% of egos whose top connection is cross-platform also rank a second cross-platform connection next). Summing across the ‘How ego knows alter’ columns, over 19\% of respondents list their top 4 connections as all being from the same connection type. 
Portfolio behavior cannot easily generate this pattern. We therefore regard a portfolio motive as, at most, a second-order influence on the rankings rather than their organizing principle—though, because observable attributes are imperfect proxies for the latent communities a respondent might have in mind, we treat this as suggestive rather than decisive evidence.

To our knowledge, no previous paper has attempted to estimate whether connections across platforms are complementary. However, a pair of papers have been able to harness experiments to measure the complementarity of different digital messaging systems. Collis and Eggers \cite{collis2022effects} conducted an RCT of university students in which half were randomly assigned to a large restriction on their use of Facebook, Instagram and Snapchat. Those in the treatment group dramatically increased their usage of WhatsApp. The authors take this as evidence of strong substitutability between WhatsApp and the bundle of social media platforms. Aridor \cite{aridorforthcoming} experimentally varied participant’s phone access to Instagram and YouTube. The author found that participants’ increased their usage of several other digital products as well as their time spent in non-digital activities. Neither of these papers speaks to the question of complementarity of specific connections across platforms. While identifying complementarity or substitution across platforms is not our primary objective, our work nonetheless provides suggestive evidence that these platforms are complements not substitutes.

\section{Results}

\subsection{Wave 1}

In Table~\ref{table:summary_stat_unfiltered1}, we present summary statistics on age, gender, ethnicity, income, and political orientation for our Wave 1 sample. At the bottom of Table~\ref{table:summary_stat_unfiltered1}, we report the average, median and standard deviation of the self-reported share of connection value that comes from the top-four connections on each platform. Consistent with later results, these results suggest strong heterogeneity in connection value, with the top four connections being valued disproportionately. Also interesting is that the share of value coming from the top four platforms is lowest for LinkedIn. This is consistent with LinkedIn being valued for introducing a large number of weak-ties for job searching. 

None of the platforms we studied release official statistics on the demographics of its US users. This makes comparison of our summary statistics to the platforms’ true user bases challenging. However, there are two general surveys of social media usage that allow us to construct comparable estimates of the shares of platform users by demographic. 

The first social media survey we compare to our sample is the 2020-2022 ANES Social Media Study \cite{ANES}. This is a three-wave panel survey conducted online. Data collection was carried out by NORC at the University of Chicago, supported by funding from Facebook and the National Science Foundation. The sample included U.S. citizens aged 18 and older and achieved a 3.2\% response rate in the initial wave. In the third wave of their survey, conducted from November 9, 2022, to January 2, 2023, the instrument asked whether respondents were current users of three of the social media platforms we studied. ANES makes the micro-data of the survey responses available, which we used to calculate platform participation rates by demographic. We combined these usage rates from US Census \cite{uscensus} data on the demographics of US persons to estimate shares of users by demographic for each platform. We estimated a 95\% confidence interval for each platform’s demographics by using the total number of respondents for questions regarding platform usage as the sample size. 

The second general survey to which we compare our sample is a 2023 survey conducted by Pew on social media usage \cite{pewsocialmedia2024}. The survey was conducted between May 19 and September 5, 2023, with a total of 5,733 respondents (2,217 online and 3,516 via paper). Participants were recruited through mailed invitations, employing address-based sampling, with materials provided in English and Spanish. To encourage participation, households received an initial incentive and respondents who completed the survey were later sent a \$10 post-completion incentive. The microdata on responses is not made available by Pew, but they do provide cross-tabulation results on the share of individuals of each demographic who use each of the platforms we studied. We combined these estimates with the same US Census data described above to create platform-level demographic share estimates. In calculating confidence intervals for this estimate, we used the Pew provided 95\% confidence interval for the full sample. 

The resulting estimates of platform demographic shares from ANES, Pew, and our sample are reported in Table \ref{PEW_and_ANES}. ANES did not ask about LinkedIn usage, so those columns are empty. 

\begin{table}[htbp]
\centering
\caption{Percent of Users by Demographic Characteristic on Each Platform: 
PEW 2023 Estimated, ANES 2022 Estimated, and this Study Sample.}
\label{PEW_and_ANES}
\resizebox{\textwidth}{!}{%
\begin{tabular}{l c c c c c c c c c c c c c c c c c c c c c c c c}
\toprule
& \multicolumn{6}{c}{Facebook}
& \multicolumn{6}{c}{Instagram}
& \multicolumn{6}{c}{LinkedIn}
& \multicolumn{6}{c}{X} \\
\cmidrule(lr){2-7}\cmidrule(lr){8-13}\cmidrule(lr){14-19}\cmidrule(lr){20-25}
& \multicolumn{2}{c}{Pew*}
& \multicolumn{2}{c}{ANES*}
& \multicolumn{2}{c}{Our Sample}
& \multicolumn{2}{c}{Pew*}
& \multicolumn{2}{c}{ANES*}
& \multicolumn{2}{c}{Our Sample}
& \multicolumn{2}{c}{Pew*}
& \multicolumn{2}{c}{ANES*}
& \multicolumn{2}{c}{Our Sample}
& \multicolumn{2}{c}{Pew*}
& \multicolumn{2}{c}{ANES*}
& \multicolumn{2}{c}{Our Sample} \\
\cmidrule(lr){2-3}\cmidrule(lr){4-5}\cmidrule(lr){6-7}
\cmidrule(lr){8-9}\cmidrule(lr){10-11}\cmidrule(lr){12-13}
\cmidrule(lr){14-15}\cmidrule(lr){16-17}\cmidrule(lr){18-19}
\cmidrule(lr){20-21}\cmidrule(lr){22-23}\cmidrule(lr){24-25}
& Val & CI\textdagger 
& Val & CI** 
& Val & CI** 
& Val & CI\textdagger 
& Val & CI** 
& Val & CI** 
& Val & CI\textdagger 
& Val & CI** 
& Val & CI** 
& Val & CI\textdagger 
& Val & CI** 
& Val & CI** \\ 
\midrule

Men (\%) 
& 43 & $\pm 1.8$
& 46 & $\pm 1.6$
& 47 & $\pm 1.2$
& 41 & $\pm 1.8$
& 41 & $\pm 1.6$
& 39 & $\pm 1.4$
& 51 & $\pm 1.8$  
& -- & --     
& 46 & $\pm 1.6$
& 57 & $\pm 1.8$
& 59 & $\pm 1.6$
& 53 & $\pm 1.4$
\\

White (\%) 
& 76 & $\pm 1.8$
& 74 & $\pm 1.4$
& 71 & $\pm 1.1$
& 69 & $\pm 1.8$
& 67 & $\pm 1.5$
& 61 & $\pm 1.4$
& 75 & $\pm 1.8$
& -- & --     
& 68 & $\pm 1.5$
& 68 & $\pm 1.8$
& 72 & $\pm 1.5$
& 64 & $\pm 1.3$
\\

65+ (\%)
& 15 & $\pm 1.8$
& 14 & $\pm 1.1$
& 7  & $\pm 0.6$
& 6  & $\pm 1.8$
& 7  & $\pm 0.8$
& 6  & $\pm 0.7$
& 7  & $\pm 1.8$
& -- & --     
& 10 & $\pm 1.0$
& 5  & $\pm 1.8$
& 9  & $\pm 0.9$
& 5  & $\pm 0.6$
\\

\bottomrule
\multicolumn{25}{l}{\textit{Note:} *Estimated US User Demographics based on Pew data or American National Election Studies (ANES) data.}\\
\multicolumn{25}{l}{\textit{Note:} **95\% confidence interval }\\
\multicolumn{25}{l}{\textit{Note:} \textdagger 95\% confidence interval for full sample as reported by Pew.}\\
\end{tabular}%
}
\end{table}

As can be seen, there is broad agreement between the different samples in the demographic breakdown of platforms. For example, Facebook is the whitest platform in both the survey estimates and in our surveys, and LinkedIn is whiter than Instagram and X. Across the sources, there is agreement on X being relatively male, and Facebook having more older users than X. 

However, the sources are not in perfect agreement. Comparing Pew to ANES, four of nine estimates are significantly different at the 95\% level. Comparing our sample to ANES, seven of nine estimates are significantly different at the 95\% level. Comparing our sample to Pew, eight of twelve estimates are significantly different at the 95\% level. 

We think these small but significantly detectable differences arise for several reasons. First, our studies were conducted in different time periods, and participation rates by demographic groups are known to change over time. Second, the ANES study is focused on US citizens, while Pew and our estimates are based on US persons and are not restricted to US citizens. Third, Pew made a special effort to include Spanish language materials, while ANES and our study did not. Fourth, there are differences in the wordings of the survey questions used, in how they solicit both demographics and current platform usage, across the studies. Finally, our confidence intervals for the Pew and ANES studies may be overly narrow, as we naively use the Pew stated confidence intervals, and for both ANES and Pew do not make additional adjustments for the fact that estimating user shares involves combining the survey data with estimates of US population shares. Overall, the broad agreement of our sample with available surveys, despite differences in their sampling methods and time periods, gives us confidence in the generalizability of our findings.

\begin{table}[!htbp] \centering 
\caption{Summary Statistics for Wave 1.}\label{summary1}
\renewcommand*{\arraystretch}{1.1}\resizebox{.8\textwidth}{!}{
\begin{tabular}{lrrrrrrrr}
\hline
 & \multicolumn{2}{c}{FACEBOOK} & \multicolumn{2}{c}{INSTAGRAM} & \multicolumn{2}{c}{LINKEDIN} & \multicolumn{2}{c}{X}  \\ 
 & \multicolumn{1}{c}{N} & \multicolumn{1}{c}{\%} & \multicolumn{1}{c}{N} & \multicolumn{1}{c}{\%} & \multicolumn{1}{c}{N} & \multicolumn{1}{c}{\%} & \multicolumn{1}{c}{N} & \multicolumn{1}{c}{\%} \\ 
\hline
 & 4149 &  & 1981 &  & 1499 &  & 2899 &  \\ 
Gender &  &  &  &  &  & &  & \\ 
-- Female & 2181 & 53\% & 1424 & 72\% & 853 & 57\% & 1236 & 43\% \\ 
-- Male & 1951 & 47\% & 543 & 27\% & 637 & 42\% & 1641 & 57\% \\ 
-- Other & 17 & 0\% & 14 & 1\% & 8 & 1\% & 22 & 1\% \\ 
\\
Age Bin &  &  &  &  &  &  &  &  \\ 
-- 18-24 years old & 646 & 16\% & 587 & 30\% & 239 & 16\% & 520 & 18\% \\ 
-- 25-34 years old & 1396 & 34\% & 713 & 36\% & 489 & 33\% & 1048 & 36\% \\ 
-- 35-44 years old & 1310 & 32\% & 388 & 20\% & 436 & 29\% & 1050 & 36\% \\ 
-- 45 years or older & 790 & 18\% & 278 & 14\% & 333 & 22\% & 277 & 10\% \\ 
\\
Ethnicity &  &  &  &  &  &  &  &  \\ 
-- White & 2952 & 71\% & 1189 & 60\% & 1014 & 68\% & 1873 & 65\% \\ 
-- Black or African American & 573 & 14\% & 354 & 18\% & 227 & 15\% & 534 & 18\% \\ 
-- Hispanic or Latino & 387 & 9\% & 278 & 14\% & 150 & 10\% & 317 & 11\% \\ 
-- Asian / Pacific Islander & 159 & 4\% & 100 & 5\% & 66 & 4\% & 106 & 4\% \\ 
-- Nat. American or Am. Indian & 36 & 1\% & 19 & 1\% & 14 & 1\% & 33 & 1\% \\ 
-- Other & 42 & 1\% & 41 & 2\% & 27 & 2\% & 36 & 1\% \\ 
\\
Income &  &  &  &  &  &  &  &  \\ 
-- Less than \$100,000 & 2859 & 69\% & 1673 & 84\% & 1119 & 75\% & 1655 & 57\% \\ 
-- \$100,000 or more & 1290 & 31\% & 308 & 16\% & 379 & 25\% & 1244 & 43\% \\ 
\\
Political Orientation&  &  &  &  &  &  &  &  \\ 
-- Liberal & 1403 & 34\% & 541 & 28\% & 435 & 29\% & 1140 & 39\%\\ 
-- Moderate & 2000 & 48\% & 1150 & 59\% & 846 & 56\% & 1290 & 44\% \\ 
-- Conservative & 746 & 18\% & 270 & 14\% & 217 & 14\% & 469 & 16\% \\ 
\\
Top 4 Connection Value (Self-reported \%) &  &  &  &  &  &  &  &  \\
-- Mean   & \multicolumn{2}{c}{53.2} & \multicolumn{2}{c}{44.9} & \multicolumn{2}{c}{43.4} & \multicolumn{2}{c}{55.7} \\
-- Median & \multicolumn{2}{c}{53}   & \multicolumn{2}{c}{44}   & \multicolumn{2}{c}{40}   & \multicolumn{2}{c}{59}   \\
-- SD     & \multicolumn{2}{c}{31.1} & \multicolumn{2}{c}{31.7} & \multicolumn{2}{c}{32.1} & \multicolumn{2}{c}{30.7} \\
%
\hline
\end{tabular}
}
\label{table:summary_stat_unfiltered1}
\end{table}

\subsubsection{Connection Value Percentile by Dyadic Characteristics}
We measured the effect of dyadic attributes on the connection value percentage (CVP) by estimating the conditional average CVP for different subpopulations to identify significant differences. Formally, we evaluated the conditional expectation and 95\% confidence intervals for 

\begin{equation}
    E[CVP_{i,j,p}| A_{i,j,p} = \textbf{A}]
\end{equation}

where $\textbf{A}$ is some dyadic attribute that an ego-alter pair could have (for example, the ego and alter are both male), where the indexes $i,j,k$ refer to the connection $i,j$ on platform $p$. 

In the main text of the paper we present the most important findings of this analysis in Figures 2 and 3. Across types of connections, one strong and consistent result is that connections between individuals that are “closer” tend to create more platform value. Across all platforms, connections seen more frequently and in more intimate relationships (e.g. family and school friends) are more valued than connections seen infrequently and in less intimate relationships (e.g. shared interest groups and those only known online). Those who are connected to the ego on multiple platforms also have higher CVP, suggesting that the negative effect of having alternative modes of communication on focal platform CVP is small, relative to the effect of closer contacts being more valued on all platforms. This result is consistent with the hypothesis that Myspace was eventually surpassed by Facebook because the latter focused on creating higher-value college friend connections, while the former fostered looser shared-interest connections \cite{aral2021hype}. Close contacts and family members are especially highly valued on Facebook, and there is a preference for same-race contacts on average.


\subsubsection{Imputing Monetary Value for Connection Characteristics}


The Connection Value Percentile (CVP) analysis in the previous section documents which dyadic attributes are associated with higher within-ego ranks,
but it does so attribute-by-attribute and on an ordinal scale. To complement that evidence, this subsection imputes a marginal monthly dollar value for each
connection characteristic, jointly estimated, by combining the Wave~1 ranking task with the platform-level estimates of total platform value (TPV) and the
share of platform value attributable to local network effects (SPV) developed in the main text. The procedure has two steps.

\paragraph{Step 1: Rank-ordered (exploded) logit on the pooled Wave 1 sample.}
We treat each ego's eight-alter ranking as the source of a sequence of choice
events---the top alter chosen from eight, the second from the remaining seven,
and so on---and estimate a rank-ordered (exploded) logit\footnote{
``Exploded'' following Beggs et al.~\cite{beggs1981assessing} means
a complete ranking of $K$ alternatives is rewritten as $K-1$ stacked
conditional-logit choices, with successively smaller choice sets at each
stage.} on the resulting data, pooling across the four
platforms~\cite{beggs1981assessing}. Letting $u_{ija}$ denote the latent
value to ego~$i$ on platform~$p$ of an alter~$a$ with characteristics
$X_{ija}$, we model
\begin{equation}
u_{ija} \;=\; X_{ija}\,\beta \;+\; \eta_{ip} \;+\; \varepsilon_{ija},
\end{equation}
where $\eta_{ip}$ is an ego$\times$ranking-stage stratum effect and
$\varepsilon_{ija}$ is i.i.d.\ Type-I extreme value, so that the probability of
observing a given complete ranking is the product of conditional logit
probabilities over successive choice sets.
The covariates in $X$ are the dyadic attributes solicited in Wave~1: how the
ego knows the alter, face-to-face contact frequency, whether the pair is also
connected on a second platform, and three same/different
indicators (same race, same age, same gender). Also included as covariates in the regression are alter gender and race directly. For each characteristic category, the
omitted baseline is the lowest-ranked level in the dyadic CVP analysis. The results of this logit regression are reported in  Table~\ref{tab:rologit_first_stage} in the form of 
odds ratios on being ranked above an alternative connection.

\paragraph{Step 2: Anchoring rank utilities to monthly dollars.}
Step~1 identifies $\beta$ only up to scale. To convert utility into dollars,
we anchor the average value of a random alter on platform~$p$ to its implied
per-connection contribution to local network value. Writing $\bar{N}_p$ for
the platform-$p$ average count of connections, the per-connection anchor is
\begin{equation}\label{eq:anchor}
\bar{V}_p \;=\; \frac{\mathrm{SPV}_p \cdot \mathrm{TPV}_p}{\bar{N}_p}\,,
\end{equation}
i.e., the share of mean monthly platform value attributable to local network
effects, divided across the typical user's connections. Our identifying
assumption is that $\bar{V}_p$ equals the average monthly dollar value, on
platform~$p$, of a \emph{random} connection in our Wave~1 design---specifically,
the four alters that each ego reported after the four closest
connections (the four top-of-mind connections). We exclude the closest-four alters from this anchoring because they are by construction atypically valuable. Combined with the additional
normalization that an alter in the omitted baseline category of every block
is worth \$0 per month, equation~(\ref{eq:anchor}) pins down the
dollars-per-utile conversion on each platform. 
We compute the sample mean of
$X_{ija}\hat{\beta}$ across the four random alters per ego on platform~$p$,
set this equal to $\bar{V}_p$, and rescale $\hat{\beta}$ accordingly. Applying
the resulting platform-specific scaling to each element of $\hat{\beta}$
yields the marginal monthly dollar values reported in
Table~\ref{tab:wta_wide}.

\paragraph{Results.}
Table~\ref{tab:wta_wide} reports the imputed monthly dollar value of each
characteristic, by platform. Before going in to the details of these results, it's important to highlight an important advantage of these results over the pure dyadic rankings. 

The combined rank-ordered logit  speaks to a question that the attribute-by-attribute CVP analysis cannot resolve. Several of the dyadic attributes we measure are correlated in the data: family members are seen
more often, are more likely to be connected on a second platform, and are more often of the same race or age as the ego. In principle, the dyadic CVP gradients documented in Figures 2 and 3 of the main text could have been driven by a single latent factor---say, relational closeness with family---with the other variables loading on that factor as proxies. The rank-ordered logit in Table~\ref{tab:rologit_first_stage} provides evidence against that single-factor interpretation. Nearly every attribute that entered the CVP gradients positively retains a positive and statistically significant coefficient when entered jointly. Contact frequency is monotone in its odds
ratio across all five tiers; ``family member or family friend,'' college connections, and shared-interest connections retain independent positive coefficients in the ``how known'' block; cross-platform connection retains an independent positive coefficient even after conditioning on contact frequency and relational category; and the homophily margins (same race,
same age) survive at conventional significance. It is therefore likely that each of these dimensions identifies a distinct source of value to the ego, rather than a common latent driver. The exception is that some of the ``how do you know'' connection characteristics, namely, ``school (K--12),'' ``work,'' and ``none of these'' are not statistically distinguishable from the online-only baseline.

Two additional patterns are worth highlighting. First,
contact frequency is the strongest predictor of platform value. An alter the ego lives with is
worth roughly four times as much per month as a non-cohabiting alter seen multiple times per week, and an order of magnitude more than an alter seen less than four times per year. Among the ``how known'' categories, family members and family friends carry a substantially larger premium than college, shared-interest, work, or K--12 connections. Cross-platform connection contributes a premium comparable in magnitude to that of a family-or-family-friend connection. Demographic and homophily characteristics---alter gender and race, same race, same age, same gender---contribute small but non-zero premia. Second, taking
the table as a whole, the premia fall into a clear three-tier hierarchy by relative importance: contact frequency at the top (with the live-with category
alone exceeding the largest premium in any other block), relational source and cross-platform connection at an intermediate tier (with family-or-family-friend status and being connected on a second platform
contributing premia of comparable magnitude), and demographic and homophily characteristics roughly an order of magnitude smaller than the intermediate
tier. The marginal monthly value of a connection on these platforms appears to derive primarily from the intensity and structural depth of the relationship rather than from demographic matching between ego and alter.

 The dollar values reported in Table~\ref{tab:wta_wide} are marginal monthly contributions of each attribute
to the value of one additional connection of a given type, conditional on the platform's per-connection anchor. Because the typical user has hundreds of connections, the value of adding a characteristic to a single connection is necessarily small in absolute terms. The largest estimates are on the order of 4 cents a month.

\begin{table}[htbp]
\centering
\caption{First-Stage Rank-Ordered Logit Estimates of Connection Value}
\label{tab:rologit_first_stage}
\begin{tabular}{lc}
\toprule
& Odds Ratio \\
\midrule
\multicolumn{2}{l}{\textit{How known (base: I only know them online)}} \\
\quad None of these                          & 1.029 \\
\quad School (K--12)                         & 1.013 \\
\quad Work                                   & 1.029 \\
\quad Through shared interest in real life   & 1.069$^{**}$ \\
\quad College or University                  & 1.068$^{**}$ \\
\quad Family member or family friend         & 1.298$^{***}$ \\
\addlinespace
\multicolumn{2}{l}{\textit{Meeting frequency (base: Never)}} \\
\quad Less than 4 times/year                 & 1.325$^{***}$ \\
\quad About once a month                     & 1.560$^{***}$ \\
\quad About once a week                      & 1.647$^{***}$ \\
\quad Multiple times per week                & 1.850$^{***}$ \\
\quad I live with this person                & 2.719$^{***}$ \\
\addlinespace
\multicolumn{2}{l}{\textit{Cross-platform connection (base: Not connected on another platform)}} \\
\quad Also connected on another platform     & 1.262$^{***}$ \\
\addlinespace
\multicolumn{2}{l}{\textit{Gender}} \\
\quad Alter female (base: alter male)                    & 1.067$^{***}$ \\
\quad Ego and alter different gender (base: same gender) & 1.024$^{**}$ \\
\addlinespace
\multicolumn{2}{l}{\textit{Age mix (base: ego younger than alter)}} \\
\quad Same age as alter                      & 1.054$^{***}$ \\
\quad Ego older than alter                   & 1.044$^{**}$ \\
\addlinespace
\multicolumn{2}{l}{\textit{Race}} \\
\quad Alter non-white (base: alter white)    & 1.011 \\
\quad Same race (base: different race)       & 1.061$^{***}$ \\
\midrule
Observations                                 & 230{,}669 \\
\bottomrule
\end{tabular}
\begin{minipage}{\linewidth}
\footnotesize
\vspace{0.5em}
\textit{Notes:} Rank-ordered exploded logit estimates of a regression of rank on all connection characteristics, pooling all platforms. The coefficients are
exponentiated and reported as odds ratios. An odds ratio greater than one indicates a
higher likelihood of being ranked above an alternative connection within
the same stratum, holding other characteristics constant. Significance:
$^{*}\,p<0.10$, $^{**}\,p<0.05$, $^{***}\,p<0.01$.
\end{minipage}
\end{table}

\begin{table}[htbp]
\centering
\caption{Imputed Marginal Dollar Value per Month of an Additional Connection,
by Characteristic and Platform}
\label{tab:wta_wide}
\sisetup{table-format=1.2e-1, table-number-alignment=center}
\begin{tabular}{l S S S S}
\toprule
Characteristic & {Facebook} & {Instagram} & {LinkedIn} & {Twitter} \\
\midrule
\multicolumn{5}{l}{\textit{Contact frequency (base: Never)}} \\
\quad Live with this person          & 0.0455   & 0.00434   & 0.0345    & 0.00389 \\
\quad Multiple times per week        & 0.0280   & 0.00267   & 0.0212    & 0.00240 \\
\quad About once a week              & 0.0227   & 0.00216   & 0.0172    & 0.00194 \\
\quad About once a month             & 0.0202   & 0.00193   & 0.0153    & 0.00173 \\
\quad Less than 4 times/year         & 0.0128   & 0.00122   & 0.00971   & 0.00110 \\
\addlinespace
\multicolumn{5}{l}{\textit{How known (base: I only know them online)}} \\
\quad Family member or family friend & 0.0119   & 0.00113   & 0.00900   & 0.00102 \\
\quad Through shared interest IRL    & 0.00305  & 0.000290  & 0.00231   & 0.000261 \\
\quad College or University          & 0.00299  & 0.000285  & 0.00227   & 0.000256 \\
\quad Work                           & 0.00132  & 0.000126  & 0.00100   & 0.000113 \\
\quad None of these                  & 0.00129  & 0.000123  & 0.000978  & 0.000110 \\
\quad School (K--12)                 & 0.000597 & 0.0000568 & 0.000452  & 0.0000510 \\
\addlinespace
\multicolumn{5}{l}{\textit{Cross-platform connection (base: Not connected on another platform)}} \\
\quad Also connected on another platform & 0.0106 & 0.00101 & 0.00803 & 0.000906 \\
\addlinespace
\multicolumn{5}{l}{\textit{Alter demographics}} \\
\quad Female (base: alter male)      & 0.00296  & 0.000282  & 0.00224   & 0.000253 \\
\quad Non-white (base: alter white)  & 0.000486 & 0.0000463 & 0.000369  & 0.0000416 \\
\addlinespace
\multicolumn{5}{l}{\textit{Dyadic characteristics}} \\
\quad Same race (base: different race)         & 0.00271  & 0.000258  & 0.00205   & 0.000232 \\
\quad Same age as alter (base: ego younger)    & 0.00239  & 0.000228  & 0.00181   & 0.000204 \\
\quad Older than alter (base: ego younger)     & 0.00197  & 0.000188  & 0.00150   & 0.000169 \\
\quad Different gender (base: same gender)     & 0.00108  & 0.000103  & 0.000822  & 0.0000927 \\
\bottomrule
\end{tabular}
\begin{minipage}{\linewidth}
\footnotesize
\vspace{0.5em}
\textit{Notes:} Each cell reports the imputed monthly dollar value of a connection to an alter possessing the row characteristic, on the column platform.
\end{minipage}
\end{table}



\subsection{Wave 2}

In Table~\ref{table:summary_stat_unfiltered2_2} we present summary statistics of network characteristics and platform valuation in each of the platforms used in Wave 2. The average monthly valuation of Facebook is comparable to the valuation found in previous research (see Table~\ref{table:wtacomparison}).\footnote{We couldn't find studies estimating valuations for Instagram, LinkedIn, and X on a national scale.} A calculation estimating monthly and yearly measures of consumer surplus generated by the four platforms is presented in Table~\ref{wtamonthlyandannual}. For monthly consumer surplus estimates, we take the average willingness to accept valuations to give up these platforms for 1 month and multiply them by their user base in the US (monthly active users).\footnote{The sources for monthly active users are \url{https://backlinko.com/social-media-users}, \url{https://sproutsocial.com/insights/linkedin-statistics/} and \url{https://news.linkedin.com/about-us#Statistics}. For linkedIn, we multiply the number of registered users (230 million) with an estimate of the percentage of users who are active monthly (48.5\%).} For yearly consumer surplus estimates, we multiply the monthly estimates by 12. Assuming that platform surplus is linear in the number of months may be an overestimate however. This is because hyperbolic discounting, rational addiction, or switching costs could make the willingness-to-pay for a year of Facebook smaller than this calculation would imply. Indeed, Allcott et al. \cite{allcott2020welfare} find some evidence of this. Individuals in their experiment who were paid to deactivate Facebook for a month valued future months of Facebook usage less. 

In Table~\ref{table:summary_stat_unfiltered2} we present summary statistics on age, gender, ethnicity, income and political orientation of the sample in Wave 2. 

\begin{table}[!htbp] \centering 
  \caption{Additional Summary Statistics for Wave 2} 
  \label{table:summary_stat_unfiltered2_2}
\begin{tabular}{@{\extracolsep{5pt}}lcccccc} 
\\[-1.8ex]\hline 
\hline \\[-1.8ex] 
\\[-1.8ex] & FB & X & IN & LK \\ 
\hline \\[-1.8ex] 
Average Number of Connections & 604.7 &  &  &  302.2 \\ 
  & & & &  \\ 
Average Number of Followers &  & 1384 & 2861.4 &    \\ 
  & & & & \\ 
Average Number of Followees &  & 648.2 & 599.4 &   \\ 
  & & & & \\ 
Average Monthly Valuation & 100.9 & 78.35 & 91.25 & 86.85  \\ 
  & & & & \\ 
\textit{N} & 2,470 & 2,238 & 2,488 & 2,200  \\ 
\hline 
\hline \\[-1.8ex] 
\end{tabular} 
\end{table} 

\begin{table}[]
\centering
\resizebox{\textwidth}{!}{%
\begin{tabular}{@{}llll@{}}
\toprule
Paper      & Sample                            & Mean WTA/month & Median WTA/month \\ \midrule
Allcott et al. (2020)      & Active Facebook users (\textgreater{}15 mins/ per day) & \$180 & \$100   \\
Brynjolfsson et al. (2024) & Representative sample of Facebook users                & N/A   & \$48.53 \\
This paper & Facebook users recruited on Lucid & \$100.90       & \$100            \\ \bottomrule
\end{tabular}%
}
\caption{Comparison of Willingness to Accept (WTA) estimates for giving up Facebook for one month}
\label{table:wtacomparison}
\end{table}

\begin{table}[]
\centering
\resizebox{\textwidth}{!}{%
\begin{tabular}{@{}lllll@{}}
\toprule
Platform & Mean WTA/month & US user base & Monthly consumer surplus & Yearly consumer surplus \\ \midrule
Facebook  & \$100.90 & 177.5 million & \$17.9 Billion & \$214.9 Billion \\
Instagram & \$91.25  & 138.5 million & \$12.6 Billion & \$151.7 Billion \\
X   & \$86.85  & 50.5 million  & \$4.4 Billion  & \$52.6 Billion  \\
LinkedIn  & \$78.35  & 111.5 million   & \$8.7 Billion   & \$104.9 Billion \\ \bottomrule
\end{tabular}%
}
\caption{Monthly and annual estimates of consumer surplus generated by the platforms}
\label{wtamonthlyandannual}
\end{table}

\begin{table}[!htbp] \centering 
\caption{Summary Statistics for Wave 2}
\renewcommand*{\arraystretch}{1.1}\resizebox{0.8\textwidth}{!}{
\begin{tabular}{lrrrrrrrr}
\hline
\hline
   & \multicolumn{2}{c}{Facebook} & \multicolumn{2}{c}{Instagram} & \multicolumn{2}{c}{LinkedIn} & \multicolumn{2}{c}{X}  \\ 
  & \multicolumn{1}{c}{N} & \multicolumn{1}{c}{\%} & \multicolumn{1}{c}{N} & \multicolumn{1}{c}{\%} & \multicolumn{1}{c}{N} & \multicolumn{1}{c}{\%} & \multicolumn{1}{c}{N} & \multicolumn{1}{c}{\%} \\ 
\hline
& 2470 &  & 2488 &  & 2200 &  & 2237 &  \\ 
gender &  &  &  &  &  &  &  &  \\ 
-- Female & 1301 & 53\% & 1259 & 51\% & 1128 & 51\% & 1132 & 51\% \\ 
-- Male & 1158 & 47\% & 1222 & 49\% & 1065 & 48\% & 1097 & 49\% \\ 
-- Other  & 11 & 0\% & 7 & 0\% & 7 & 0\% & 8 & 0\% \\ 
\\
Age &  &  &  &  &  &  &  &  \\ 
-- 18-24 years old & 406 & 16\% & 399 & 16\% & 312 & 14\% & 331 & 15\% \\ 
-- 25-34 years old & 538 & 22\% & 429 & 17\% & 299 & 14\% & 388 & 17\% \\ 
-- 35-44 years old & 412 & 17\% & 504 & 20\% & 443 & 20\% & 466 & 21\% \\ 
-- 45 years or older & 1111 & 45\% & 1153 & 46\% & 1145 & 52\% & 1050 & 47\% \\ 
\\
Ethnicity &  &  &  &  &  &  &  &  \\ 
-- White & 1762 & 71\% & 1534 & 62\% & 1497 & 68\% & 1432 & 64\% \\ 
-- Black or African American & 421 & 17\% & 417 & 17\% & 352 & 16\% & 438 & 20\% \\
-- Hispanic or Latino & 180 & 7\% & 318 & 13\% & 173 & 8\% & 235 & 11\% \\ 
-- Asian or Pacific Islander & 72 & 3\% & 119 & 5\% & 114 & 5\% & 84 & 4\% \\  
-- Nat. American or Am. Indian & 18 & 1\% & 28 & 1\% & 26 & 1\% & 21 & 1\% \\ 
-- Other & 17 & 1\% & 72 & 3\% & 38 & 2\% & 27 & 1\% \\ 
\\
Income &  &  &  &  &  &  &  &  \\ 
-- Less than \$100,000 & 925 & 37\% & 1111 & 45\% & 935 & 43\% & 895 & 40\% \\ 
-- \$100,000 or more & 1545 & 63\% & 1377 & 55\% & 1265 & 57\% & 1342 & 60\% \\ 
\\
Political Orientation &  &  &  &  &  &  &  &  \\ 
-- Liberal & 568 & 23\% & 519 & 21\% & 523 & 24\% & 515 & 23\% \\ 
-- Moderate & 1479 & 60\% & 1512 & 61\% & 1243 & 56\% & 1306 & 58\% \\ 
-- Conservative & 423 & 17\% & 457 & 18\% & 434 & 20\% & 416 & 19\% \\ 
\\
Total Platform Valuation &  &  &  &  &  &  &  &  \\
Median   & \multicolumn{2}{c}{100} & \multicolumn{2}{c}{100} & \multicolumn{2}{c}{90} & \multicolumn{2}{c}{70}  \\
Mean   & \multicolumn{2}{c}{101} & \multicolumn{2}{c}{91.3} & \multicolumn{2}{c}{86.8} & \multicolumn{2}{c}{78.4}  \\
SD   & \multicolumn{2}{c}{60.5} & \multicolumn{2}{c}{58.5} & \multicolumn{2}{c}{59.4} & \multicolumn{2}{c}{58.7}  \\
\hline
\hline
\hline
\end{tabular}
}
\label{table:summary_stat_unfiltered2}
\end{table}



\subsubsection{Predictors of Total Platform Valuation}\label{network_effects_model}

We implement two sets of models that predict Total Platform Value. The first set predicts this value with connection counts being included as linear or log-linear predictive terms. This approach is atheoretical, and the specifications were pre-registered.

After describing these specifications and results, we discuss a theoretically motivated model of Total Platform Value used to estimate the Share of Platform Value created by local network effects, which is reported in the manuscript Figure 1b. 

\subsubsection{Pre-registered Models Total Platform Value}

For Wave 2 we report all pre-registered analysis as published at: \newline \url{https://www.socialscienceregistry.org/trials/12300}.

The first set of results are on the predictors of Total Platform Valuation. In Tables~\ref{first_pre_table} to~\ref{logols_LI} we present OLS results where monthly Total Platform Valuation is explained by the number of connections on the platform. They are estimated using variations on the following equation:
\begin{equation}
    V_i = \beta_1 C_i + \beta_2 \chi + \alpha
\end{equation}
Where $V$ is Total Platform Valuation, $C_i$ is some measure of the number of connections the ego has on the platform, and $\alpha$ is an intercept term. $\chi$ corresponds to a vector of controls, which varies by specification. Regressions are run separately for each platform. We are primarily interested in the coefficient $\beta_1$, which estimates the relationship between platform value and connection count. We are also interested in the predictors of $V_i$, as well as the predictability of $V_i$ overall.



The main econometric challenge in interpreting these results is the possibility of omitted variables bias. As we do not have experimental variation in the number of contacts for each consumer, these correlations should not necessarily be interpreted as causal. Popularity in the real world is one example of a problematic omitted variable. Plausibly, individuals who are more popular in the real world will find it easier to accumulate contacts online. But it may also be that these popular people have a higher opportunity cost of their time, and therefore value social media less. This omitted variable could bias the magnitude of the effect of contacts on platform value downwards. 

The interpretation of these results also faces two other econometric challenges. First, we do not know the exact functional form of the relationship between connection count and platform value---whether it is linear, log-linear, or something else. Second is the role that anchoring may play in setting ego's Total Platform Valuations. Our pre-registration therefore calls for estimating several different functional forms, as well as possible interaction terms and controlling for different potential confounders.

Our baseline preregistered specification is reported in
Table ~\ref{log_ols}. In this specification, Total Platform Valuation is explained by the log of the number of connections, for each platform. For platforms where connections are split by followers and followees, we have three specifications: one with each type of connection included individually, and ones where both types of connection are included. This table shows significant positive relationships between overall contacts and the log of follower count on all platforms. The effects are largest for Facebook connections, with a doubling of connection count associated with a \$5.04 increase in total valuation. For platforms where followers and followees are split, including either predictor alone leads to a strong and significant correlation between the logged connection count and Total Platform Valuation. However, for both X and Instagram, when both connection types are included, followers is the predictor that has a larger and more significant positive point estimate. 

Next, \ref{first_pre_table} estimates the effect of the count of connections rather than their logarithm. The effect of total connections on monthly platform value remains positive and significant, except on X. In Tables ~\ref{logols_FB} through ~\ref{logols_LI}, we report the estimated effect of log connection counts on platform value with a variety of additional controls. For Facebook and LinkedIn, the first specification adds controls for ego characteristics. The second column includes controls for the maximum platform value users were allowed to report---this is meant to investigate and account for possible anchoring effects. The final column includes both previous sets of controls, as well as controls for platform usage. Arguably this last set of controls is over-controlling, as platform value and platform usage intensity may both be positively caused by the number of platform contacts. 

Even with this wide battery of controls, we continue to find positive and highly significant relationships between the number of platform connections and platform value. While adding platform usage controls does attenuate the size of the effect, adding ego characteristic controls does so only marginally, ameliorating the concern that omitted variables are biasing the estimates. For X and Instagram we conduct the same exercise, but with six specifications as we split followers and followees. Similarly, the estimated effect of log connections on Total Platform Valuation remains significant and positive, regardless of which controls are added, although platform usage controls attenuate the effect more. The robustness of the correlation leads us to strongly suspect a positive causal role of connections on platform value, increasing valuation by between one to six dollars a month for a doubling in connections.

\begin{figure}[htpb]
    \centering
\includegraphics[width=\textwidth]{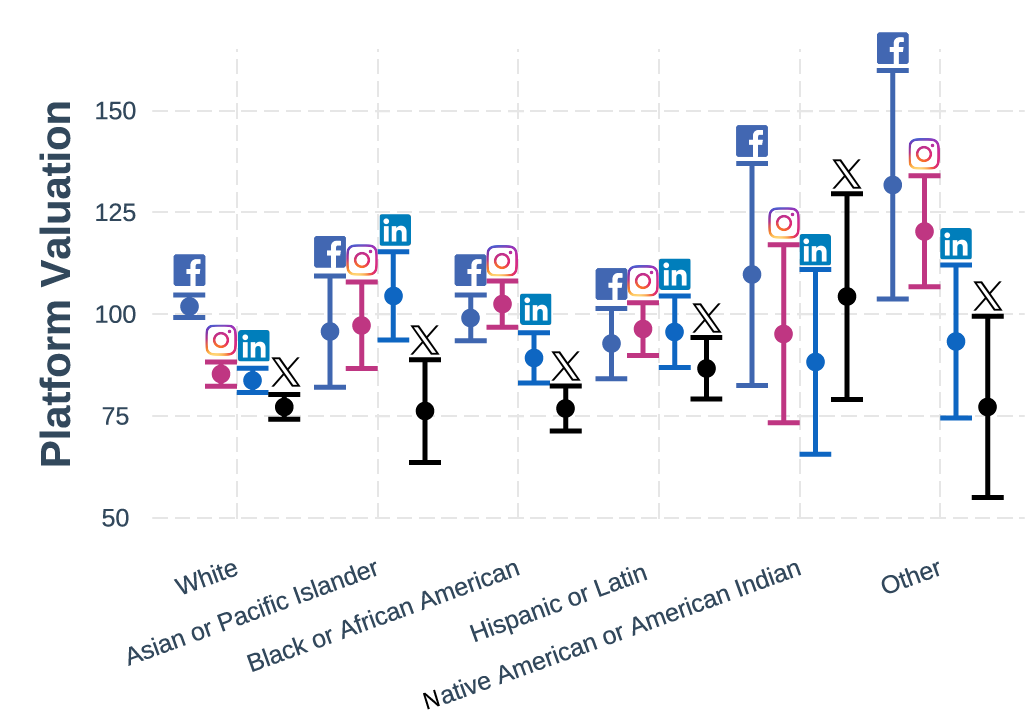}
    \caption{Conditional average valuations by ethnicity. Preregistered analysis for Wave 2.}
    \label{prereg-1}
\end{figure}

\begin{figure}[htpb]
    \centering
\includegraphics[width=0.8\textwidth]{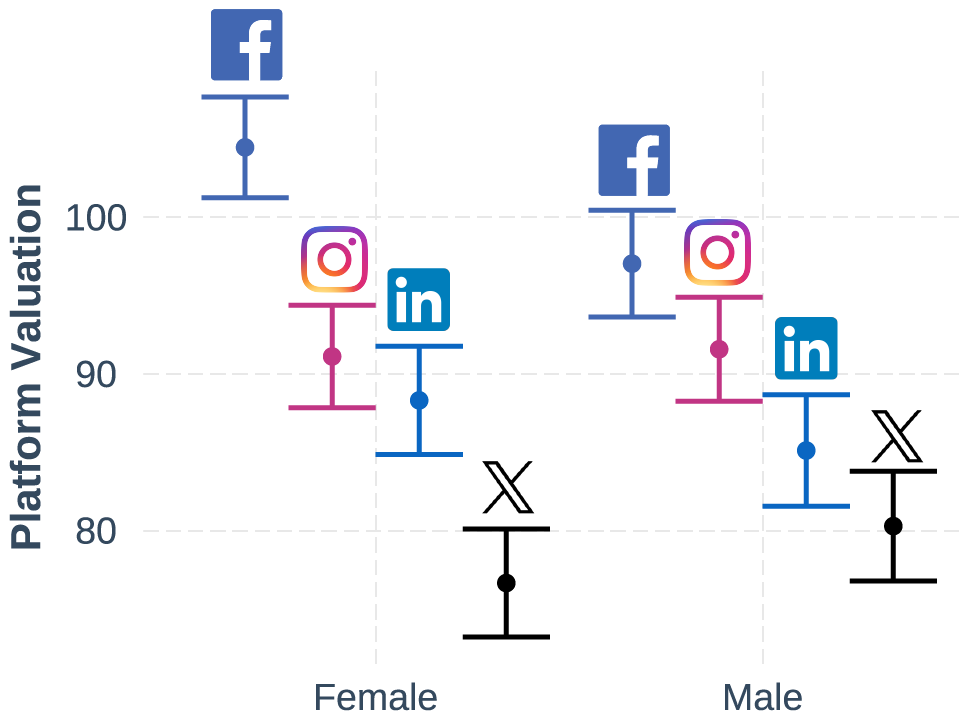}
    \caption{Conditional average valuations by gender. Preregistered analysis for Wave 2.}
    \label{prereg-2}
\end{figure}

\begin{figure}[htpb]
    \centering
\includegraphics[width=\textwidth]{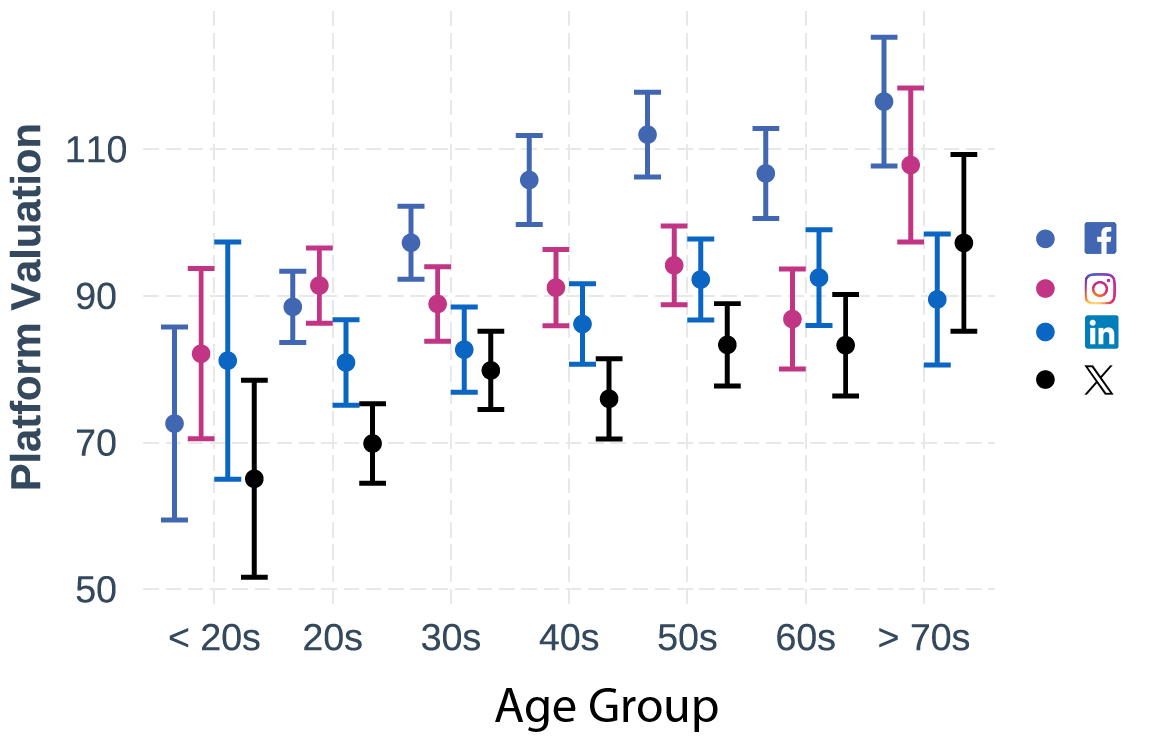}
    \caption{Conditional average valuations by age group. Preregistered analysis for Wave 2.}
    \label{prereg-3}
\end{figure}

\begin{figure}[htpb]
    \centering
    \includegraphics[width=\textwidth]{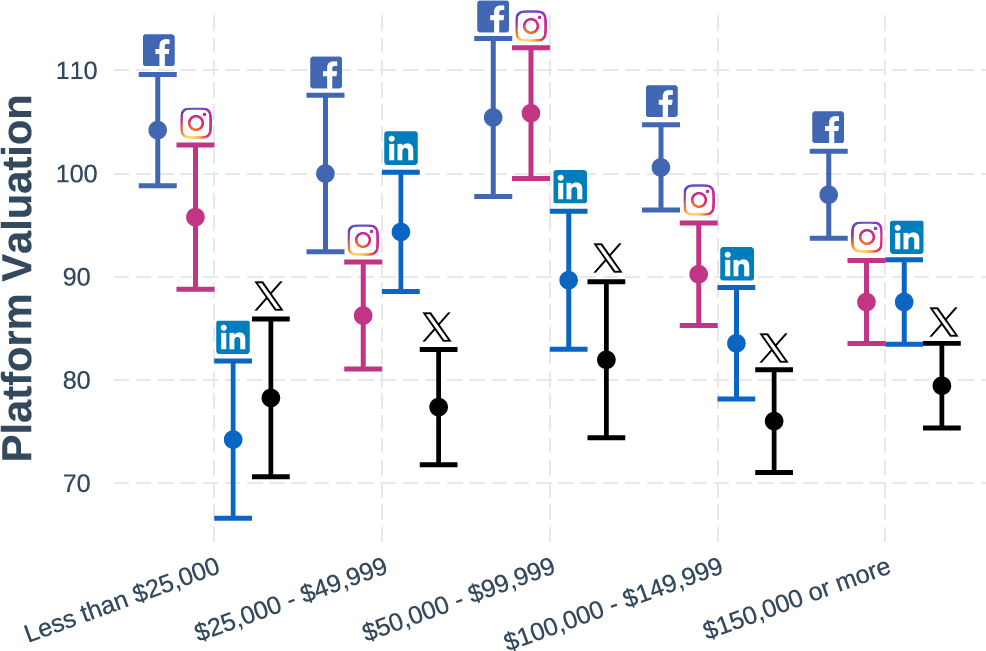}
    \caption{Conditional average valuations by income. Preregistered analysis for Wave 2.}
    \label{prereg-4}
\end{figure}

\begin{figure}[htpb]
    \centering
    \includegraphics[width=\textwidth]{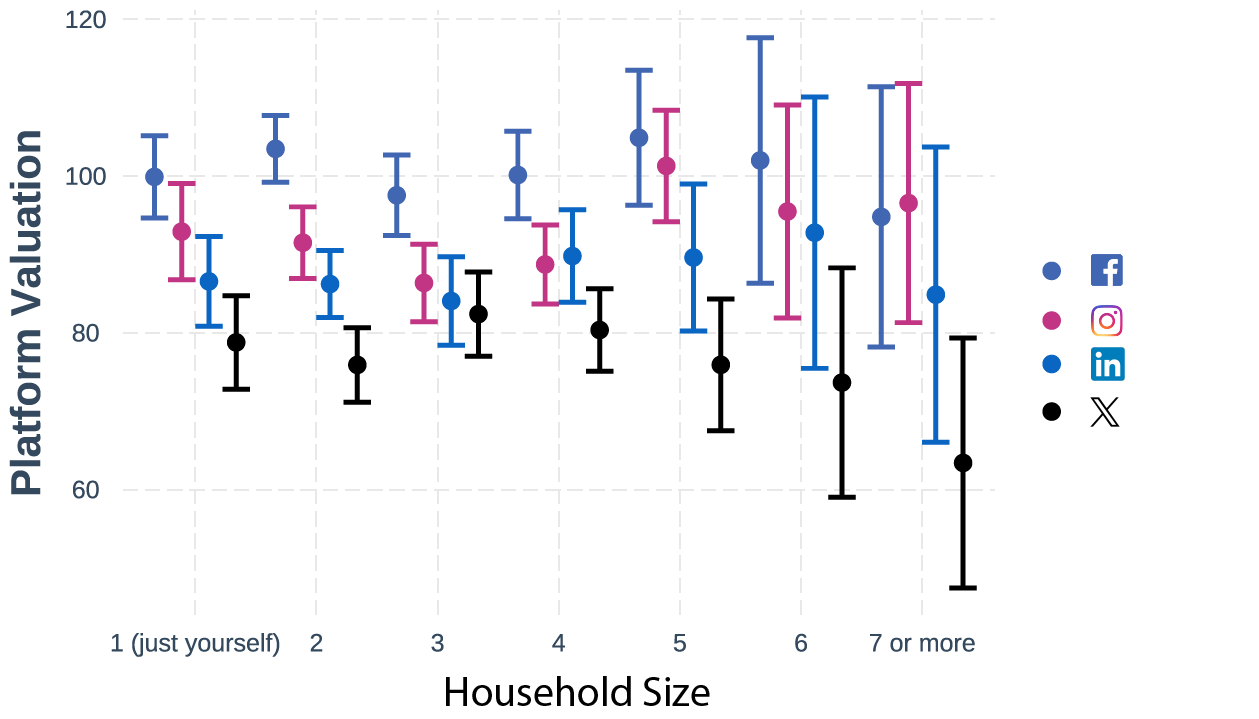}
    \caption{Conditional average valuations by number in household. Preregistered analysis for Wave 2.}
    \label{prereg-5}
\end{figure}

\begin{figure}[htpb]
    \centering
    \includegraphics[width=\textwidth]{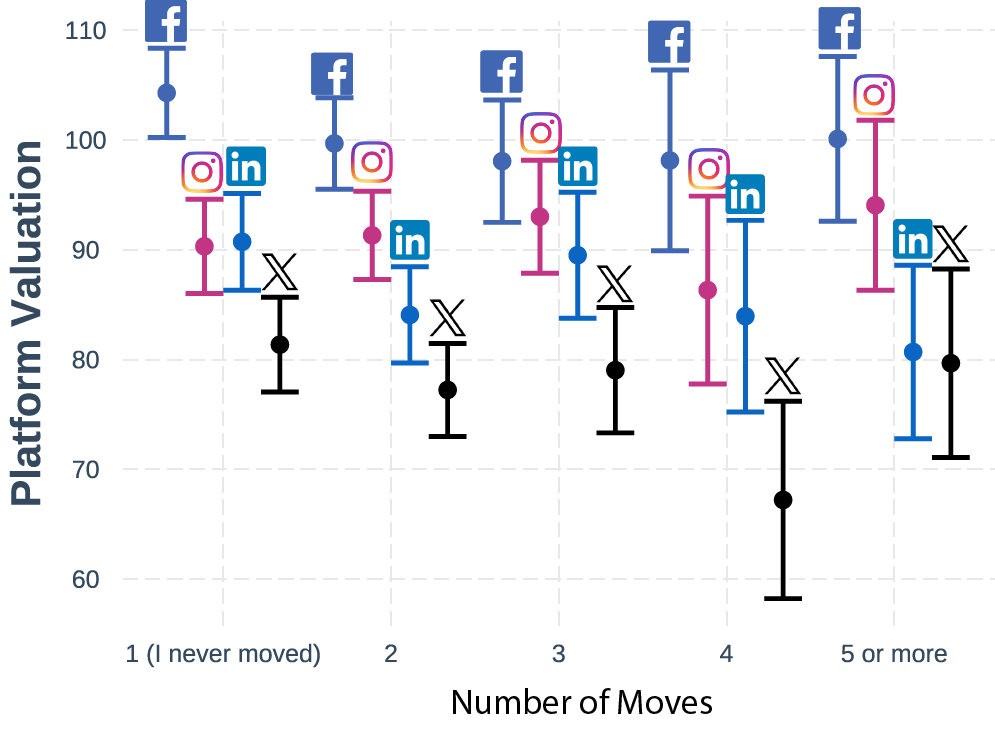}
    \caption{Conditional average valuations by number of moves. Preregistered analysis for Wave 2.}
    \label{prereg-6}
\end{figure}

\begin{figure}[htpb]
    \centering
    \includegraphics[width=\textwidth]{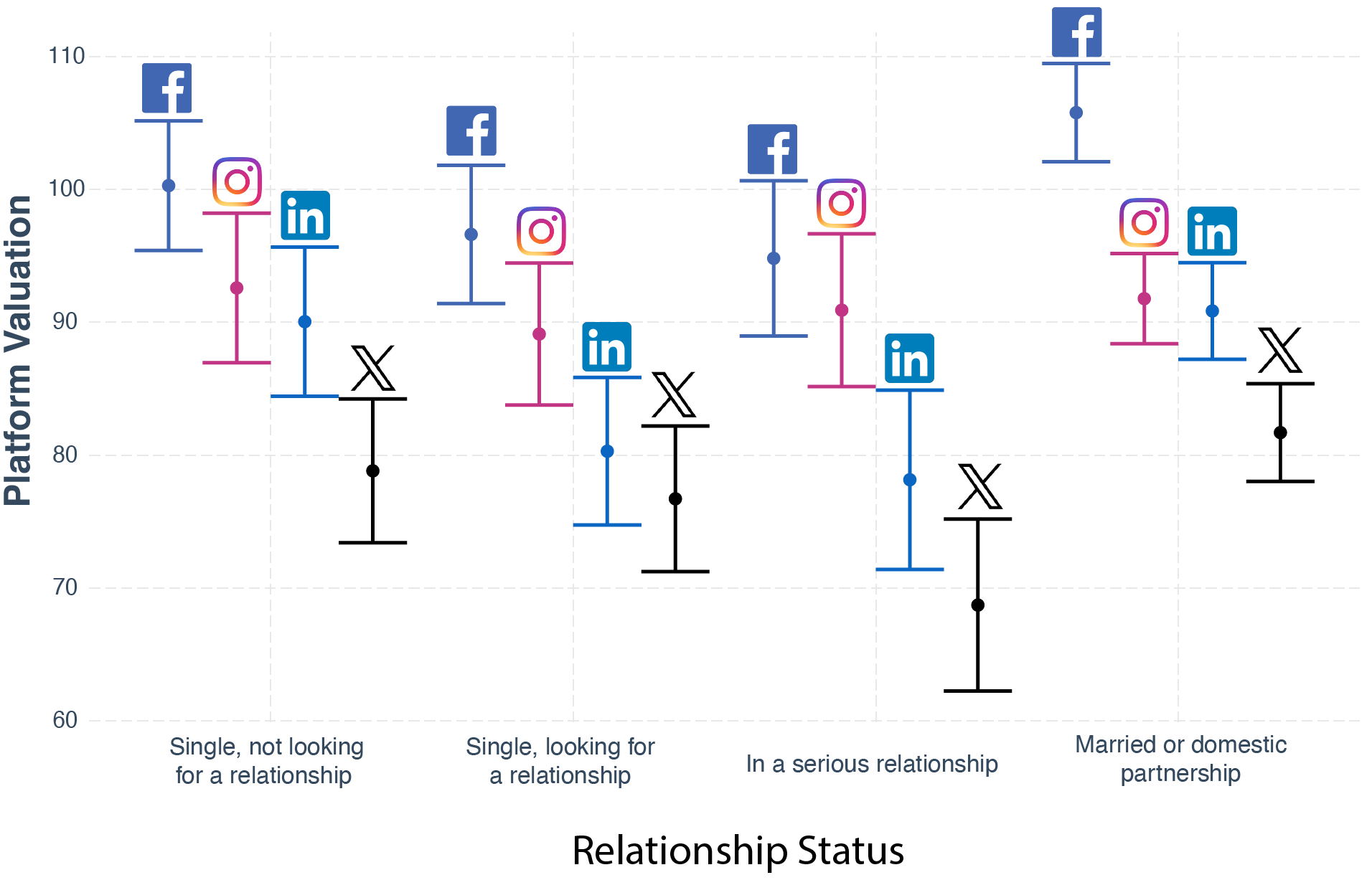}
    \caption{Conditional average valuations by relationship status. Preregistered analysis for Wave 2.}
    \label{prereg-7}
\end{figure}

\begin{figure}[htpb]
    \centering
\includegraphics[width=0.8\textwidth]{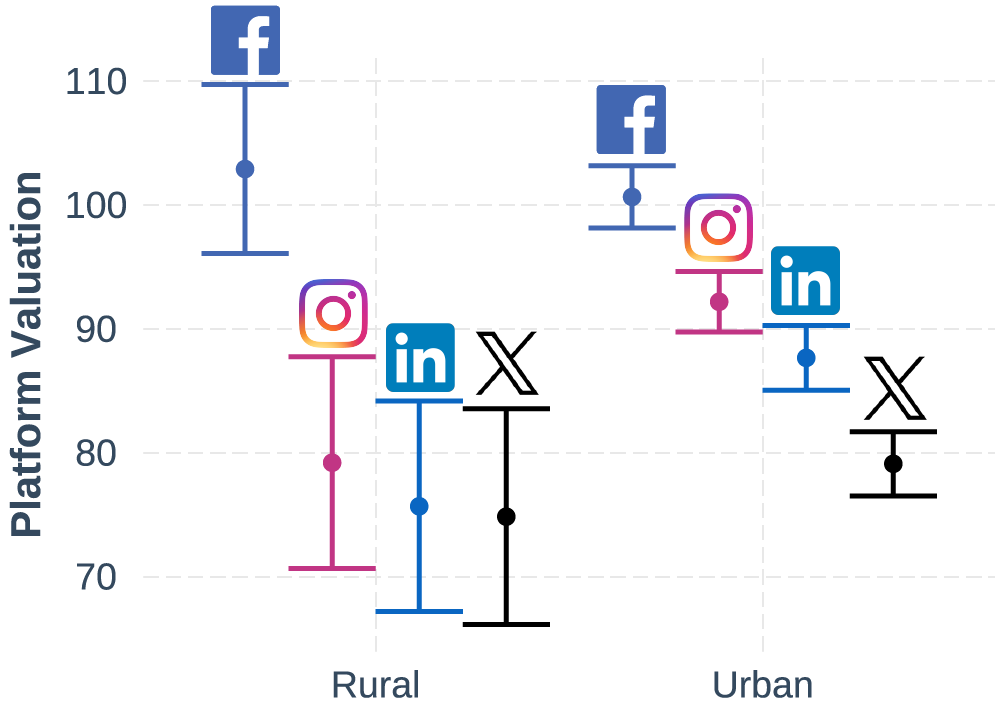}
    \caption{Conditional average valuations by urban vs. rural location. Preregistered analysis for Wave 2.}
    \label{prereg-8}
\end{figure}

\begin{figure}[!htpb]
    \centering
\includegraphics[width=\textwidth]{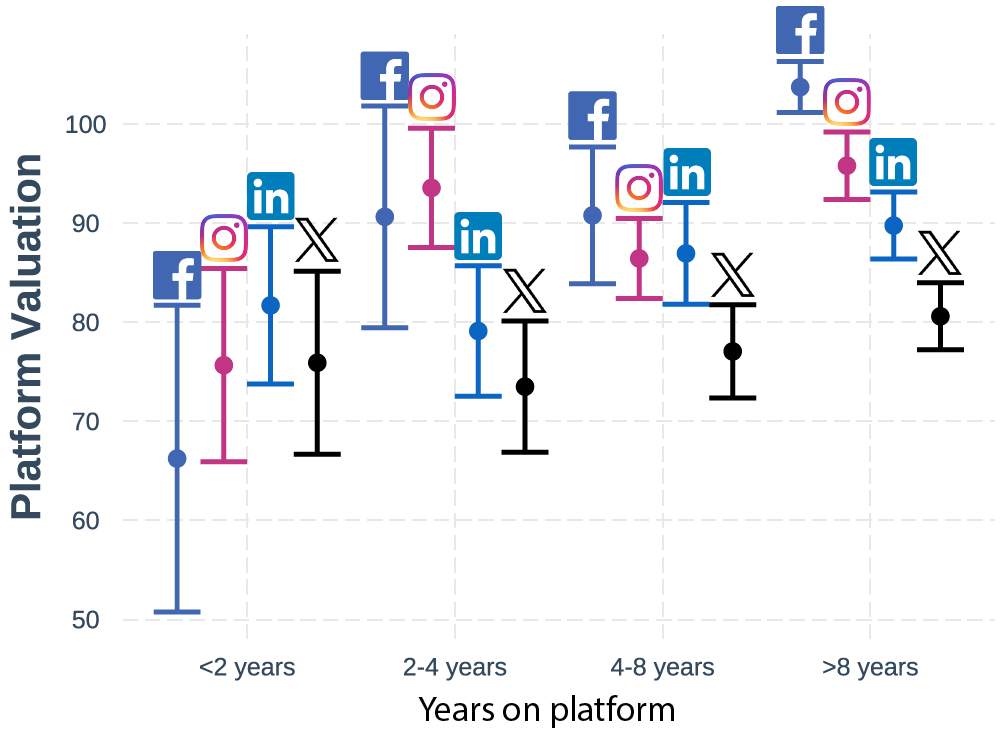}
    \caption{Conditional average valuations by years using platform. Preregistered analysis for Wave 2.}
    \label{prereg-9}
\end{figure}

\begin{figure}[!htpb]
    \centering
\includegraphics[width=\textwidth]{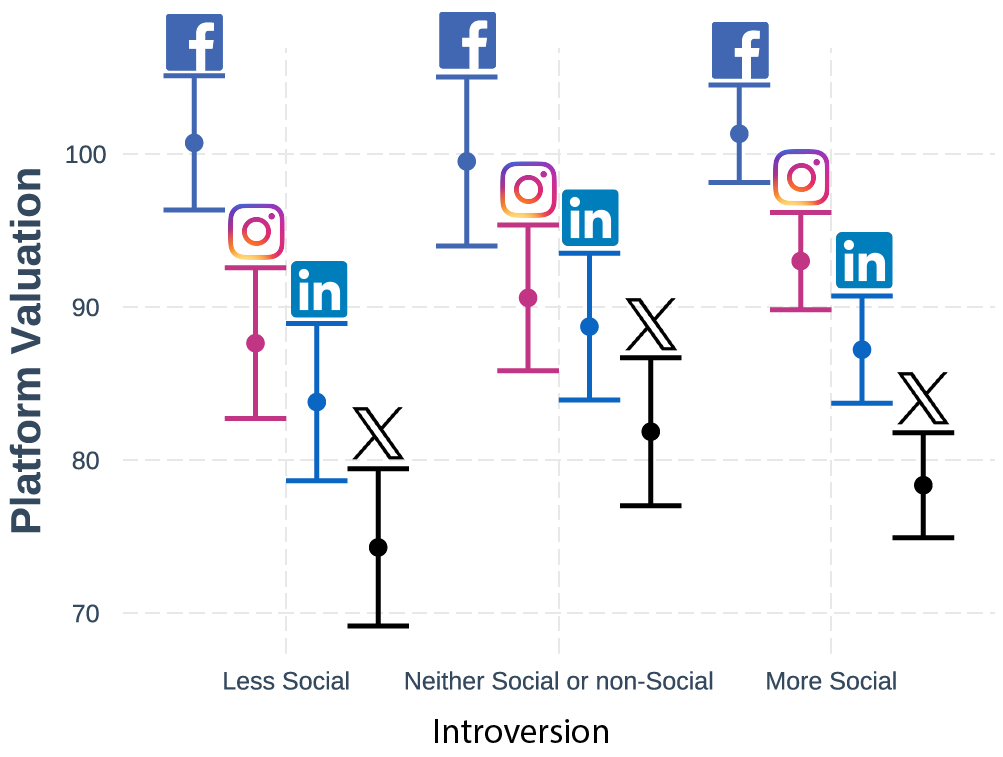}
    \caption{Conditional average valuations by self-reported introversion. The index is the average of the two introversion likert questions in the instrument. Preregistered analysis for Wave 2.}
    \label{prereg-last}
\end{figure}

We are also interested in other ego characteristics that are correlated with platform value. Even if we cannot make causal claims about these relationships, because of omitted variables concerns, they are suggestive for future research. Figures~\ref{prereg-1} to~\ref{prereg-last} report conditional average Total Platform Valuations for Wave 2 survey respondents in different groups, following the preregistration. Figure 1 in the main text reports four variations on conditional average total platform value: by platform time use, political orientation, relationship status, and job seeking status. These show a positive relationship between time use and platform value. They also show significantly higher valuations reported by those looking for a job for LinkedIn users, but not for users of other platforms.  Figure 1 also reports the empirical CDF of platform valuations for all four platforms (panel A) as well as the share of value on each platform imputed to be due to connections (panel B). Panel B is generated by dividing average platform value by the constant term estimated in a regression of platform value on the number of connections (and an intercept) in a redundant signal model as reported in Table \ref{table:fig1b_table} to calculate the contribution of other factors, while the remainder is attributed to connections.

In the main text, Figure 4 reports how total platform value varies as a function of feature use. Panel A reports, for each platform, the coefficients on feature use from a regression of total platform value as a function of using different platform features. Features found to have a significant positive effect on platform value on Facebook are using Messenger, positing on the news feed, and engaging with others' content on the news feed; on Instagram posting Reels and posting photos; on LinkedIn using Groups and updating one's profile; and on X using Spaces and bookmarking tweets. No feature was associated with a significant negative effect on value. 

Figure 4 Panel B reports how platform value varies with feature use and ego income. The conditional average platform values, for users and non-users of the feature, are plotted for egos of different income groups. The split for high vs. low income is drawn at one hundred thousand dollars. \ref{fig4btable}, Column 1 reports the difference between conditional average total platform value for high and low income egos who use the feature. P-values for this difference being different than zero are also reported. Column 2 reports the difference in differences between conditional average total platform value for high income feature users and non-users minus the difference between low-income feature users and non-users.  This quantity measures the difference in marginal contributions to total platform value of feature use for high vs. low income egos. P-values for this difference of differences being different from zero are also reported. These p-values were calculated by first determining the standard error of this difference using the standard errors of the individual groups. The t-statistic was obtained by dividing the difference in differences by its standard error. Finally, the p-value was calculated, conservatively, by using the t-distribution with degrees of freedom equal to the minimum of the two groups.

\begin{table}[h]
\centering
\caption{Summary of total platform value by feature use and ego income group for selected features.   
\\$\Delta_1 = \text{Value (High income \& uses feature)} - \text{Value (Low income \& uses feature)}$
\\$\Delta_2 = [\text{Value (High income \& uses feature)} - \text{Value (High income \& doesn't use feature)}] - [\text{Value (Low income \& uses feature)} - \text{Value (Low income \& doesn't use feature)}]$}\label{fig4btable}
\begin{tabular}{l|l|l}

\textbf{Feature} & $\Delta_1$ (p-value) & $\Delta_2$ (p-value) \\
\hline
FB Messenger & $\Delta_1 = -6.3157$ (p = 0.02143) & $\Delta_2 = -12.23003$ (p = 0.06125) \\
\hline
Instagram Reels & $\Delta_1 = -10.69087$ (p = 0.00556) & $\Delta_2 = -7.48214$ (p = 0.12209) \\
\hline
LinkedIn Groups & $\Delta_1 = -8.70992$ (p = 0.03359) & $\Delta_2 = -11.54328$ (p = 0.02670) \\
\hline
X Lists & $\Delta_1 = -21.05425$ (p = $2.621 \times 10^{-5}$) & $\Delta_2 = -27.34639$ (p = $6.176 \times 10^{-6}$) \\

\end{tabular}
\end{table}

All pre-registered figures not appearing in the main text appear here. A few highlight results from these figures include that white respondents and women have the strongest relative preference for Facebook versus the other platforms (~\ref{prereg-1} and ~\ref{prereg-2}) and that older people have a higher valuation for all platforms overall, which may be confounded by income (~\ref{prereg-3}). We found no significant relationship between introversion, family size, or number of recent moves and social media willingness-to-pay (~\ref{prereg-5}, \ref{prereg-6}, \ref{prereg-last}). On the other hand, we were able to detect some significant differences across platforms by relationship status, income, and years on the platform. Facebook was more valued by married people than single people (~\ref{prereg-7}); LinkedIn was more valued by our highest income group than our lowest (~\ref{prereg-4}); and platform value was increasing in the number of years spent on the platform, especially for Facebook (~\ref{prereg-9}). 

Tables ~\ref{LassoFB} to~\ref{LassoLI} report the results of our pre-registered LASSO analysis of the predictors of monthly platform value for each platform. The purpose of this analysis is to understand how predictable platform value is overall, and to estimate the most robust predictors of platform value. We used LASSO, with a lambda penalty selected through cross-fold validation, to select the most robust predictors. All ego-level characteristics, as well as their two- and three-way interactions were included as potential predictors. An OLS regression was then run regressing platform value on all the selected regressors. 

For each platform, two to three variations on this analysis were conducted, differing in the number of candidate predictors and their interactions. In run 1, for LinkedIn and Facebook, predictors from the following five categories were included: connection count, log of connection count, maximum platform value random group assignment (100, 150 or 200), ego characteristics\footnote{specifically, the response to any of the following: gender, age, ethnicity, urban/rural status, marital status, number in household, income, number of moves, looked.for.job, political orientation, and two introversion measuring questions} and digital platform use characteristics\footnote{specifically, the response to any of the following: years on platform, platform time use, [list of platform specific feature use] and use of one of the other main platforms or TikTok or Snapchat}. This means that up to a five-way interaction of the above factors was explored. For Instagram and X, this regression was split into two versions: Run 1 measured connections through number of followers, and run 2 measured connections through the number of followees. Run 2 for LinkedIn and Facebook was parallel to run 3 for X and Instagram. In these runs we excluded platform usage characteristics, including the number of connections and platform feature use variables because these are more endogenous than demographic characteristics. The specific predictors and their interactions included maximum value assignment and ego characteristics (the same list of these as run 1). This means in run 2 (for Facebook and LinkedIn)  and run 3 (for X and Instagram) up to two-way interactions were explored.


The first thing to note about these results is that a significant portion of the variation in platform value was explained by our predictors. In the normal OLS regressions (Tables ~\ref{logols_FB} through ~\ref{logols_LI}), the specifications with the most predictors explained between .138 (for X) and .218 (for Facebook) of the variation, based on adjusted $R^2$s. The LASSO models (~\ref{LassoFB} to~\ref{LassoLI}), for the procedure with the greatest number of predictors, had adjusted $R^2$ of between  .133 (X) and .254 (Facebook). This predictable heterogeneity provides opportunities for platforms to third-degree price discriminate and extract a larger share of the surplus from their platforms. For X, the predictive power of the model using follower count was much greater than the other two specifications, while for Instagram, the model using the count of followees was slightly more predictive than the one using followers. 

The second thing to note about the LASSO models is in the second half of each Table. In this section, coefficients for several of the selected predictors is reported. Across all platforms, many interactions of connections/followers with ego and platform use were found to positively predict platform value. This suggests that connection count is indeed a robust predictor of platform value. On Facebook, terms that were found, in combination with log of number of connections, to positively predict platform value included age, assignment to a high valuation group, and engaging on the news feed. For X it was high value group assignment, age, and using X for many hours and use of bookmarking. For Instagram and LinkedIn, age and feature use combined with the log of the number of connections strongly predicted platform value. One interesting result for Instagram is a large positive estimate of the interaction between log of followees and Urban status. This is consistent with an Instagram following being more useful for urban users. Across all platforms, the linear measure of connection count was never included as a predictor, suggesting our focus on the log of connection count in our main regression is appropriate. In the final specifications (runs 2 and 3), which do not include platform use, one robust predictor that appeared across two platforms was the interaction between the maximum value assignment and Asian ethnicity, suggesting that this group may be particularly sensitive to anchoring.

\begin{table}[!htbp] \centering 
  \caption{Preregistered OLS regressions of total valuation on number of social media contacts, Wave 2} 
  \label{first_pre_table} 
\begin{tabular}{@{\extracolsep{5pt}}lcccccc} 
\\[-1.8ex]\hline 
\hline \\[-1.8ex] 
\\[-1.8ex] & \multicolumn{6}{c}{Monthly Total Platform Valuation} \\ 
\\[-1.8ex] & FB & X(1) & X(2) & IN(1) & IN(2) & LK\\ 
\hline \\[-1.8ex] 
Connections & 0.004$^{***}$ &  &  &  &  & 0.004$^{***}$ \\ 
  & (0.001) &  &  &  &  & (0.001) \\ 
  & & & & & & \\ 
Followers &  & $-$0.00004 &  & 0.0002$^{***}$ &  &  \\ 
  &  & (0.0001) &  & (0.0001) &  &  \\ 
  & & & & & & \\ 
Followees &  &  & 0.00000 &  & 0.005$^{***}$ &  \\ 
  &  &  & (0.0002) &  & (0.001) &  \\ 
  & & & & & & \\ 
 Constant & 98.405$^{***}$ & 78.401$^{***}$ & 78.339$^{***}$ & 90.753$^{***}$ & 88.335$^{***}$ & 85.719$^{***}$ \\ 
  & (1.420) & (1.245) & (1.249) & (1.178) & (1.353) & (1.304) \\ 
  & & & & & & \\ 
\textit{N} & 2,470 & 2,238 & 2,238 & 2,488 & 2,488 & 2,200 \\ 
R$^{2}$ & 0.004 & 0.0002 & 0.000 & 0.005 & 0.007 & 0.005 \\ 
Adj. R$^{2}$ & 0.004 & $-$0.0002 & $-$0.0004 & 0.004 & 0.007 & 0.005 \\ 
Res. SE & 60.349  & 58.720  & 58.726  & 58.320  & 58.249  & 59.245  \\ 
F Statistic & 11.061$^{***}$  & 0.460 & 0.00003  & 12.123$^{***}$ & 18.222$^{***}$  & 12.089$^{***}$  \\ 
\hline 
\hline \\[-1.8ex] 
\textit{Notes:} & \multicolumn{6}{r}{$^{***}$Significant at the 1 percent level.} \\ 
 & \multicolumn{6}{r}{$^{**}$Significant at the 5 percent level.} \\ 
 & \multicolumn{6}{r}{$^{*}$Significant at the 10 percent level.} \\ 
\end{tabular} 
\end{table} 

\pagebreak

\begin{table}[!htbp] \centering 
 \scriptsize
  \caption{Preregistered OLS regressions of total valuation on the log of the number of social media contacts, Wave 2} 
  \label{log_ols} 
\begin{tabular}{@{\extracolsep{5pt}}lcccccccc} 
\\[-1.8ex]\hline 
\hline \\[-1.8ex] 
\\[-1.8ex] & \multicolumn{8}{c}{Monthly Total Platform Valuation} \\ 
\\[-1.8ex] & FB & X 1 & X 2 & X 3 & IN 1 & IN 2 & IN 3 & LK\\ 
\hline \\[-1.8ex] 
Connection  (log)  & 5.037$^{***}$ &  &  &  &  &  &  & 4.044$^{***}$ \\ 
  & (0.664) &  &  &  &  &  &  & (0.584) \\ 
  & & & & & & & & \\ 
Followers  (log) &  & 2.707$^{***}$ &  & 2.019$^{***}$ & 5.077$^{***}$ &  & 4.365$^{***}$ &  \\ 
  &  & (0.558) &  & (0.764) & (0.593) &  & (0.728) &  \\ 
  & & & & & & & & \\ 
Followees  (log)&  &  & 2.919$^{***}$ & 1.230 &  & 4.355$^{***}$ & 1.418$^{*}$ &  \\ 
  &  &  & (0.683) & (0.934) &  & (0.691) & (0.843) &  \\ 
  & & & & & & & & \\ 
 Constant & 74.637$^{***}$ & 65.993$^{***}$ & 63.835$^{***}$ & 63.014$^{***}$ & 64.138$^{***}$ & 68.101$^{***}$ & 60.402$^{***}$ & 71.590$^{***}$ \\ 
  & (3.659) & (2.830) & (3.614) & (3.622) & (3.371) & (3.853) & (4.036) & (2.533) \\ 
  & & & & & & & & \\ 
\textit{N} & 2,470 & 2,237 & 2,237 & 2,237 & 2,488 & 2,488 & 2,488 & 2,200 \\ 
R$^{2}$ & 0.023 & 0.010 & 0.008 & 0.011 & 0.029 & 0.016 & 0.030 & 0.021 \\ 
Adjusted R$^{2}$ & 0.022 & 0.010 & 0.008 & 0.010 & 0.028 & 0.015 & 0.029 & 0.021 \\ 
Res. Std. Error & 59.791 & 58.429 & 58.498 & 58.419 & 57.618 & 58.000 & 57.597 & 58.769  \\ 
F Statistic & 57.551$^{***}$ & 23.573$^{***}$ & 18.280$^{***}$ & 12.658$^{***}$  & 73.317$^{***}$  & 39.730$^{***}$ & 38.100$^{***}$ & 48.027$^{***}$  \\ 
\hline 
\hline \\[-1.8ex] 
\textit{Notes:} & \multicolumn{8}{r}{$^{***}$Significant at the 1 percent level.} \\ 
 & \multicolumn{8}{r}{$^{**}$Significant at the 5 percent level.} \\ 
 & \multicolumn{8}{r}{$^{*}$Significant at the 10 percent level.} \\ 
\end{tabular} 
\end{table}

 \newgeometry{left=3cm,top=0.5cm, bottom=0.1cm} 

\begin{table}[!htbp] \centering 
\tiny
  \caption{{\bf Facebook}: Preregistered OLS regressions of total valuation on the number of social media contacts and other controls, Wave 2} 
  \label{logols_FB} 
\begin{tabular}{@{\extracolsep{5pt}}lccc} 
\\[-1.8ex]\hline 
\hline \\[-3.5ex] 
\\[-1.8ex] & \multicolumn{3}{c}{Monthly Facebook Valuation} \\ 
\\[-2.5ex] & 1b(i) & 1b(ii) & 1b(iii)\\ 
\hline \\[-1.8ex] 
 Facebook connection count (log) & 5.520$^{***}$ & 5.340$^{***}$ & 3.620$^{***}$ \\
\end{tabular} 
\end{table}

\begin{table}[!htbp] \centering 
       \setlength{\tabcolsep}{0pt}
 \tiny
  \caption{{\bf X}: Preregistered OLS regressions of total valuation on the number of social media contacts and other controls, Wave 2} 
  \label{logols_X} 
\begin{tabular}{@{\extracolsep{5pt}}lcccccc} 
\\[-1.8ex]\hline 
\hline \\[-3ex] 
\\[-1.8ex] & \multicolumn{6}{c}{Monthly X Valuation} \\ 
\\[-1.8ex] & 1b(i) & 1b(i) & 1b(ii) & 1b(ii) & 1b(iii) & 1b(iii)\\ 
\hline \\[-1.8ex] 
X Followers count (log) & 2.814$^{***}$ &  & 2.771$^{***}$ &  & 1.597$^{***}$ &  \\
\end{tabular} 
\end{table}

\begin{table}[!htbp] \centering 
      \setlength{\tabcolsep}{0pt}
 \tiny
  \caption{{\bf Instagram}: Preregistered OLS regressions of total valuation on the number of social media contacts and other controls, Wave 2} 
  \label{logols_inst} 
\begin{tabular}{@{\extracolsep{5pt}}lcccccc} 
\\[-1.8ex]\hline 
\hline \\[-3ex] 
\\[-1.8ex] & \multicolumn{6}{c}{Monthly Instagram Valuation} \\ 
\\[-2ex] & 1b(i) & 1b(i) & 1b(ii) & 1b(ii) & 1b(iii) & 1b(iii)\\ 
\hline \\[-2ex] 
Insta Followers count (log) & 4.909$^{***}$ &  & 5.091$^{***}$ &  & 3.194$^{***}$ &  \\
\end{tabular} 
\end{table}

\begin{table}[!htbp] \centering 
       \setlength{\tabcolsep}{0pt}
 \tiny
 
  \caption{{\bf LinkedIn}: Preregistered OLS regressions of total valuation on the number of social media contacts and other controls, Wave 2} 
  \label{logols_LI} 
\begin{tabular}{@{\extracolsep{5pt}}lccc} 
\\[-1.8ex]\hline 
\hline \\[-3ex] 
\\[-1.8ex] & \multicolumn{3}{c}{Monthly LinkedIn Valuation} \\ 
\\[-2.5ex] & 1b(i) & 1b(ii) & 1b(iii)\\ 
\hline \\[-1.8ex] 
LinkedIn connections count (log) & 3.740$^{***}$ & 4.160$^{***}$ & 3.254$^{***}$ \\
\end{tabular} 
\end{table}  

\newgeometry{left=3cm,top=1cm, bottom=1.7cm}

\begin{table}[!htpb]
\centering
\begin{tabular}{l c c}
\toprule
\textbf{Characteristic} & \textbf{Run 1} & \textbf{Run 2} \\
\midrule
Observations & 2,428 & 2,428 \\
R-squared & 0.262 & 0.167 \\
Adjusted R-squared & 0.254 & 0.166 \\
F Statistic & 30.473*** & 162.106*** \\
Number of Coefficients & 28 & 3 \\
\midrule
log(connections):age & 0.011 & -- \\
 & (0.015) & \\
log(connections):group\_assignment150usd:age & 0.082*** & -- \\
 & (0.021) & \\
log(connections):group\_assignment200usd:age & 0.038 & -- \\
 & (0.037) & \\
log(connections):group\_assignment200usd:Messenger & -1.922* & -- \\
 & (1.117) & \\
log(connections):age:Engaging\_on\_your\_news\_feed & 0.022** & -- \\
 & (0.010) & \\
log(connections):Urban\_RuralUrban:FB.10--30 minutes per day & 1.306* & -- \\
 & (0.667) & \\
log(connections):group\_assignment200usd:Slightly conservative & -- & 0.299** \\
 & & (0.135) \\
\bottomrule
\end{tabular}
\begin{flushleft}
\small Note: *p$<$0.1; **p$<$0.05; ***p$<$0.01
\end{flushleft}
\caption{Summarized Facebook OLS Regression Results: Facebook Value on LASSO Covariates}\label{LassoFB}
\end{table}

\begin{table}[!htpb]
\centering
\resizebox{\textwidth}{!}{%
\begin{tabular}{l c c c}
\toprule
\textbf{Characteristic} & \textbf{Run 1} & \textbf{Run 2} & \textbf{Run 3} \\
\midrule
Observations & 2,187 & 2,187 & 2,187 \\
R-squared & 0.141 & 0.094 & 0.070 \\
Adjusted R-squared & 0.133 & 0.091 & 0.068 \\
F Statistic & 18.724*** & 32.263*** & 54.435*** \\
Number of Coefficients & 19 & 7 & 3 \\
\midrule
log(followers):group\_assignment150usd:age & 0.057*** & -- & -- \\
 & (0.017) & & \\
log(followers):group\_assignment200usd:age & 0.0004 & -- & -- \\
 & (0.046) & & \\
log(followers):group\_assignment200usd:Bookmark\_tweets & 3.013*** & -- & -- \\
 & (0.790) & & \\
log(followers):group\_assignment200usd:age:X.2-3 hours per day & 0.072*** & -- & -- \\
 & (0.022) & & \\
log(followers):group\_assignment200usd:age:Like\_reply\_retweet\_tweets & 0.030 & -- & -- \\
 & (0.050) & & \\
log(followers):group\_assignment150usd:age:Direct\_messages & 0.033* & -- & -- \\
 & (0.019) & & \\
log(followings):group\_assignment200usd:age & -- & 0.063* & -- \\
 & & (0.020) & \\
group\_assignment200usd:Married or domestic partnership & -- & -- & 10.161** \\
 & & & (4.416) \\
\bottomrule
\end{tabular}%
}
\caption{Summarized X OLS Regression Results: X Value on LASSO Covariates}\label{LassoX}
\begin{flushleft}
\footnotesize{Note: *p$<$0.1; **p$<$0.05; ***p$<$0.01}
\end{flushleft}
\end{table}

\begin{table}[!htpb]
\centering
\resizebox{\textwidth}{!}{%
\begin{tabular}{l c c c}
\toprule
\textbf{Characteristic} & \textbf{Run 1} & \textbf{Run 2} & \textbf{Run 3} \\
\midrule
Observations & 2,440 & 2,440 & 2,440 \\
R-squared & 0.204 & 0.207 & 0.141 \\
Adjusted R-squared & 0.194 & 0.197 & 0.135 \\
F Statistic & 20.551*** & 21.639*** & 22.062*** \\
Number of Coefficients & 30 & 29 & 18 \\
\midrule
log(followers):group\_assignment150usd:age & 0.066*** & -- & -- \\
 & (0.020) & & \\
log(followers):group\_assignment200usd:age & -0.003 & -- & -- \\
 & (0.025) & & \\
log(followers):group\_assignment150usd:Urban\_RuralUrban & 0.796 & -- & -- \\
 & (1.063) & & \\
log(followers):age:Post\_Instagram\_reels & 0.011 & -- & -- \\
 & (0.013) & & \\
log(followers):Urban\_RuralUrban:Post\_photos\_as\_a\_post & 0.462 & -- & -- \\
 & (0.520) & & \\
log(followers):group\_assignment200usd:age:Like\_and\_comment\_on\_stories & 0.018 & -- & -- \\
 & (0.030) & & \\
log(followings):Urban\_Rural-Urban & -- & 1.357** & -- \\
 & & (0.608) & \\
group\_assignment200usd:ethnicity - Asian or Pacific Islander & -- & -- & 38.307*** \\
 & & & (8.968) \\
\bottomrule
\end{tabular}%
}
\caption{Summarized Instagram OLS Regression Results: Instagram Value on LASSO Covariates}\label{LassoINST}
\begin{flushleft}
\footnotesize{Note: *p$<$0.1; **p$<$0.05; ***p$<$0.01}
\end{flushleft}
\end{table}

\begin{table}[!htpb]
\centering
\resizebox{\textwidth}{!}{%
\begin{tabular}{l c c}
\toprule
\textbf{Characteristic} & \textbf{Run 1} & \textbf{Run 2} \\
\midrule
Observations & 2,159 & 2,159 \\
R-squared & 0.216 & 0.141 \\
Adjusted R-squared & 0.200 & 0.137 \\
F Statistic & 13.255*** & 32.036*** \\
Number of Coefficients & 44 & 11 \\
\midrule
log(connections):group\_assignment150usd:age & 0.050* & -- \\
 & (0.026) & \\
log(connections):group\_assignment200usd:age & -0.023 & -- \\
 & (0.035) & \\
log(connections):group\_assignment200usd:Facebook & -2.090 & -- \\
 & (2.330) & \\
log(connections):age:LinkedIn\_2-3 hours per day & 0.039* & -- \\
 & (0.020) & \\
log(connections):age:LinkedIn\_groups & 0.028** & -- \\
 & (0.013) & \\
log(connections):group\_assignment150usd:age:Keep\_profile\_up\_to\_date & 0.299** & -- \\
 & (0.135) & \\
group\_assignment200usd:ethnicity - Asian or Pacific Islander & -- & 36.604*** \\
 & & (8.698) \\
\bottomrule
\end{tabular}%
}
\caption{Summarized LinkedIn OLS Regression Results: LinkedIn Value on LASSO Covariates}\label{LassoLI}
\begin{flushleft}
\footnotesize{Note: *p$<$0.1; **p$<$0.05; ***p$<$0.01}
\end{flushleft}
\end{table}

\pagebreak

 \newgeometry{left=1in,top=1in, bottom=1in} 
 
\subsubsection{The Redundant Signal Model of Local Network Effect Share of Platform Value}
\doublespacing

In the manuscript, we use a different, theoretically motivated model for relating the number of connections to platform value. This model is used for calculating the estimates of the Share of Total Platform Value caused by network effects reported in Figure 1b. The equation is reported in equation \ref{redundant_signals}. We now briefly motivate the use of this model. 

Previous papers have studied the value of two types of network effects, which are similar, but not identical, to the phenomena of our primary interest. First are studies of undifferentiated network effects, such as Metcalfe's Law, examined theoretically ~\cite{rohlfs1974theory} or empirically ~\cite{bramoulle2009identification}. These studies are not of the value of specific connections to specific users, but rather of how a large user base leads to high platform valuations, on average, through a variety of mechanisms. A second set of studies is sociological, looking into how the value of different sorts of immediate connections, in person or online, creates value for users. See, for example, the seminal work by Granovetter~\cite{granovetter1973strength} proposing the importance of weak ties for accessing information and opportunities, and recent work by Rajkumar et al.~\cite{rajkumar2022causal} testing these theories on LinkedIn. 

In this paper, we are interested in how individual-level platform value varies with the number and type of connections. To do so, we conceive of a model of platform value where connections are formed to transmit information, both to the focal user and to the platform, to improve the service for the focal user, for example, through more useful news feeds, direct messages, or other personalized content. 

In this framework, platform users make connections because they believe it will lead to them either receiving or communicating data that is useful to them.

For example, consider someone on social media who wants to hear important information. This could be personal information, like the birth of a colleague's son, or professional information, like a job opening in an industry important to the user. The connection could also create value from being able to communicate the same information outwards. For example, the information that the individual is newly single and looking for a date. The value gained from these communications are examples of local network effects. This is because our user would be highly unlikely to gain these benefits from the platform if he or she had not made the connection. 

This framework has several conceptual and practical advantages. First, it motivates the fact that value from information conveyance on the platform faces decreasing returns, rather than simply assuming it. Second, due to incentive compatibility constraints (we cannot credibly commit to payments in the thousands of dollars), our measure of platform value has a maximum, so it makes sense to have a model of platform value where value saturates as the number of connection increases. Finally, it is plausible that the value of information received on a platform has a heavy tail, with most value being conveyed by a small minority of information, making a model where one or a few signals create most of connection value plausible.  

In this framework, the value of a particular connection may differ for several reasons. First, certain connections may be more likely to send signals of different types. Second, certain connections may be more redundant, giving information that the user is likely to know. Finally, the value of a type of signal may be complementary to whether a signal of another type has arisen: for example, the information that a job is available in a certain city is complementary to information that the city has certain amenities. 

Our vision of the value of local network effects -- as connections that increase the likelihood of me receiving valuable communications -- leads us to model platform value through a redundant signal or latent topics communication model. Messages across topics can be complementary, because knowing one piece of information increases the value of another. More commonly, messages are redundant, and not useful once a message has been received. This leads to the following model of the total value to the user from all platforms:

\begin{equation}\label{redundant_signals}
TPV_i \;=\;  \sum_{p=1}^P \delta_p
\;+\; \sum_{k=1}^K \lambda_k \Big(1 - e^{-X_{ik}}\Big)
\;+\; \sum_{k=1}^K \sum_{\ell = k+1}^K 
   \theta_{k\ell}\,\Big(1 - e^{-X_{ik}}\Big)\,\Big(1 - e^{-X_{i\ell}}\Big)
\;+\; \varepsilon_i ,
\end{equation}

and
\[
\qquad 
X_{ik} \;=\; \sum_{p=1}^P\sum_{t=1}^T z_{itkp}\,n_{itp}, 
\qquad 
z_{itkp} \equiv -\log(1 - \pi_{itkp}) .
\]

Where $TPV_i$ represents total platform value for user $i$ across all platforms $p$; $\lambda_k$ represents the value that $i$ gets from hearing message $k$, assuming they receive no other messages; $1 - e^{-X_{ik}}$ corresponds to the probability that the user receives message $k$. It is a function of $n_{itp}$ -- the number of connections the user has of type $t$ on platform $p$, as well as $\pi_{itkp}$, which is the probability a contact of type $t$ provides the information of type $k$ on platform $p$;  $\theta_{k,\ell}$ represents the level of complementarity (or redundancy) between information of type $k$ and $\ell$. 

This framework suggests several directions for analysis: understanding the share of value that comes from the platform without connections ($\delta$) vs. local network effects (the share of platform value from local network effects, or SPV) is one. Another direction is to understand the relative value of different sorts of connections, both at the margin and on average. We are also interested in how different sorts of information might complement each other. Finally, we might imagine that the types of information shared on different networks (the topics $k$) might be different even for the same user. We focus in particular on heterogeneity in the value of connections of different types in the rest of the paper. 

Figure 1b in the manuscript uses the model for the first question. Unfortunately, we do not observe the share of connections by type at the individual level in our WTA analysis. We also do not observe multiple platforms or data streams per user. However, we can still estimate $\delta$ in a simplified specification with homogeneous connection types and messages. This yields:

\[
TPV_{p,i} \;=\; \delta_p \;+\; \lambda_p \left(1 - e^{-z_p\,C_{p,i}}\right) + \varepsilon_i ,
\]

\[
z_p \;\equiv\; -\log(1-\pi_p), 
\qquad z \geq 0, 
\qquad \lambda \geq 0 .
\]

In the context of platforms with both followers and followees, we allow for these to have two $\lambda_{k,p}$ and two $\pi_{k,p}$ terms, representing a separate value (and cumulative probability) for an outgoing and incoming signal -- essentially treating these as different topics. $\theta_{k,\ell}$ is set so these values are perfect substitutes.

\begin{table}
\caption{Estimate of the value of local network effects following the redundant signal model, underlying Figure 1b}
\label{table:fig1b_table}
\begin{center}
\begin{tabular}{l c c c c}
\hline
 & Facebook & X & Instagram & LinkediN \\
\hline\\
alpha          & $77.006^{***}$ & $72.007^{***}$ & $79.043^{***}$ & $79.119^{***}$ \\
               & $(4.074)$      & $(1.837)$      & $(2.639)$      & $(1.760)$      \\
lambda (connection count)        & $30.499^{***}$ &                &                & $36.926^{***}$ \\
               & $(4.163)$      &                &                & $(7.217)$      \\
p (connection count)             & $-0.015^{**}$  &                &                & $-0.002^{*}$   \\
               & $(0.005)$      &                &                & $(0.001)$      \\
lambda1 (followees)        &                & $22.647$       & $19.658^{***}$ &                \\
               &                & $(12.664)$     & $(3.563)$      &                \\
p1 (followees)             &                & $-0.000$       & $-0.003^{*}$   &                \\
               &                & $(0.000)$      & $(0.001)$      &                \\
lambda2 (followers)       &                & $13.026^{*}$   & $71.256^{***}$ &                \\
               &                & $(6.205)$      & $(21.477)$     &                \\
p2 (followers)             &                & $-0.001$       & $-0.000$       &                \\
               &                & $(0.001)$      & $(0.000)$      &                \\
\hline
Sigma          & $59.736$       & $58.281$       & $57.531$       & $58.619$       \\
Convergence    & $1$            & $1$            & $1$            & $1$            \\
Tolerance      & $0.000$        & $0.000$        & $0.000$        & $0.000$        \\
Log Likelihood & $-13605.422$   & $-12265.680$   & $-13609.988$   & $-12076.511$   \\
AIC            & $27218.844$    & $24543.360$    & $27231.977$    & $24161.021$    \\
BIC            & $27242.092$    & $24577.637$    & $27266.892$    & $24183.806$    \\
Deviance       & $8803257.076$  & $7581341.312$  & $8218205.717$  & $7549421.848$  \\
DF Resid.      & $2467$         & $2232$         & $2483$         & $2197$         \\
nobs           & $2470$         & $2237$         & $2488$         & $2200$         \\
\hline
\multicolumn{5}{l}{\scriptsize{$^{***}p<0.001$; $^{**}p<0.01$; $^{*}p<0.05$}}
\end{tabular}
\end{center}
\end{table}

\pagebreak

\section{Robustness}\label{robustness_section}

\normalsize{}{
\doublespacing

This section reports the results of several robustness checks on the validity of our results. Section \ref{FVP_robust} focuses on the predictors of connection value percentile after restricting attention to only the top four connections. Section \ref{S42} reports analysis of the predictors of total platform value. Our main results on total platform value are derived from Wave 2. Here, where possible, we look to see whether the same patterns held in Wave 1 data. Section \ref{S422} reports regressions of total platform value on the number of platform connections and other scraped platform data from Wave 1. Section \ref{S423} reports whether predictors of platform value investigated in Wave 2 also held in Wave 1, focusing on platform time use and political orientation. The Wave 1 sample retains the strong relationship between hours of use and total platform value observed in Wave 2. 

A final possible concern about our analyses may be the sensitivity of our findings to the maximum value egos are allowed to report for total platform value using the BDM mechanism. In Wave 1, respondents could enter a free-response answer if their monthly total platform value was greater than \$100. In Wave 2, users are randomly assigned to a maximum value of \$100, \$150 or \$200. We investigated the robustness of our main results to additional controls for the Wave 2 maximum value assignment and found these controls do not attenuate our large and significant estimates of the relationship of log connection count and total monthly platform value (see Tables \ref{logols_FB} through \ref{logols_LI}). As a second investigation of the importance of the long-tail of high-value platform valuers, we report the empirical CDF of total platform value by ego respondents along two variations: with the Wave 1 sample and a winsorization at \$100, vs using the free response valuations. We found that less than 1\% of respondents value the platform at more than \$1000 in Wave 1. One implication of this finding is that, given the small fraction of high valuations, our total platform value results from Wave 1 and Wave 2 should be interpreted as representative of typical valuations rather than fully describing the distribution of valuations.

\subsection{Robustness of Predictors of Connection Value Percentile}\label{FVP_robust}
\doublespacing


Previous studies investigating the determinants of relationship importance have used a variety of methods for soliciting the list of contacts to be investigated. In Marsden and Campbell\cite{marsden1984measuring}, interviewers solicited the names of the subjects' three closest friends. In Fischer\cite{fischer1982what}, 
interviewers asked open-ended questions about the subject's life, generating a list of 2-26 acquaintances per subject. Our approach in the main data was a hybrid of these approaches. We asked the egos to report their four closest connections (who may or may not be on the platform) as well as four other top-of-mind contacts on the platform. 

There are advantages and disadvantages to different approaches for soliciting connections. A focus on the closest connections helps ensure that the subject will have good knowledge of and strong preferences over the relationships. On the other hand, the relationships we have with close connections are distinctive. The things that endear us to our closest connections may repel us in more distant relations. To cover a more complete range of relationships, it is important to include less close relations in the sample.

For robustness, we re-analyzed our CVP results for just the ego's closest connections. These are the connections that consumers are likely to have the strongest and most consistent preferences over, making these robustness results most closely comparable to the results reported by Marsden and Campbell\cite{marsden1984measuring}. Figures \ref{top4_seen}, \ref{top4_friendsource} and \ref{top4_otherplatform} report conditional average CVPs, restricting attention to the four closest contacts. Figure   \ref{top4_seen} shows that our first main result, that people seen more often had higher CVPs, was replicated in the top four connections. Just like in the main analysis Figure 2B, for Facebook the CVP of alters who live with egos was significantly higher than on X. Figure \ref{top4_friendsource} broadly replicates the main analysis of family and college friends having the highest CVP. Unlike in the main analysis here, for Facebook, the point estimate for the CVP for `none of these describe how I know this person' was larger than the CVPs for school and work, which is suggestive of the value of top connections. \ref{top4_otherplatform} is also very similar to the main result in Figure 2C, with the closest four connections having significantly higher CVP when connected to the ego on a second platform, for all focal platforms, replicating the suggestive evidence of a lack of substitution between platforms. 

\begin{figure}[!h]
    \centering
\includegraphics[width=\textwidth]{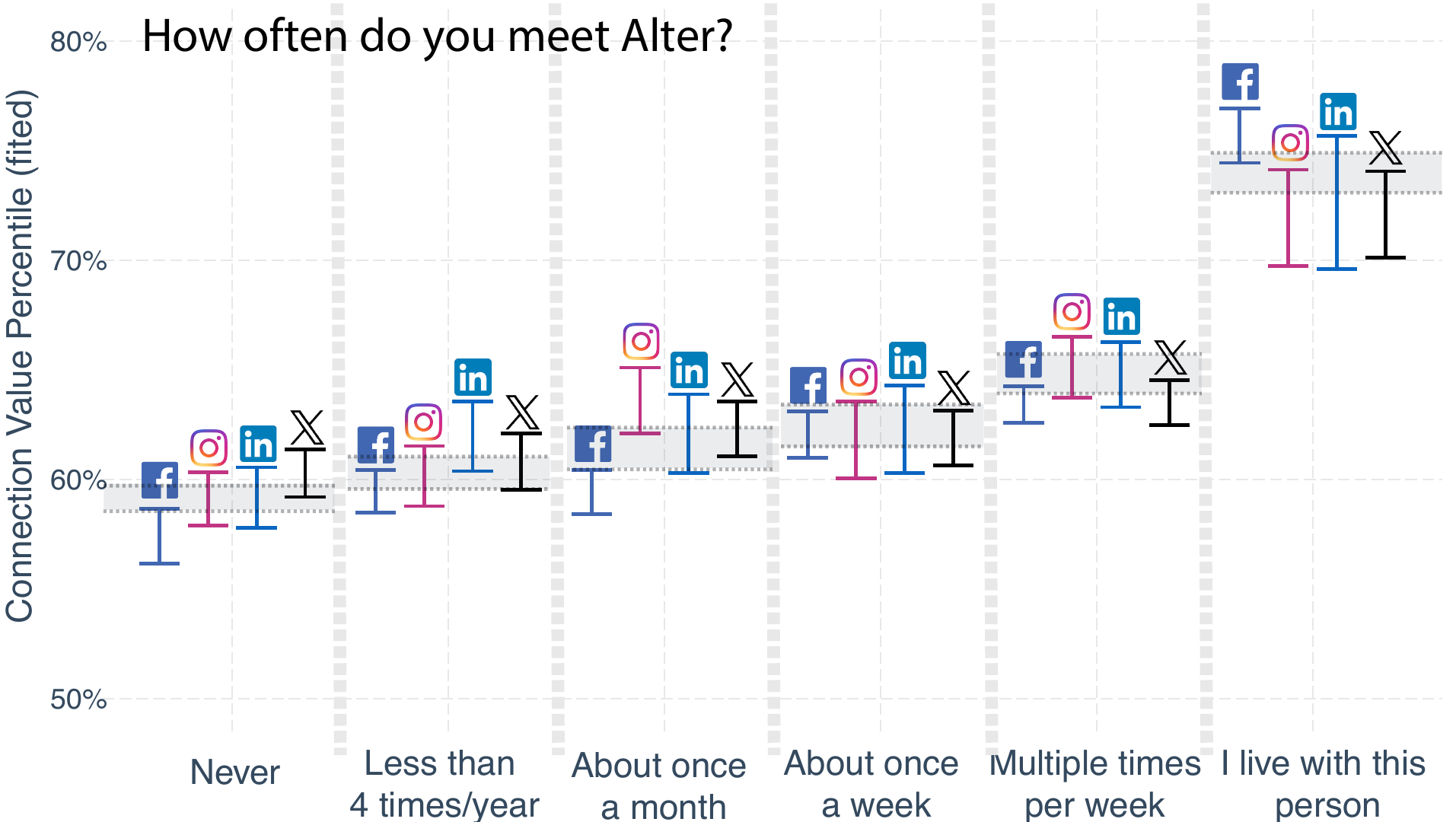}
    \caption{Connection value percentile as a function of how often seen. Wave 1, four closest connections only.}
    \label{top4_seen}
\end{figure}

\begin{figure}[!h]
    \centering
\includegraphics[width=\textwidth]{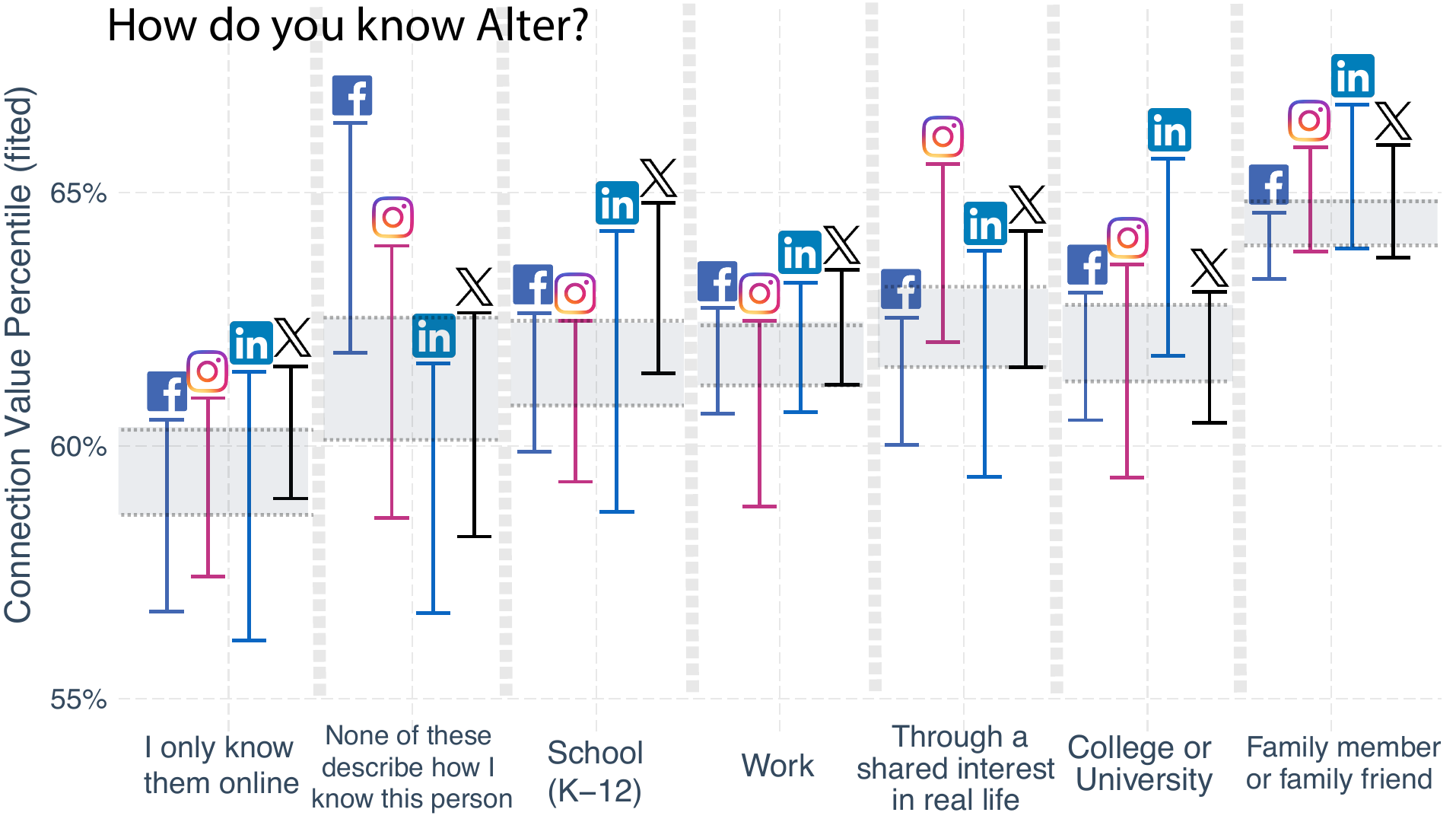}
    \caption{Connection value percentile as a function of connection source. Wave 1, four closest connections only.}
    \label{top4_friendsource}
\end{figure}

\begin{figure}[!h]
    \centering
\includegraphics[width=\textwidth]{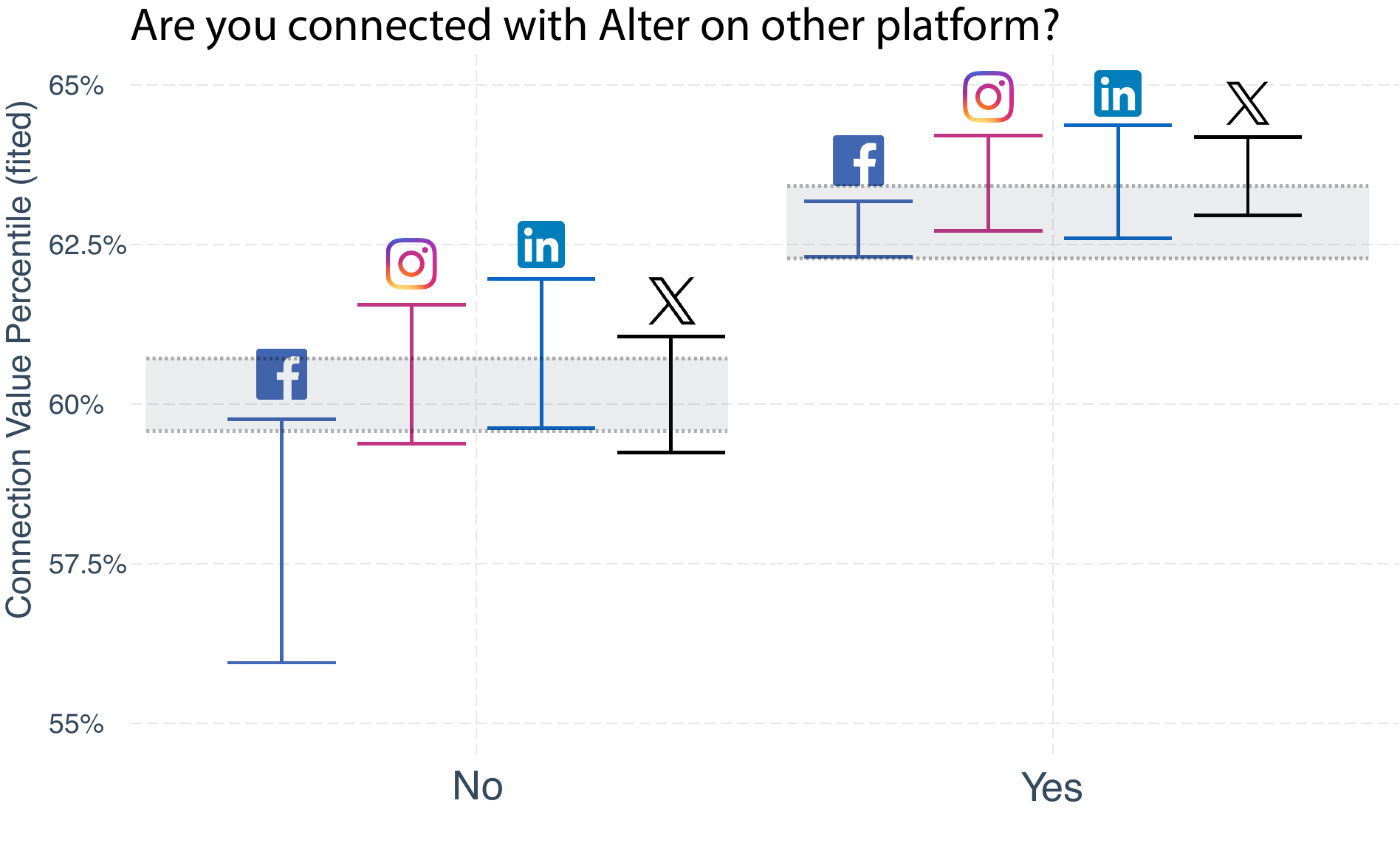}
    \caption{Connection value percentile as a function of connection source. Wave 1, four closest connections only.}
    \label{top4_otherplatform}
\end{figure}

\subsection{Robustness of Predictors of Total Platform Value}\label{S42}

While Wave 2 focused on Total Platform Valuation, we examine whether similar patterns hold in Wave 1 data.

\subsubsection{Measuring the Value of Connections with Wave 1 Data}\label{S422}

We used Wave 1 data to establish the robustness of our results on the value of connections across platforms. For X, Facebook, and LinkedIn, we were able to collect additional data on platform use from a subset of survey participants. We were unable to collect additional information from Instagram users because of platform scraping restrictions. We were able to collect additional data from about  $\sim$25\% of respondents overall. For each of these respondents, we were able to collect connection count information directly from the platform.  

Table \ref{connection_value_robustness_combined} reports the results of a regression of total platform valuation, winsorized at \$1,000, on the log of connection counts scraped from Wave 1 survey participants. While not significant, the point estimates are consistent with those from Wave 2 focused on whether platform value is increasing in log(connections) for Facebook and LinkedIn. For X, where followers and followees are distinguished, log(followers) entered the analysis positively and log(followees) negatively. The lack of significance in these estimates is almost certainly due to the small sample size of our scraped data.

\begin{table}[!htbp] 
\centering
\footnotesize
\begin{tabular}{l c c c} 
\toprule
 & \multicolumn{3}{c}{\textbf{Total Monthly Platform Valuation}} \\ 
\cmidrule(lr){2-4}
 & \textbf{Facebook} & \textbf{X} & \textbf{LinkedIn} \\ 
\midrule
$\log(\text{Connections})$ & 2.108 &  &  \\ 
 & (2.716) &  &  \\ 
[0.5em]
$\log(\text{Followers})$ &  & 3.874 &  \\ 
 &  & (5.111) &  \\ 
[0.5em]
$\log(\text{Followees})$ &  & $-$3.030 &  \\ 
 &  & (6.649) &  \\ 
[0.5em]
$\log(\text{Connections})$ &  &  & 4.913 \\ 
 &  &  & (5.311) \\ 
[0.5em]
Constant & 114.980$^{***}$ & 106.880$^{***}$ & 73.120$^{***}$ \\ 
 & (14.534) & (22.839) & (26.182) \\ 
\midrule
Observations & 944 & 599 & 344 \\ 
$R^{2}$ & 0.001 & 0.001 & 0.002 \\ 
Adjusted $R^{2}$ & $-$0.0004 & $-$0.002 & $-$0.0004 \\ 
Residual SE & 165.870 & 155.840 & 130.790 \\ 
$F$ Statistic & 0.602 & 0.318 & 0.856 \\ 
\bottomrule
\end{tabular}  
\caption{OLS Regression for the Total Monthly Valuation on Log-Connection Count (Wave 1 Main Text Filtering)} 
\label{connection_value_robustness_combined} 
\begin{flushleft}
\footnotesize{Standard Errors in parenthesis. Note: *p$<$0.1; **p$<$0.05; ***p$<$0.01}
\end{flushleft}
\end{table}

Interestingly, the result that log follower count positively predicts total platform value while log followee count negatively predicts value is sensitive to the level at which platform valuations were winsorized (the positive relationship between total connection count and platform value for Facebook and LinkedIn was not sensitive to this decision). This finding is reported in Table  \ref{x_connectval_winsorizations}.

This result is consistent with the `attention bartering’ theory of X value in Filippas et. al. \cite{production_of_social_media}. According to this theory, intensive users of X get positive value from being followed, but negative value from following all but a small subset of accounts. However, a close correlation between the count of followers and followees emerges because users trade following a third party that they would not follow otherwise in return for that party following their own accounts. They document this correlation in their study population of economists. In our data, log followers and log followees were also highly correlated---for example, on X the correlation was over 80\%. 

Under the attention bartering theory, a change in the relative importance of followers and followees as total platform value winsorization increases would emerge if there are two types of users. We’ll call these `lurkers’ and `broadcasters’. The lurkers do not care about being followed (or may even dislike it) and follow only accounts they directly value. This group has a low total platform value, but their platform value is increasing in the number of accounts followed (followees). Broadcasters have high total platform value, which is increasing in the number of accounts that follow them (followers). In order to increase their number of followers they barter following uninteresting accounts that they do not directly value. Broadcasters may also value following some accounts, but most of their followees actually give them negative value directly. When platform values are winsorized at low levels, it is the preferences of lurkers which dominate the correlation between connection count and platform value. When platform values are winsorized at high levels, the preferences of the broadcasters are more influential in determining the correlations. 

\begin{table}[!htbp] 
\centering
\footnotesize
\begin{tabular}{lcccccc} 
\toprule
 & \multicolumn{6}{c}{\textbf{Total Monthly Platform Valuation (X)}} \\ 
\cmidrule(lr){2-7}
 & Win. = 100 & Win. = 200 & Win. = 300 & Win. = 400 & Win. = 500 & Win. = 1000 \\ 
\midrule
$\log(\text{Followers})$ & $-0.504$ & $-0.444$ & $0.042$ & $0.691$ & $1.343$ & $3.874$ \\ 
 & $(1.129)$ & $(1.689)$ & $(2.247)$ & $(2.721)$ & $(3.139)$ & $(5.111)$ \\ 
[0.5em]
$\log(\text{Followees})$ & $2.414$ & $1.783$ & $1.192$ & $0.522$ & $-0.089$ & $-3.030$ \\ 
 & $(1.468)$ & $(2.197)$ & $(2.922)$ & $(3.539)$ & $(4.083)$ & $(6.649)$ \\ 
[0.5em]
Constant & $59.655^{***}$ & $73.253^{***}$ & $80.609^{***}$ & $85.751^{***}$ & $89.578^{***}$ & $106.880^{***}$ \\ 
 & $(5.044)$ & $(7.548)$ & $(10.038)$ & $(12.157)$ & $(14.024)$ & $(22.839)$ \\ 
\midrule
Observations & 599 & 599 & 599 & 599 & 599 & 599 \\ 
$R^{2}$ & $0.008$ & $0.002$ & $0.001$ & $0.001$ & $0.001$ & $0.001$ \\ 
Adjusted $R^{2}$ & $0.005$ & $-0.002$ & $-0.003$ & $-0.003$ & $-0.003$ & $-0.002$ \\ 
Residual SE & $34.417$ & $51.501$ & $68.496$ & $82.951$ & $95.691$ & $155.840$ \\ 
$F$ Statistic & $2.387^{*}$ & $0.534$ & $0.247$ & $0.202$ & $0.233$ & $0.318$ \\ 
\bottomrule
\end{tabular}  
\caption{OLS Regression of the Total Monthly Valuation for X on Log Followers and Followees (Different Winsorizations of Total Platform Value)} 
\label{x_connectval_winsorizations}
\begin{flushleft}
\footnotesize{Notes: Standard Errors in parenthesis. *p$<$0.1; **p$<$0.05; ***p$<$0.01}
\end{flushleft}
\end{table}

In addition to connection count, for LinkedIn users we were able to collect information on labor force participation. Table \ref{connection_value_robustness_LI} shows a positive point estimate for connection count on LinkedIn platform value. Furthermore, individuals with a long tenure at their current workplace valued LinkedIn less, which is consistent with LinkedIn being used for job search. This finding is also consistent with our finding from Wave 2 that people who are searching for a job value LinkedIn more.

\begin{table}[!htbp] 
\centering
\footnotesize
\begin{tabular}{lccc} 
\toprule
 & \multicolumn{3}{c}{\textbf{LinkedIn Monthly Valuation}} \\ 
\cmidrule(lr){2-4}
 & (1) & (2) & (3) \\ 
\midrule
$\log(\text{Connections})$ & $4.913$ &  & $7.447$ \\ 
 & $(5.311)$ &  & $(5.821)$ \\ 
[0.5em]
Currently working $>1$ year &  & $-14.460$ & $-16.458$ \\ 
 &  & $(15.475)$ & $(15.538)$ \\ 
[0.5em]
Constant & $73.120^{***}$ & $103.820^{***}$ & $68.913^{**}$ \\ 
 & $(26.182)$ & $(10.037)$ & $(29.070)$ \\ 
\midrule
Observations & 344 & 309 & 309 \\ 
$R^{2}$ & $0.002$ & $0.003$ & $0.008$ \\ 
Adjusted $R^{2}$ & $-0.0004$ & $-0.0004$ & $0.002$ \\ 
Residual SE & $130.790$ & $134.290$ (df = 307) & $134.150$ \\ 
$F$ Statistic & $0.856$ & $0.873$ & $1.256$ \\ 
\bottomrule
\end{tabular}  
\caption{OLS Regression for Total Monthly Valuation on Connections Count and Tenure of Current Employment} 
\label{connection_value_robustness_LI}
\begin{flushleft}
\footnotesize{Notes: Standard Errors in parenthesis. *p$<$0.1; **p$<$0.05; ***p$<$0.01}
\end{flushleft}
\end{table}

Table \ref{connection_value_robustness_TW} looks at other aspects of X use and platform value. This information was downloaded from the X API for a subset of Wave 1 participants. We found that the number of reposts the account has had in its history is a significant positive predictor of Total Platform Valuation, with every repost raising the monthly value of the platform by 5.1 cents, and quotes by 29.8 cents. On the other hand, receiving replies is predictive of lower platform value, by 2.3 cents each. This result is again consistent with that idea that broadcasting information is of higher value than consuming it.

\begin{table}[!htbp] 
\centering
\footnotesize
\begin{tabular}{lcc} 
\toprule
 & \multicolumn{2}{c}{\textbf{X Monthly Valuation}} \\ 
\cmidrule(lr){2-3}
 & (1) & (2) \\ 
\midrule
$\log(\text{Followers})$ & $3.874$ &  \\ 
 & $(5.111)$ &  \\ 
[0.5em]
$\log(\text{Followees})$ & $-3.030$ &  \\ 
 & $(6.649)$ &  \\ 
[0.5em]
Total retweets &  & $0.051^{***}$ \\ 
 &  & $(0.013)$ \\ 
[0.5em]
Total quotes &  & $0.298^{***}$ \\ 
 &  & $(0.065)$ \\ 
[0.5em]
Total replies &  & $-0.023^{***}$ \\ 
 &  & $(0.009)$ \\ 
[0.5em]
Favorite count &  & $0.005^{**}$ \\ 
 &  & $(0.003)$ \\ 
[0.5em]
Constant & $106.880^{***}$ & $87.902^{***}$ \\ 
 & $(22.839)$ & $(7.738)$ \\ 
\midrule
Observations & $599$ & $571$ \\ 
$R^{2}$ & $0.001$ & $0.093$ \\ 
Adjusted $R^{2}$ & $-0.002$ & $0.087$ \\ 
Residual SE & $155.840$ & $142.450$ \\ 
$F$ Statistic & $0.318$ & $14.523^{***}$ \\ 
\bottomrule
\end{tabular}  
\caption{OLS Regression of Total Monthly Valuation on Social Metrics (Platform Value Winsorized at \$1,000)} 
\label{connection_value_robustness_TW}
\begin{flushleft}
\footnotesize{Notes: Standard Errors in parenthesis. *p$<$0.1; **p$<$0.05; ***p$<$0.01}
\end{flushleft}
\end{table}

\subsubsection{Other Predictors of Total Platform Value from Wave 1}\label{S423}

Table \ref{other_tpv1}  reports conditional average Total Platform Valuations by self-reported political orientation from Wave 1. The figure confirms the Wave 2 result that individuals who use a platform more value it more highly. 

\begin{figure}
    \centering
    \includegraphics[width=0.5\linewidth]{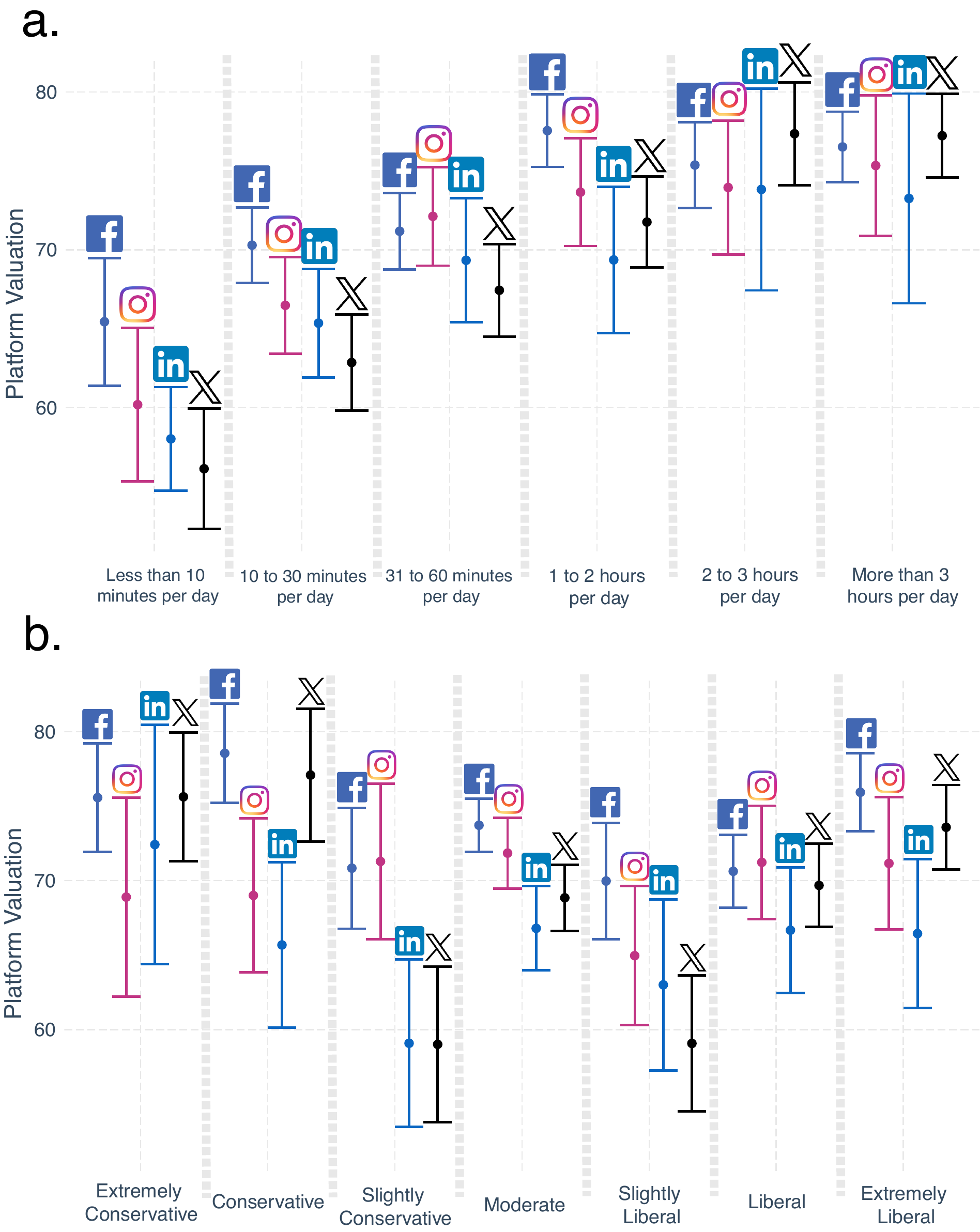}
    \caption{Conditional average valuations by self-reported political orientation and platform usage. Wave 1 main analysis data. }
    \label{other_tpv1}
\end{figure}

\subsubsection{Sensitivity to Alternate Maximum Platform Value}\label{S421}

For our total platform value results, we gathered data using the BDM method. A potential weakness of the BDM method for soliciting willingness-to-pay is that results may be less reliable for those with very large valuations. There are a few reasons for this. First, a low-attention survey taker may select a very large willingness to pay out of carelessness. To control for this possibility we executed attention checks, described earlier, and filtered out respondents found to display carelessness. Second, all BDM mechanisms rely on the subject having a positive expectation that they will be offered a follow up contract. As true valuations get larger, users may find it increasingly unlikely that the BDM mechanism will actually be enforced, and therefore be more careless in reporting their true valuation if it is large. These challenges are common and apply to all incentive compatible valuations and BDM mechanisms. Here we describe their implications for our results and our work to ensure robustness in the face of these challenges.

The inability to precisely probe valuations for users who have very high WTPs for social media has important consequences, especially for studies trying to make aggregate claims about the contribution of social media to total welfare. This is because there plausibly exists a thick-tail of platform users with extremely high valuations -- which would be consistent with the large count of users who use social media for a large share of their waking hours. For example, one nationally representative Gallup survey found that 29\% of teens used social media for 6 hours or more a day on average \cite{Rothwell2023}. Therefore it poses a specific challenge for work like that of Bursztyn et al.\cite{bursztyn2023product} who focus on the question of aggregate welfare. 

Other BDM studies of the value of social media have approached this problem in different ways. Allcott et al.\cite{allcott2020welfare} allow for a free response in their BDM value solicitation, but winsorize maximum replies at \$170. This value is selected as the maximum because it is the maximum payment that they allow in their randomized BDM offer, though this cutoff was not made known to participants. Bursztyn et al.\cite{bursztyn2023product} have a multiple price list soliciting their BDM, with a maximum set at \$200. In a robustness section, they fit a thin-tailed curve through their results, allowing for imputed values above \$200 and argue this does not change their findings. Brynjolfsson et al.\cite{brynjolfsson2019using} focus on median valuations to avoid this issue entirely. 

In our context, we are primarily interested in determinants of CVP, which are not impacted by this issue, as well as how connections predict total platform value. Estimation of total platform value is impacted by BDM measurement challenges, but less so as the results can simply be interpreted as predictors of platform value for typical users rather than all users. Regardless, in our two study Waves, we took different approaches to this challenge. In Wave 1, we used a slider up to \$100 and then a free-response box for values larger than \$100. Our main results were winsorized at the \$100 dollar level. In Wave 2 we used a multiple price list, with users randomly assigned to \$100, \$150 and \$200 maximum values. We used this Wave 2 data for our main total platform value results and showed the sensitivity of results to controls for these maximum values in Tables \ref{logols_FB} thorough \ref{logols_LI}. The point estimates of the effect of log(connections) on total platform value were not sensitive to these controls, indicating the robustness of our main results. 

In this section we report further robustness checks using Wave 1 data to understand what percentage of responses were impacted by the presence of a maximum value. Figure \ref{ecdf_wave_1_capped} A reports the empirical CDF of Total Platform Valuations in our Wave 1 study with a winsorization of \$100. Figure  \ref{ecdf_wave_1_capped}B reports the same, but increases the \$100 winsorization to \$1,000,000. In Panel B, log total platform values are reported on the x-axis. 

In the Wave 1 data, pooling all platforms, less than 1\% of respondents reported valuations above \$1,000. X and Instagram were the platforms most impacted by winsorization in the filtered data, suggesting that the long tail of high valuers may be particularly important for these platforms. These facts suggest a need for some caution when interpreting the aggregate welfare implications of the results derived from our valuation methods, if a large share of value is driven by a small number of high valuers who are imprecisely measured. However, the fact that less than 1\% of our sample are high valuers makes us retain confidence in our total platform valuation estimates. Furthermore, winsorization and maximum values do not impact our CVP results or connection valuations for typical platform users, giving us more confidence in our results and interpretations.

\begin{figure}[htb]
        \centering
        \includegraphics[width=\linewidth]{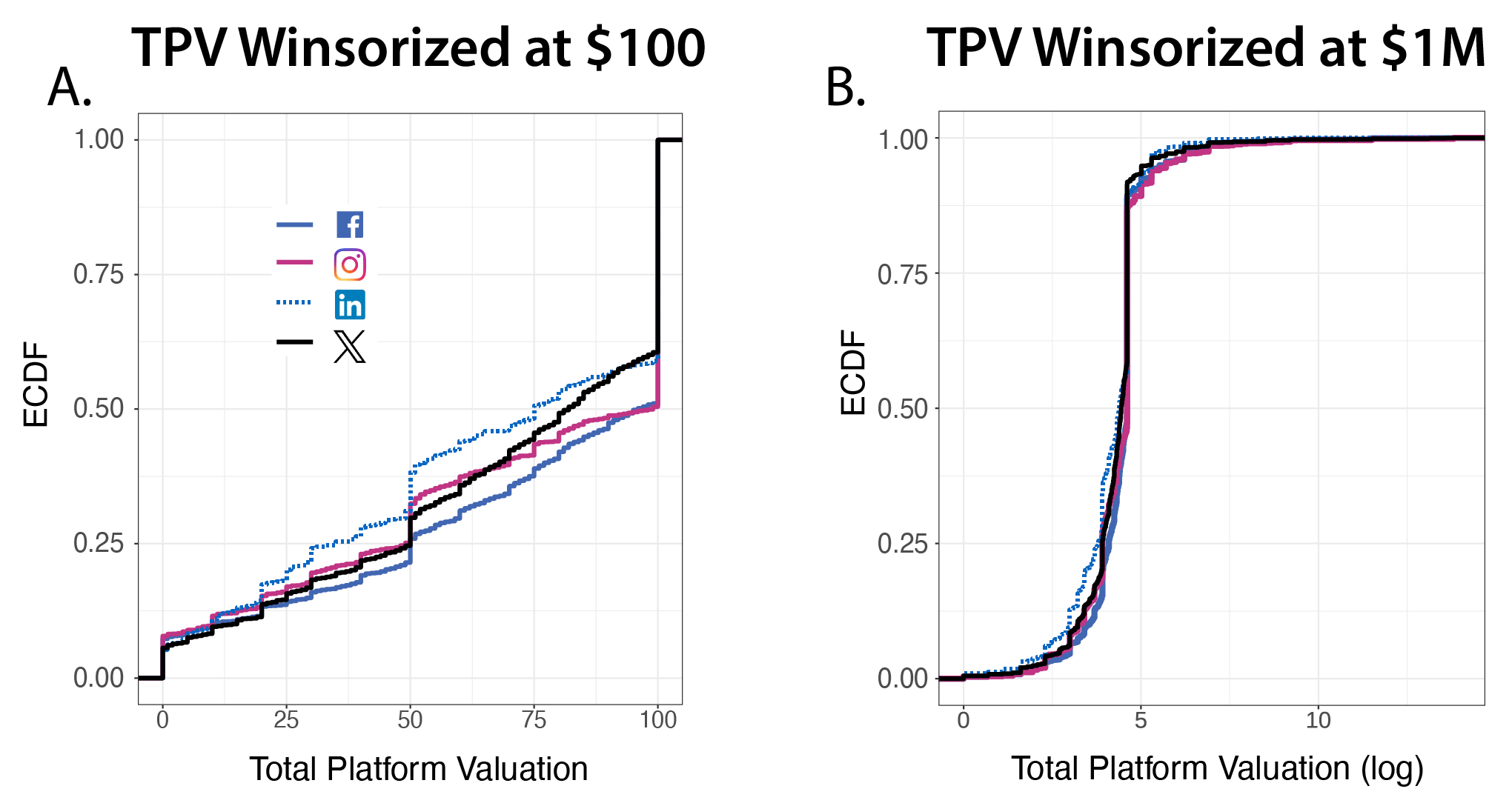} 

    \caption{Empirical CDFs of total platform value from Wave 1. A. Responses capped at \$100. B.  Responses capped
at \$1,000,000. }
    \label{ecdf_wave_1_capped}
\end{figure}

\clearpage
\newpage
\pagebreak

\section{Additional Data Collection Tables}\label{Additional_data_collect_tables}

\footnotesize
\renewcommand{\arraystretch}{1} 
\setlength{\tabcolsep}{3pt} 

\begin{longtable}{|p{0.4\textwidth}|p{0.65\textwidth}|}
\hline
\textbf{Description} & \textbf{Details} \\
\hline
User's id by number & Unique numerical identifier for a user \\
\hline
Tweet id by number & Unique numerical identifier for a tweet \\
\hline
Tweet creation date & Timestamp when the tweet was posted \\
\hline
User's id by screen name & User's handle on X \\
\hline
Tweet content & Text of the tweet \\
\hline
App used for writing the tweet & Source application (e.g., Android, iPhone, Web App) \\
\hline
Number of characters used in the tweet & Displayed tweet width in characters \\
\hline
Tweet's id to which the user responded & ID of the tweet being replied to (NA if not a reply) \\
\hline
User's id to which the user responded & Numerical ID of the user being replied to (NA if not a reply) \\
\hline
User's id by screen name to which the user responded & Screen name of the user being replied to (NA if not a reply) \\
\hline
TRUE if quoting another tweet & Indicates if the tweet is a quote \\
\hline
TRUE if retweeting another tweet & Indicates if the tweet is a retweet \\
\hline
Number of "favorite" reactions & Count of likes/favorites on the tweet \\
\hline
Number of retweets & Count of retweets for the tweet \\
\hline
Number of quotes & Count of quotes for the tweet \\
\hline
Number of responses & Count of replies to the tweet \\
\hline
Words after the hashtag symbol & Hashtags used in the tweet \\
\hline
Symbols used in the tweet  \\
\hline
URLs in the tweet & List of URLs used \\
\hline
Media used in the tweet & Links and types of media (e.g., photos, videos) \\
\hline
Mentions in the tweet & Users mentioned by screen name or ID \\
\hline
Language of the tweet & Detected language of the content \\
\hline
Information about quoted tweets & Includes ID, content, creation date, and engagement metrics \\
\hline
Information about retweeted tweets & Includes ID, content, creation date, and engagement metrics \\
\hline
Location data & Place name, coordinates, and country if available \\
\hline
User's profile data & Includes name, location, description, URLs, and account status \\
\hline
Account statistics & Followers, friends, statuses, and favorites counts \\
\hline
Verification and protection & Whether the account is verified or protected \\
\hline
User's media links & Profile images, banner URLs, and background themes \\
\hline
\caption{Twitter/X Additional Ego Timeline Data Collection: This data is available for a subset of Twitter/X study participants through November 2021. For regressions in Table \ref{connection_value_robustness_TW} some variables were aggregated to the account level. }
\end{longtable}

\normalsize{}


\newpage

\newpage
\begin{longtable}{|p{0.45\textwidth}|p{0.45\textwidth}|}
\hline
\endfirsthead

\hline
\endhead
{Account ID} & UTC datetime of account creation\\ \hline
whether `contributor mode' enabled& user description \\ \hline
User display name& number of Tweets user has liked \\ \hline
User follower count & Authenticating user has issued follow request \\ \hline
User followee count & Geotagging Enabled\\ \hline
Participant in Twitter translator community & User language\\ \hline
Number of public lists user is on & User defined location \\ \hline
Indicator for protected tweets & User handle \\ \hline
Count of Tweets and retweets & User Timezone \\ \hline
User URL & User UTC offset \\ \hline
User verified indicator & Countries user is witheld from \\ \hline

\caption{Twitter/X Account level data collected for a subset of Wave 1 study participants and their non-private followers and followees.}
\label{tab:twitter_ego_data}
\end{longtable}

\newpage

\footnotesize
\renewcommand{\arraystretch}{1}

\begin{longtable}{|p{0.3\textwidth}|p{0.65\textwidth}|}
\hline
\textbf{Field Name} & \textbf{Description} \\ \hline
\endfirsthead

\textbf{Field Name} & \textbf{Description} \\ \hline
\endhead

linkedinProfile & URL of the LinkedIn profile \\ \hline
email & Email address \\ \hline
headline & User's professional headline \\ \hline
location & User's location \\ \hline
imgUrl & URL of the user's profile picture \\ \hline
firstName, lastName, fullName & User's first name, last name, and full name \\ \hline
subscribers & Number of followers or subscribers \\ \hline
connectionsCount &  Total number of connections \\ \hline
company, company2 & Names of current and previous company \\ \hline
companyUrl, companyUrl2 & URLs of current and previous company LinkedIn pages \\ \hline
jobTitle, jobTitle2 & Titles of current and previous job \\ \hline
jobDescription, jobDescription2 & Descriptions of current and previous job \\ \hline
jobLocation, jobLocation2 & Locations of current and previous job \\ \hline
jobDateRange, jobDateRange2 & Date ranges of current and previous job \\ \hline
school, school2 & Names of attended schools \\ \hline
schoolUrl, schoolUrl2 & URLs of attended schools' LinkedIn pages \\ \hline
schoolDegree, schoolDegree2 & Degrees obtained from schools \\ \hline
schoolDateRange, schoolDateRange2 & Date ranges of attendance at schools \\ \hline
schoolDescription, schoolDescription2 & Descriptions of attended schools \\ \hline
skill1 - skill6; endorsement1-endorsement6 & Up to 6 listed skills count of `endorsements' by third parties for each \\ \hline
allskills & any self-reported skills including beyond the first 6 \\ \hline
timestamp & Timestamp of data scraping \\ \hline
website, twitter, facebookUrl & User's website, Twitter handle, and Facebook URL \\ \hline
birthday & User's birthday \\ \hline

\caption{Resume information collected from a subset of Wave 1 LinkedIn Study Participants from their profile page. }
\label{tab:linkedin_data_condensed}
\end{longtable}

\normalsize{}

\clearpage
\newpage
\pagebreak

\section{Full Statistical Reporting of the Findings }\label{statistical_reporting}
For completeness and in the interest of maximum clarity, this section presents estimated differences, and associated p-values, for the results in Figures 1 through 4 of the main text. In section \ref{grouped_comparisons}, we report estimates for selected groups which are aggregated at a higher level than in the main text figures. In section \ref{all_comparisons} we report all estimated differences between pairs of estimates in the main text figures.

\subsection{Grouped Comparisons} \label{grouped_comparisons}
In this section we present estimated differences, and associated p-values, for selected aggregations of respondents. 

For example, Figure 1C reports estimates of platform valuations, conditional on platform type and six bins indicating the amount of time spent on the platform.  Table \ref{tab:first_grouped_table} reports corresponding estimates based on this figure. In particular, here we  report the estimated difference in platform value between pairs of conditions but aggregating to only two participation levels: above and below one hour a day. The first row of the table indicates that users who use Facebook less than 1 hour a day value it $\$8.848$ a month less than those who use it 1 hour a day or more. The row is bolded, indicating the difference is significantly different from zero at the 99\% confidence level, having an associated p-value of .006. 

\subsubsection{Figure 1C: Total Platform Valuation by Platform \& Usage, Aggregated to Two Usage Levels}

\begin{table}[htbp]
\centering
\scriptsize
\renewcommand{\arraystretch}{0.53} 
\begin{tabular}{@{\extracolsep{5pt}} cccccc} 
\\[-4.8ex]\hline 
\hline \\[-1.8ex] 
 & contrast & estimate & SE & t-stat & p-value \\ 
\hline \\[-1.8ex] 
\textbf{1} & \textbf{Less than 1 hour per day Facebook - More than 1 hour per day Facebook} & \textbf{-8.848} & \textbf{2.401} & \textbf{-3.685} & \textbf{0.006} \\ 
\textbf{2} & \textbf{Less than 1 hour per day Facebook - Less than 1 hour per day Instagram} & \textbf{11.193} & \textbf{2.173} & \textbf{5.152} & \textbf{0.00001} \\ 
3 & Less than 1 hour per day Facebook - More than 1 hour per day Instagram & -2.837 & 2.473 & -1.147 & 0.946 \\ 
\textbf{4} & \textbf{Less than 1 hour per day Facebook - Less than 1 hour per day LinkedIn} & \textbf{11.906} & \textbf{2.097} & \textbf{5.678} & \textbf{0.00000} \\
\end{tabular} 
\caption{Full reporting of the comparative statistics of Figure~1C in the main text where we group usage levels to More or Less than 1 hour. \textbf{Highlighted contrasts} are significant at the 1\% level. DF = 9387.} 
\label{tab:first_grouped_table}
\end{table}

\subsubsection{Figure 1C: Total Platform Valuation by Platform \& Usage, Aggregated over all platforms}

\begin{center}
  \renewcommand{\arraystretch}{0.63}   
    \footnotesize
\begin{tabular}{@{\extracolsep{5pt}} cccccc} 
\\[-4.8ex]\hline 
\hline \\[-1.8ex] 
 & contrast & estimate (\$) & SE & t-stat & p-value \\ 
\hline \\[-1.8ex] 
\textbf{1} & \textbf{Less than 10 minutes per day - 10–30 minutes per day} & \textbf{-27.546} & \textbf{2.067} & \textbf{-13.329} & \textbf{0} \\ 
\textbf{2} & \textbf{Less than 10 minutes per day - 31–60 minutes per day} & \textbf{-11.478} & \textbf{1.739} & \textbf{-6.602} & \textbf{0} \\ 
\textbf{3} & \textbf{Less than 10 minutes per day - 1–2 hours per day} & \textbf{-27.037} & \textbf{2.609} & \textbf{-10.361} & \textbf{0} \\ 
\textbf{4} & \textbf{Less than 10 minutes per day - 2–3 hours per day} & \textbf{-18.408} & \textbf{1.925} & \textbf{-9.565} & \textbf{0} \\
\end{tabular} 
  \captionof{table}{Full reporting of the comparative statistics of Figure 1C in the main text:
          Pairwise Comparisons of the average over all platforms TPV for the different Usage Levels. 
          	\textbf{Highlighted contrasts} 
          are significant at the 1\% level. DF $=$ 9389.} 
\end{center}

\subsubsection{Figure 1D: Total Platform Valuation by Platform \& Political Orientation, Aggregated to Three Orientations}

\begin{center}
  \scriptsize
  \renewcommand{\arraystretch}{0.57} 
\begin{tabular}{@{\extracolsep{5pt}} cccccc} 
\\[-4.8ex]\hline 
\hline \\[-1.8ex] 
 & contrast & estimate (\$) & SE & t-stat & p-value \\ 
\hline \\[-1.8ex] 
1 & Liberal Facebook - Moderate Facebook & -2.910 & 2.923 & -0.995 & 0.998 \\ 
2 & Liberal Facebook - Conservative Facebook & 6.328 & 3.803 & 1.664 & 0.885 \\ 
3 & Liberal Facebook - Liberal Instagram & 11.417 & 3.596 & 3.175 & 0.066 \\ 
4 & Liberal Facebook - Moderate Instagram & 7.452 & 2.914 & 2.557 & 0.305 \\
\end{tabular} 
\captionof{table}{Full reporting of the comparative statistics of Figure 1D in the main text where extremely conservative and conservatives are grouped as conservative and extremely Liberals are groups with 
  Liberals. Everything else is grouped as moderates. \textbf{Highlighted
contrasts} are significant at the 1\% level. DF $=$ 9383.} 
\end{center}

\subsubsection{Figure 1D: Total Platform Valuation by Platform \& Political Orientation, Aggregated over all platforms}

\begin{center}
  \renewcommand{\arraystretch}{0.63}   
\begin{tabular}{@{\extracolsep{5pt}} cccccc} 
\\[-4.8ex]\hline 
\hline \\[-1.8ex] 
 & contrast & estimate (\$) & SE & t-stat & p-value \\ 
\hline \\[-1.8ex] 
1 & Extremely liberal - Liberal & $$-$3.727$ & $2.621$ & $$-$1.422$ & $0.790$ \\ 
2 & Extremely liberal - Slightly liberal & $$-$7.983$ & $3.147$ & $$-$2.537$ & $0.146$ \\ 
3 & Extremely liberal - Moderate & $$-$1.395$ & $2.511$ & $$-$0.556$ & $0.998$ \\ 
4 & Extremely liberal - Slightly conservative & $$-$6.748$ & $2.233$ & $$-$3.022$ & $0.040$ \\
\end{tabular} 
  \captionof{table}{Full reporting of the comparative statistics of Figure 1D in the main text: Pairwise Comparisons of average over the four platforms TPV for the different Political Groups. \textbf{Highlighted contrasts} are significant at the 1\% level. DF$=$9388.} 
\end{center}

\clearpage

\subsubsection{Figure 1E: Total Platform Valuation by Platform \& Relationship Status, Aggregated to Married and Unmarried}

  \begin{center}
  \scriptsize
  \renewcommand{\arraystretch}{0.65} 
\begin{tabular}{@{\extracolsep{5pt}} cccccc} 
\\[-4.8ex]\hline 
\hline \\[-1.8ex] 
 & contrast & estimate (\$) & SE & t-stat & p-value \\ 
\hline \\[-1.8ex] 
1 & Married or domestic partnership Facebook - Not Married Facebook & 8.166 & 2.435 & 3.354 & 0.018 \\ 
\textbf{2} & \textbf{Married or domestic partnership Facebook - Married or domestic partnership Instagram} & \textbf{13.997} & \textbf{2.564} & \textbf{5.458} & \textbf{0.00000} \\ 
\textbf{3} & \textbf{Married or domestic partnership Facebook - Not Married Instagram} & \textbf{14.984} & \textbf{2.495} & \textbf{6.005} & \textbf{0.00000} \\ 
\textbf{4} & \textbf{Married or domestic partnership Facebook - Married or domestic partnership LinkedIn} & \textbf{14.923} & \textbf{2.650} & \textbf{5.631} & \textbf{0.00000} \\
\end{tabular} 
 \captionof{table}{Full reporting of the comparative statistics of Figure 1E in the main text: Pairwise Comparisons of TPV for the different Platforms and different Relationship Status. Relationship statuses are grouped into Married or Not Married. \textbf{Highlighted
contrasts} are significant at the 1\% level. DF $=$ 9387.} 
\end{center}

\subsubsection{Figure 1E: Total Platform Valuation by Platform \& Relationship Status, Aggregated over all platforms}

\begin{center}
  \renewcommand{\arraystretch}{0.63}   
  \footnotesize
\begin{tabular}{@{\extracolsep{5pt}} cccccc} 
\\[-4.8ex]\hline 
\hline \\[-1.8ex] 
 & contrast & estimate (\$) & SE & t-stat & p-value \\ 
\hline \\[-1.8ex] 
1 & Single, not looking for a relationship - Single, looking for a relationship & 4.913 & 1.956 & 2.512 & 0.058 \\ 
\textbf{2} & \textbf{Single, not looking for a relationship - In a serious relationship} & \textbf{6.659} & \textbf{2.102} & \textbf{3.168} & \textbf{0.008} \\ 
3 & Single, not looking for a relationship - Married or domestic partnership & -1.491 & 1.661 & -0.897 & 0.806 \\ 
4 & Single, looking for a relationship - In a serious relationship & 1.746 & 2.107 & 0.828 & 0.841 \\ 
\textbf{5} & \textbf{Single, looking for a relationship - Married or domestic partnership} & \textbf{-6.404} & \textbf{1.668} & \textbf{-3.840} & \textbf{0.001} \\ 
\textbf{6} & \textbf{In a serious relationship - Married or domestic partnership} & \textbf{-8.149} & \textbf{1.837} & \textbf{-4.437} & \textbf{0.0001} \\ 
\hline \\[-1.8ex] 
\end{tabular} 
  \captionof{table}{Full reporting of the comparative statistics of Figure 1E in the main text: Pairwise Comparisons of the average over all platforms TPV for the different Relationship Statuses. 
          	\textbf{Highlighted contrasts} 
          are significant at the 1\% level. DF $=$ 9289.} 
\end{center}

\clearpage

\subsubsection{Figure 2A: Connection Value Percentile by Platform \& How Ego Knows Alter, Aggregated to Three Relationship Types}
  \begin{center}
  \scriptsize
  \renewcommand{\arraystretch}{0.55}  
\begin{tabular}{@{\extracolsep{5pt}} cccccc} 
\\[-3.8ex]\hline 
\hline \\[-1.8ex] 
 & contrast & estimate & SE & t.ratio & p.value \\ 
\hline \\[-1.8ex] 
\textbf{1} & \textbf{Family member or family friend Facebook - Work, school or university Facebook} & \textbf{5.428} & \textbf{0.362} & \textbf{14.995} & \textbf{0 } \\
\textbf{2} & \textbf{Family member or family friend Facebook - Online or shared interest Facebook} & \textbf{7.392} & \textbf{0.427} & \textbf{17.331} & \textbf{0 } \\
\textbf{3} & \textbf{Family member or family friend Facebook - Family member or family friend Instagram} & \textbf{-2.701} & \textbf{0.554} & \textbf{-4.878} & \textbf{0.0001 } \\
\textbf{4} & \textbf{Family member or family friend Facebook - Work, school or university Instagram} & \textbf{2.154} & \textbf{0.537} & \textbf{4.009} & \textbf{0.004 } \\
\end{tabular}
  \captionof{table}{Full reporting of the comparative statistics of Figure 2A in the main text where we group the way Ego knows Alter into 3 bins: A. Online or shared interest,
          B. Work, school or university and C. Family member or family friend. \textbf{Highlighted
contrasts} are significant at the 1\% level. DF = 70755.} 
\end{center} 

\subsubsection{Figure 2A: Connection Value Percentile by Platform \& How Ego Knows Alter, Aggregated over all platforms}

  \begin{center}
  \renewcommand{\arraystretch}{0.65}   
  \scriptsize
\begin{tabular}{@{\extracolsep{5pt}} cccccc} 
\\[-1.8ex]\hline 
\hline \\[-1.8ex] 
 & contrast & estimate(\%) & SE & t-stat & p-value \\ 
\hline \\[-1.8ex] 
1 & I only know them online - None of these describe how I know this person & -1.279 & 0.615 & -2.081 & 0.364 \\ 
\textbf{2} & \textbf{I only know them online - School (K-12)} & \textbf{-2.317} & \textbf{0.484} & \textbf{-4.782} & \textbf{0.00004 } \\
\textbf{3} & \textbf{I only know them online - Work} & \textbf{-2.412} & \textbf{0.421} & \textbf{-5.726} & \textbf{0.00000 } \\
\textbf{4} & \textbf{I only know them online - Through a shared interest in real life} & \textbf{-2.904} & \textbf{0.459} & \textbf{-6.332} & \textbf{0 } \\
\end{tabular} 
  \captionof{table}{Full reporting of the comparative statistics of Figure 2A in the main text:
          Pairwise Comparisons of the average over all platforms CVP for the different ways Ego know Alter. 
          	\textbf{Highlighted contrasts} are significant at the 1\% level. DF = 70760.} 
\end{center}

\subsubsection{Figure 2B: Connection Value Percentile by Platform \& Frequency of Meeting Alter, Aggregated to Three Meeting Frequencies}

   \begin{center}
  \scriptsize
  \renewcommand{\arraystretch}{0.58}   
\begin{tabular}{@{\extracolsep{5pt}} ccccccc} 
\\[-3.8ex]\hline 
\hline \\[-1.8ex] 
 & contrast & estimate(\%) & SE & df & t-stat & p-value \\ 
\hline \\[-1.8ex] 
\textbf{\textbf{1}} & \textbf{\textbf{Never Facebook - I live with this person Facebook}} & \textbf{\textbf{-22.322}} & \textbf{\textbf{0.688}} & \textbf{\textbf{70,755}} & \textbf{\textbf{-32.463}} & \textbf{0 } \\
\textbf{\textbf{2}} & \textbf{\textbf{Never Facebook - Sometimes Facebook}} & \textbf{\textbf{-8.080}} & \textbf{\textbf{0.455}} & \textbf{\textbf{70,755}} & \textbf{\textbf{-17.775}} & \textbf{0 } \\
\textbf{\textbf{3}} & \textbf{\textbf{Never Facebook - Never Instagram}} & \textbf{\textbf{-8.448}} & \textbf{\textbf{0.716}} & \textbf{\textbf{70,755}} & \textbf{\textbf{-11.805}} & \textbf{0 } \\
\textbf{\textbf{4}} & \textbf{\textbf{Never Facebook - I live with this person Instagram}} & \textbf{\textbf{-21.667}} & \textbf{\textbf{1.136}} & \textbf{\textbf{70,755}} & \textbf{\textbf{-19.080}} & \textbf{0 } \\
\end{tabular} 
  \captionof{table}{Full reporting of the comparative statistics of Figure 2B in the main text where we group frequency of meets with alter to 
          Never or Sometimes or I Live with Alter. \textbf{Highlighted
contrasts} are significant at the 1\% level. DF = 70755.} 
\end{center} 

\subsubsection{Figure 2B: Connection Value Percentile by Platform \& Frequency of Meeting Alter, Aggregated Across All Platforms}

    \begin{center}
  \renewcommand{\arraystretch}{0.61}  
  \small
\begin{tabular}{@{\extracolsep{5pt}} cccccc} 
\\[-3.8ex]\hline 
\hline \\[-1.8ex] 
 & contrast & estimate(\%) & SE & t-stat & p-value \\ 
\hline \\[-1.8ex] 
\textbf{1} & \textbf{Never - Less than 4 times/year} & \textbf{-2.468} & \textbf{0.347} & \textbf{-7.105} & \textbf{0 } \\
\textbf{2} & \textbf{Never - About once a month} & \textbf{-4.027} & \textbf{0.358} & \textbf{-11.253} & \textbf{0 } \\
\textbf{3} & \textbf{Never - About once a week} & \textbf{-4.531} & \textbf{0.369} & \textbf{-12.273} & \textbf{0 } \\
\textbf{4} & \textbf{Never - Multiple times per week} & \textbf{-6.145} & \textbf{0.330} & \textbf{-18.645} & \textbf{0 } \\
\end{tabular} 
  \captionof{table}{Full reporting of the comparative statistics of Figure 2B in the main text:
          Pairwise Comparisons of the average over all platforms CVP for the different frequencies of meeting with Alter. 
          	\textbf{Highlighted contrasts} 
          are significant at the 1\% level. DF = 70761.} 
\end{center}

\subsubsection{Figure 2C: Connection Value Percentile by Whether You Are Connected with Alter on Another Platform, Aggregated Across All Platforms}

     \begin{center}
  \renewcommand{\arraystretch}{0.60}   
\begin{tabular}{@{\extracolsep{5pt}} cccccc} 
\\[-3.8ex]\hline 
\hline \\[-1.8ex] 
 & contrast & estimate & SE & t-stat & p-value \\ 
\hline \\[-1.8ex] 
\textbf{1} & \textbf{No - Yes} & \textbf{-2.392} & \textbf{0.272} & \textbf{-8.798} & \textbf{0  } \\
\hline \\[-1.8ex] 
\end{tabular} 
  \captionof{table}{Full reporting of the comparative statistics of Figure 2C in the main text: Pairwise Comparisons of the average over all platforms CVP for whether Ego is connected to Alter on other platform or Not. \textbf{Highlighted contrasts} are significant at the 1\% level. DF = 70765.} 
\end{center} 

\clearpage

\subsubsection{Figure 3A: Connection Value Percentile by Platform \& Gender, Aggregated by Alter Gender}

\begin{center}
  \renewcommand{\arraystretch}{0.55}   
  \small
\begin{tabular}{@{\extracolsep{5pt}} cccccc} 
\\[-3.8ex]\hline 
\hline \\[-1.8ex] 
 & contrast & estimate (\%) & SE & t-stat & p-value \\ 
\hline \\[-1.8ex] 
1 & Alter Female Facebook - Alter Male Facebook & 1.009 & 0.325 & 3.101 & 0.041 \\ 
\textbf{2} & \textbf{Alter Female Facebook - Alter Female Instagram} & \textbf{-3.389} & \textbf{0.449} & \textbf{-7.548} & \textbf{0 } \\
\textbf{3} & \textbf{Alter Female Facebook - Alter Male Instagram} & \textbf{-2.081} & \textbf{0.507} & \textbf{-4.104} & \textbf{0.001 } \\
\textbf{4} & \textbf{Alter Female Facebook - Alter Female LinkedIn} & \textbf{-1.905} & \textbf{0.503} & \textbf{-3.791} & \textbf{0.004 } \\
\end{tabular} 
  \captionof{table}{Full reporting of the comparative statistics of Figure 3A in the main text where we group gender mix into 2 bins: A. Alter Female,
          B. Alter Male. \textbf{Highlighted contrasts} are significant at the 1\% level. DF = 67575.} 
\end{center} 

\subsubsection{Figure 3A: Connection Value Percentile by Whether You relative Ego and Alter Gender, Aggregated Across All Platforms}
  \begin{center}
  \renewcommand{\arraystretch}{0.63}
  \small
\begin{tabular}{@{\extracolsep{5pt}} cccccc} 
\\[-3.8ex]\hline 
\hline \\[-1.8ex] 
 & contrast & estimate (\%) & SE & t-stat & p-value \\ 
\hline \\[-1.8ex] 
\textbf{1} & \textbf{(Ego Male / Alter Male) - (Ego Female / Alter Female)} & \textbf{-1.719} & \textbf{0.274} & \textbf{-6.266} & \textbf{0 } \\
\textbf{2} & \textbf{(Ego Male / Alter Male) - (Ego Female / Alter Male)} & \textbf{-1.873} & \textbf{0.318} & \textbf{-5.892} & \textbf{0.00000 } \\
\textbf{3} & \textbf{(Ego Male / Alter Male) - (Ego Male / Alter Female)} & \textbf{-2.863} & \textbf{0.333} & \textbf{-8.608} & \textbf{0 } \\
4 & (Ego Female / Alter Female) - (Ego Female / Alter Male) & -0.154 & 0.318 & -0.485 & 0.962 \\ 
\textbf{5} & \textbf{(Ego Female / Alter Female) - (Ego Male / Alter Female)} & \textbf{-1.145} & \textbf{0.332} & \textbf{-3.443} & \textbf{0.003 } \\
6 & (Ego Female / Alter Male) - (Ego Male / Alter Female) & -0.990 & 0.369 & -2.682 & 0.037 \\ 
\hline \\[-3.8ex] 
\end{tabular} 
 \captionof{table}{Full reporting of the comparative statistics of Figure 3A in the main text:
          Pairwise Comparisons of the average over all platforms CVP for the different Gender mix of Ego and Alter. 
          	\textbf{Highlighted contrasts} are significant at the 1\% level. DF = 67579.} 
\end{center}

\subsubsection{Figure 3B: Connection Value Percentile by Platform \& Race, Aggregated to Alter Race}
\begin{center}
  \renewcommand{\arraystretch}{0.60}   
  \footnotesize
\begin{tabular}{@{\extracolsep{5pt}} cccccc} 
\\[-1.8ex]\hline 
\hline \\[-1.8ex] 
 & contrast & estimate (\%) & SE & t-stat & p-value \\ 
\hline \\[-1.8ex] 
1 & Alter White Facebook - (Alter non-White Facebook) & 0.447 & 0.342 & 1.307 & 0.897 \\ 
\textbf{2} & \textbf{Alter White Facebook - Alter White Instagram} & \textbf{-3.472} & \textbf{0.441} & \textbf{-7.880} & \textbf{0 } \\
\textbf{3} & \textbf{Alter White Facebook - (Alter non-White Instagram)} & \textbf{-2.923} & \textbf{0.500} & \textbf{-5.847} & \textbf{0.00000 } \\
4 & Alter White Facebook - Alter White LinkedIn & -0.813 & 0.426 & -1.907 & 0.546 \\
\end{tabular} 
  \captionof{table}{Full reporting of the comparative statistics of Figure 3B in the main text where we group gender mix into 2 bins: A. Alter White,
          B. Alter non-White \textbf{Highlighted
contrasts} are significant at the 1\% level. DF = 66770} 
\end{center} 

\subsubsection{Figure 3B: Connection Value Percentile by Whether You relative Race of Ego and Alter, Aggregated Across All Platforms}

  \begin{center}
  \renewcommand{\arraystretch}{0.63}
  \scriptsize
\begin{tabular}{@{\extracolsep{5pt}} cccccc} 
\\[-3.8ex]\hline 
\hline \\[-1.8ex] 
 & contrast & estimate (\%) & SE & t-stat & p-value \\ 
\hline \\[-1.8ex] 
\textbf{1} & \textbf{Ego} \& \textbf{Alter both Whites - (Ego White - Alter Non-White)} & \textbf{1.354} & \textbf{0.330} & \textbf{4.105} & \textbf{0.0002 } \\
2 & Ego \& Alter both Whites - (Ego Non-White - Alter White) & 1.203 & 0.445 & 2.702 & 0.035 \\ 
3 & Ego \& Alter both Whites - (Both Ego and Alter Non-Whites) & -0.669 & 0.273 & -2.453 & 0.067 \\ 
4 & (Ego White - Alter Non-White) - (Ego Non-White - Alter White) & -0.151 & 0.512 & -0.295 & 0.991 \\ 
\textbf{5} & \textbf{(Ego White - Alter Non-White) - (Both Ego and Alter Non-Whites)} & \textbf{-2.023} & \textbf{0.373} & \textbf{-5.429} & \textbf{0.00000 } \\
\textbf{6} & \textbf{(Ego Non-White - Alter White) - (Both Ego and Alter Non-Whites)} & \textbf{-1.871} & \textbf{0.478} & \textbf{-3.918} & \textbf{0.001 } \\
\hline \\[-2.8ex] 
\end{tabular} 
  \captionof{table}{Full reporting of the comparative statistics of Figure 3B in the main text:
          Pairwise Comparisons of the average over all platforms CVP for the different Race mix between Alter and Ego groups. 
          \textbf{Highlighted contrasts} are significant at the 1\% level. DF = 66774.} 
\end{center} 

\subsubsection{Figure 3C: Connection Value Percentile by Whether You relative Age of Ego and Alter, Aggregated Across All Platforms}

 \begin{center}
  \renewcommand{\arraystretch}{0.63}
\begin{tabular}{@{\extracolsep{5pt}} cccccc} 
\\[-3.8ex]\hline 
\hline \\[-1.8ex] 
 & contrast & estimate (\%) & SE & t-stat & p-value \\ 
\hline \\[-1.8ex] 
1 & ego younger than alter - ego same age as alter & -0.241 & 0.270 & -0.893 & 0.645 \\ 
2 & ego younger than alter - ego older than alter & -0.176 & 0.293 & -0.600 & 0.820 \\ 
3 & ego same age as alter - ego older than alter & 0.065 & 0.261 & 0.251 & 0.966 \\ 
\hline \\[-2.8ex] 
\end{tabular} 
\end{center} 
  \captionof{table}{Full reporting of the comparative statistics of Figure 3C in the main text: Pairwise Comparisons of 
          Connection Value Percentile between different Age mix of Ego and Alter for the different platforms. \textbf{Highlighted
contrasts} are significant at the 1\% level. DF = 67965.}

\subsubsection{Figure 3D: Connection Value Percentile by Platform \& Alter Age, Aggregated to Alter Over or Under Age 18}

\begin{center}
  \renewcommand{\arraystretch}{0.60}   
  \scriptsize
\begin{tabular}{@{\extracolsep{5pt}} cccccc} 
\\[-3.8ex]\hline 
\hline \\[-1.8ex] 
 & contrast & estimate (\%) & SE & t-stat & p-value \\ 
\hline \\[-1.8ex] 
1 & Alter under 18 years old Facebook- Alter over 18 years old Facebook& 0.724 & 0.935 & 0.775 & 0.994 \\ 
\textbf{2} & \textbf{Alter under 18 years old Facebook- Alter under 18 years old Instagram} & \textbf{-6.907} & \textbf{1.590} & \textbf{-4.345} & \textbf{0.0004 } \\
3 & Alter under 18 years old Facebook- Alter over 18 years old Instagram & -2.497 & 0.966 & -2.585 & 0.161 \\ 
4 & Alter under 18 years old Facebook- Alter under 18 years old LinkedIn & -1.105 & 2.576 & -0.429 & 1.000 \\
\end{tabular} 
  \captionof{table}{Full reporting of the comparative statistics of Figure 3D in the main text where we group Alter Age into 2 bins: A. Below 18 yo,
          B. Alter over 18 yo. \textbf{Highlighted
contrasts} are significant at the 1\% level. DF = 70759} 
\end{center} 

\subsubsection{Figure 3D: Connection Value Percentile by Platform \& 
Alter Age, Aggregated Across All Platforms}

\begin{center}
  \renewcommand{\arraystretch}{0.63}
\begin{tabular}{@{\extracolsep{5pt}} cccccc} 
\\[-3.8ex]\hline 
\hline \\[-1.8ex] 
 & contrast & estimate (\%) & SE & t-stat & p-value \\ 
\hline \\[-1.8ex] 
1 & Under 18 years old - (18-24 years old) & 1.644 & 0.669 & 2.459 & 0.137 \\ 
\textbf{2} & \textbf{Under 18 years old - (25-34 years old)} & \textbf{2.214} & \textbf{0.653} & \textbf{3.393} & \textbf{0.009 } \\
\textbf{3} & \textbf{Under 18 years old - (35-44 years old)} & \textbf{2.353} & \textbf{0.664} & \textbf{3.545} & \textbf{0.005 } \\
\textbf{4} & \textbf{Under 18 years old - (45-54 years old)} & \textbf{3.135} & \textbf{0.705} & \textbf{4.445} & \textbf{0.0001 } \\
5 & Under 18 years old - 55 years or older & 1.466 & 0.724 & 2.025 & 0.328 \\ 
6 & (18-24 years old) - (25-34 years old) & 0.570 & 0.308 & 1.854 & 0.431 \\ 
7 & (18-24 years old) - (35-44 years old) & 0.710 & 0.331 & 2.144 & 0.265 \\ 
\textbf{8} & \textbf{(18-24 years old) - (45-54 years old)} & \textbf{1.491} & \textbf{0.408} & \textbf{3.658} & \textbf{0.003 } \\
9 & (18-24 years old) - 55 years or older & -0.178 & 0.440 & -0.404 & 0.999 \\ 
10 & (25-34 years old) - (35-44 years old) & 0.139 & 0.297 & 0.468 & 0.997 \\ 
11 & (25-34 years old) - (45-54 years old) & 0.921 & 0.381 & 2.418 & 0.150 \\ 
12 & (25-34 years old) - 55 years or older & -0.748 & 0.415 & -1.803 & 0.463 \\ 
13 & (35-44 years old) - (45-54 years old) & 0.781 & 0.400 & 1.954 & 0.369 \\ 
14 & (35-44 years old) - 55 years or older & -0.887 & 0.432 & -2.052 & 0.313 \\ 
\textbf{15} & \textbf{(45-54 years old) - 55 years or older} & \textbf{-1.669} & \textbf{0.493} & \textbf{-3.381} & \textbf{0.009 } \\
\hline \\[-1.8ex] 
\end{tabular} 
  \captionof{table}{Full reporting of the comparative statistics of Figure 3D in the main text:
          Pairwise Comparisons of the average over all platforms CVP for the different Alter Age groups. 
          	\textbf{Highlighted contrasts} are significant at the 1\% level. DF = 67971} 
\end{center}

\clearpage

\subsection{All Statistical Comparisons}\label{all_comparisons}
In this section we present estimated differences, and associated p-values, for every pair of respondent types in Figures 1 through 4. 

\subsubsection{Figure 1B: Average Total Platform Valuation by Platform and Average Total Platform Valuation that is attributed to the local network effects by Platform}

 \begin{center}
  \renewcommand{\arraystretch}{0.60}   
\begin{tabular}{@{\extracolsep{5pt}} cccccc} 
\\[-1.8ex]\hline 
\hline \\[-1.8ex] 
 & contrast & estimate (\$) & SE & t-test & p-value \\ 
\hline \\[-1.8ex] 
\textbf{1} & \textbf{Facebook - Instagram} & \textbf{9.598} & \textbf{1.684} & \textbf{5.701} & \textbf{0.00000 } \\
\textbf{2} & \textbf{Facebook - LinkedIn} & \textbf{14.007} & \textbf{1.738} & \textbf{8.061} & \textbf{0 } \\
\textbf{3} & \textbf{Facebook - X} & \textbf{22.500} & \textbf{1.730} & \textbf{13.005} & \textbf{0 } \\
4 & Instagram - LinkedIn & 4.409 & 1.735 & 2.541 & 0.054 \\ 
\textbf{5} & \textbf{Instagram - X} & \textbf{12.901} & \textbf{1.727} & \textbf{7.470} & \textbf{0 } \\
\textbf{6} & \textbf{LinkedIn - X} & \textbf{8.493} & \textbf{1.780} & \textbf{4.772} & \textbf{0.00001 } \\
\hline \\[-3.8ex] 
\end{tabular} 
  \captionof{table}{Full comparative statistics of Figure 1B in the main text:
          Pairwise Comparisons of \textbf{the average TPV} for the different platforms. 
          	\textbf{Highlighted contrasts} 
          are significant at the 1\% level. DF = 9391.} 
\end{center}

\begin{center}
  \renewcommand{\arraystretch}{0.60}   
\begin{tabular}{@{\extracolsep{5pt}} cccccc} 
\\[-1.8ex]\hline 
\hline \\[-1.8ex] 
 & contrast & estimate (\$) & SE & t-stat & p-value \\ 
\hline \\[-1.8ex] 
1 & Facebook - Instagram & -4.637 & 3.175 & -1.460 & 0.144 \\ 
\textbf{2} & \textbf{Facebook} - \textbf{LinkedIn}  & \textbf{10.960} & \textbf{2.945} & \textbf{3.721} & \textbf{0.0002 } \\
\textbf{3} & \textbf{Facebook} - \textbf{X}  & \textbf{10.877} & \textbf{3.121} & \textbf{3.485} & \textbf{0.0005 } \\
\textbf{4} & \textbf{Instagram} - \textbf{LinkedIn}  & \textbf{15.597} & \textbf{3.001} & \textbf{5.197} & \textbf{0.00000 } \\
\textbf{5} & \textbf{Instagram} - \textbf{X}   & \textbf{15.513} & \textbf{3.174} & \textbf{4.888} & \textbf{0.00000 } \\
6 & LinkedIn - X  & 0.083 & 2.943 & 0.028 & 0.977 \\ 
\hline \\[-3.8ex] 
\end{tabular} 
  \captionof{table}{Full comparative statistics of Figure 1B in the main text:
          Pairwise Comparisons of the \textbf{average TPV that is attributed to local network effects} for the different platforms. 
          	\textbf{Highlighted contrasts} 
          are significant at the 1\% level. DF = 9391.} 
\end{center}

\begin{center}
  \renewcommand{\arraystretch}{0.60}
\begin{tabular}{@{\extracolsep{5pt}} cccccc} 
\\[-1.8ex]\hline 
\hline \\[-1.8ex] 
 & contrast  & estimate (\%) & SE & t-stat & p-value \\ 
\hline \\[-1.8ex] 
1 & Facebook - Instagram & -7.815 & 3.367 & -2.321 & 0.020 \\ 
\textbf{2} & \textbf{Facebook} - \textbf{LinkedIn} & \textbf{8.427} & \textbf{3.161} & \textbf{2.666} & \textbf{0.008 } \\
3 & Facebook - X & 6.417 & 3.593 & 1.786 & 0.074 \\ 
\textbf{4} & \textbf{Instagram} - \textbf{LinkedIn} & \textbf{16.243} & \textbf{3.398} & \textbf{4.780} & \textbf{0.00000 } \\
\textbf{5} & \textbf{Instagram} - \textbf{X} & \textbf{14.232} & \textbf{3.803} & \textbf{3.743} & \textbf{0.0002 } \\
6 & LinkedIn - X & -2.010 & 3.622 & -0.555 & 0.579 \\ 
\hline \\[-3.8ex] 
\end{tabular} 
  \captionof{table}{Full comparative statistics of Figure 1B in the main text:
          Pairwise Comparisons of the \textbf{\% of TPV attributed to the local network effects} for the different platforms. 
          	\textbf{Highlighted contrasts} 
          are significant at the 1\% level. DF = 9391.} 
\end{center}

\clearpage

\subsubsection{Figure 1C: Total Platform Valuation by Platform \& Usage Levels}

  \begin{center}
  \renewcommand{\arraystretch}{0.60}   
  \small
\begin{tabular}{@{\extracolsep{5pt}} ccccccc} 
\\[-1.8ex]\hline 
\hline \\[-1.8ex] 
 & contrast & usage & estimate (\$) & SE & t-stat & p-value \\ 
\hline \\[-1.8ex] 
1 & Facebook - Instagram & Less than 10 minutes per day & 11.745 & 4.519 & 2.599 & 0.046 \\ 
2 & Facebook - LinkedIn & Less than 10 minutes per day & 3.889 & 3.760 & 1.034 & 0.729 \\ 
\textbf{3} & \textbf{Facebook - X} & \textbf{Less than 10 minutes per day} & \textbf{14.676} & \textbf{4.317} & \textbf{3.399} & \textbf{0.004} \\ 
4 & Instagram - LinkedIn & Less than 10 minutes per day & -7.857 & 3.805 & -2.065 & 0.165 \\
\end{tabular} 
  \captionof{table}{Full reporting of the comparative statistics of Figure 1C in the main text: Pairwise Comparisons of TPV for the different Platforms within Usage Levels. \textbf{Highlighted
contrasts} are significant at the 1\% level. DF $=$ 9371.}  
\end{center}

\begin{center}
  \renewcommand{\arraystretch}{0.63}   
    \scriptsize
\begin{tabular}{@{\extracolsep{5pt}} ccccccc} 
\\[-4.8ex]\hline 
\hline \\[-1.8ex] 
 & contrast & platform & estimate (\$) & SE & t-stat & p-value \\ 
\hline \\[-1.8ex] 
\textbf{1} & \textbf{Less than 10 minutes per day - 10–30 minutes per day} & \textbf{Facebook} & \textbf{-34.148} & \textbf{4.240} & \textbf{-8.054} & \textbf{0 } \\
\textbf{2} & \textbf{Less than 10 minutes per day - 31–60 minutes per day} & \textbf{Facebook} & \textbf{-13.902} & \textbf{3.905} & \textbf{-3.560} & \textbf{0.005 } \\
\textbf{3} & \textbf{Less than 10 minutes per day - 1–2 hours per day} & \textbf{Facebook} & \textbf{-30.799} & \textbf{4.796} & \textbf{-6.422} & \textbf{0 } \\
\textbf{4} & \textbf{Less than 10 minutes per day - 2–3 hours per day} & \textbf{Facebook} & \textbf{-16.677} & \textbf{4.099} & \textbf{-4.068} & \textbf{0.001 } \\
\end{tabular} 
  \captionof{table}{Full reporting of the comparative statistics of Figure 1C in the main text: Pair-
wise Comparisons of TPV for the different Usage Levels within Platforms. 	\textbf{Highlighted
contrasts} are significant at the 1\% level. DF= 9371.} 
\end{center}

\subsubsection{Figure 1D: Total Platform Valuation by Platform \& Political Orientation}

\begin{center}
  \renewcommand{\arraystretch}{0.63}   
    \small
\begin{tabular}{@{\extracolsep{5pt}} ccccccc} 
\\[-4.8ex]\hline 
\hline \\[-1.8ex] 
 & contrast & political & estimate (\$) & SE & t-stat & p-value \\ 
\hline \\[-1.8ex] 
1 & Facebook - Instagram & Extremely liberal & 6.435 & 5.423 & 1.187 & 0.635 \\ 
2 & Facebook - LinkedIn & Extremely liberal & 10.433 & 5.410 & 1.929 & 0.216 \\ 
\textbf{3} & \textbf{Facebook - X} & \textbf{Extremely liberal} & \textbf{20.973} & \textbf{5.552} & \textbf{3.777} & \textbf{0.001} \\ 
4 & Instagram - LinkedIn & Extremely liberal & 3.998 & 5.632 & 0.710 & 0.893 \\
\end{tabular} 
  \captionof{table}{Full reporting of the comparative statistics of Figure 1D in the main text: Pairwise Comparisons of TPV for the different Platforms Within Political Groups. \textbf{Highlighted contrasts} are significant at the 1\% level. DF$=$9367.} 
\end{center}

\begin{center}
  \renewcommand{\arraystretch}{0.505}   
    \scriptsize
\begin{tabular}{@{\extracolsep{5pt}} ccccccc} 
\\[-4.8ex]\hline 
\hline \\[-1.8ex] 
 & contrast & platform & estimate (\$) & SE & t-stat & p-value \\ 
\hline \\[-1.8ex] 
1 & Extremely liberal - Liberal & Facebook & $$-$9.152$ & $4.986$ & $$-$1.836$ & $0.524$ \\ 
2 & Extremely liberal - Slightly liberal & Facebook & $$-$9.952$ & $5.633$ & $$-$1.767$ & $0.571$ \\ 
3 & Extremely liberal - Moderate & Facebook & $$-$1.209$ & $4.909$ & $$-$0.246$ & $1.000$ \\ 
4 & Extremely liberal - Slightly conservative & Facebook & $$-$9.754$ & $4.140$ & $$-$2.356$ & $0.218$ \\
\end{tabular} 
  \captionof{table}{Full reporting of the comparative statistics of Figure 1D in the main text: Pairwise Comparisons of TPV for the different Political Groups within Platforms. \textbf{Highlighted contrasts} are significant at the 1\% level. DF$=$9367.} 
\end{center}

\clearpage

\subsubsection{Figure 1E: Total Platform Valuation by Platform \& Relationship Status}

\begin{center}
  \renewcommand{\arraystretch}{0.60}   
  \footnotesize
\begin{tabular}{@{\extracolsep{5pt}} ccccccc} 
\\[-4.8ex]\hline 
\hline \\[-1.8ex] 
 & contrast & Relationship Status & estimate (\$) & SE & t-stat & p-value \\ 
\hline \\[-1.8ex] 
1 & Facebook - Instagram & Single, not looking for a relationship & 7.701 & 3.799 & 2.027 & 0.178 \\ 
2 & Facebook - LinkedIn & Single, not looking for a relationship & 10.237 & 3.793 & 2.699 & 0.035 \\ 
\textbf{3} & \textbf{Facebook - X} & \textbf{Single, not looking for a relationship} & \textbf{21.463} & \textbf{3.717} & \textbf{5.774} & \textbf{0.00000} \\ 
4 & Instagram - LinkedIn & Single, not looking for a relationship & 2.536 & 4.054 & 0.625 & 0.924 \\
\end{tabular} 
  \captionof{table}{Full reporting of the comparative statistics of Figure 1E in the main text: Pairwise Comparisons of TPV for the different Platforms within Relationship Status. \textbf{Highlighted
contrasts} are significant at the 1\% level. DF = 9277.} 
\end{center} 

\newpage

\begin{center}
  \renewcommand{\arraystretch}{0.63}   
  \scriptsize
\begin{tabular}{@{\extracolsep{5pt}} ccccccc} 
\\[-4.8ex]\hline 
\hline \\[-1.8ex] 
 & contrast & Platform & estimate (\$) & SE & t-stat & p-value \\ 
\hline \\[-1.8ex] 
1 & Single, not looking for a relationship - Single, looking for a relationship & Facebook & $3.671$ & $3.638$ & $1.009$ & $0.744$ \\ 
2 & Single, not looking for a relationship - In a serious relationship & Facebook & $5.475$ & $3.883$ & $1.410$ & $0.493$ \\ 
3 & Single, not looking for a relationship - Married or domestic partnership & Facebook & $$-$5.491$ & $3.123$ & $$-$1.758$ & $0.294$ \\ 
4 & Single, looking for a relationship - In a serious relationship & Facebook & $1.804$ & $3.991$ & $0.452$ & $0.969$ \\
\end{tabular} 
  \captionof{table}{Full reporting of the comparative statistics of Figure 1E in the main text: Pairwise Comparisons of TPV for the different Relationship Statuses within Platforms. 	\textbf{Highlighted
contrasts} are significant at the 1\% level. DF $=$ 9277.}
\end{center}

\subsubsection{Figure 1F: Total Platform Valuation by Platform \& Job Seeking}

\begin{center}
  \renewcommand{\arraystretch}{0.63}   
\begin{tabular}{@{\extracolsep{5pt}} ccccccc} 
\\[-1.8ex]\hline 
\hline \\[-1.8ex] 
 & contrast & platform & estimate (\$) & SE & t-stat & p-value \\ 
\hline \\[-1.8ex] 
\textbf{1} & \textbf{Job Seekers - Non Job Seekers} & \textbf{Facebook} & \textbf{-9.472} & \textbf{2.623} & \textbf{-3.612} & \textbf{0.0003} \\ 
2 & Job Seekers - Non Job Seekers & Instagram & 0.052 & 2.542 & 0.020 & 0.984 \\ 
\textbf{3} & \textbf{Job Seekers - Non Job Seekers} & \textbf{LinkedIn} & \textbf{8.330} & \textbf{2.612} & \textbf{3.189} & \textbf{0.001} \\ 
4 & Job Seekers - Non Job Seekers & X & -2.528 & 2.735 & -0.924 & 0.355 \\ 
\hline \\[-1.8ex] 
\end{tabular} 
  \captionof{table}{Full reporting of the comparative statistics of Figure 1F in the main text: Pairwise Comparisons of the TPV between job seekers and non job seekers for the different platforms. 
          	\textbf{Highlighted contrasts} 
          are significant at the 1\% level. DF $=$ 9387.} 
\end{center}

\subsubsection{Figure 2A: Connection Value Percentile by Platform \& How You Know Alter}

\begin{center}
  \renewcommand{\arraystretch}{0.63}   
  \tiny
\begin{tabular}{@{\extracolsep{5pt}} ccccccc} 
\\[-3.8ex]\hline 
\hline \\[-1.8ex] 
 & contrast & Platform & estimate(\%) & SE & t-stat & p-value \\ 
\hline \\[-1.8ex] 
\textbf{1} & \textbf{I only know them online - None of these describe how I know this person} & \textbf{Facebook} & \textbf{-3.937} & \textbf{1.061} & \textbf{-3.712} & \textbf{0.004 } \\
\textbf{2} & \textbf{I only know them online - School (K-12)} & \textbf{Facebook} & \textbf{-4.109} & \textbf{0.814} & \textbf{-5.048} & \textbf{0.00001 } \\
\textbf{3} & \textbf{I only know them online - Work} & \textbf{Facebook} & \textbf{-4.826} & \textbf{0.748} & \textbf{-6.453} & \textbf{0 } \\
\textbf{4} & \textbf{I only know them online - Through a shared interest in real life} & \textbf{Facebook} & \textbf{-4.419} & \textbf{0.789} & \textbf{-5.603} & \textbf{0.00000 } \\
\end{tabular} 
  \captionof{table}{Full reporting of the comparative statistics of Figure 2A in the main text: Pairwise Comparisons of 
          Connection Value Percentile between different ways that Ego knows Alter for the different platforms. \textbf{Highlighted
contrasts} are significant at the 1\% level. DF = 70739} 
\end{center}

\begin{center}
  \renewcommand{\arraystretch}{0.65}   
\footnotesize
\begin{tabular}{rllrrrr}
  \hline
 & contrast & How do you know Alter? & estimate(\%) & SE & t-stat & p-value \\ 
  \hline
\textbf{1} & \textbf{Facebook - Instagram} & \textbf{I only know them online} & \textbf{-6.54} & \textbf{1.04} & \textbf{-6.28} & \textbf{0.00 } \\
  2 & Facebook - LinkedIn & I only know them online & -3.09 & 1.21 & -2.55 & 0.05 \\ 
\textbf{  3} & \textbf{Facebook - X} & \textbf{I only know them online} & \textbf{-5.65} & \textbf{0.84} & \textbf{-6.76} & \textbf{0.00 } \\
  4 & Instagram - LinkedIn & I only know them online & 3.44 & 1.31 & 2.62 & 0.04 \\ 
  5 & Instagram - X & I only know them online & 0.89 & 0.97 & 0.91 & 0.80 \\ 
  6 & LinkedIn - X & I only know them online & -2.55 & 1.16 & -2.21 & 0.12 \\ 
\textbf{  7} & \textbf{Facebook - Instagram} & \textbf{None of these describe how I know this person} & \textbf{-4.87} & \textbf{1.55} & \textbf{-3.14} & \textbf{0.01 } \\
\hline
\end{tabular}
\captionof{table}{Full reporting of the comparative statistics of Figure 2A in the main text: Pairwise Comparisons of Connection Value Percentile for the different Platforms for different ways that Ego knows Alter. \textbf{Highlighted
contrasts} are significant at the 1\% level. DF = 70739.} 
\end{center}

\subsubsection{Figure 2B: Connection Value Percentile by Platform \& Frequency of Meeting Alter}

  \begin{center}
  \renewcommand{\arraystretch}{0.65}   
  \scriptsize
\begin{tabular}{rllrrrr}
  \hline
 & contrast & Platform & estimate(\%) & SE & t-stat & p-value \\ 
  \hline
\textbf{1} & \textbf{Never - Less than 4 times/year)} & \textbf{Facebook} & \textbf{-5.12} & \textbf{0.54} & \textbf{-9.41} & \textbf{0.00 } \\
\textbf{  2} & \textbf{Never - About once a month} & \textbf{Facebook} & \textbf{-7.25} & \textbf{0.56} & \textbf{-12.84} & \textbf{0.00 } \\
\textbf{  3} & \textbf{Never - About once a week} & \textbf{Facebook} & \textbf{-9.02} & \textbf{0.58} & \textbf{-15.50} & \textbf{0.00 } \\
\textbf{  4} & \textbf{Never - Multiple times per week} & \textbf{Facebook} & \textbf{-10.61} & \textbf{0.53} & \textbf{-20.11} & \textbf{0.00 } \\
\textbf{  5} & \textbf{Never - I live with this person} & \textbf{Facebook} & \textbf{-22.32} & \textbf{0.69} & \textbf{-32.50} & \textbf{0.00 } \\
\textbf{  6} & \textbf{Less than 4 times/year) - About once a month} & \textbf{Facebook} & \textbf{-2.13} & \textbf{0.52} & \textbf{-4.11} & \textbf{0.00 } \\
\textbf{  7} & \textbf{Less than 4 times/year) - About once a week} & \textbf{Facebook} & \textbf{-3.89} & \textbf{0.54} & \textbf{-7.26} & \textbf{0.00 } \\
\hline
\end{tabular}
\captionof{table}{Full reporting of the comparative statistics of Figure 2B in the main text: 
    Pairwise Comparisons of Connection Value Percentile for different Platforms within different bins for the frequency in which ego meets alter. 
    \textbf{Highlighted contrasts} are significant at the 1\% level. DF = 70743.}
\end{center}

    \begin{center}
  \renewcommand{\arraystretch}{0.63}   
\begin{tabular}{rllrrrr}
  \hline
 & contrast & How Freq. you meet Alter & estimate(\%) & SE & t-stat & p-value \\ 
  \hline
\textbf{1} & \textbf{Facebook - Instagram} & \textbf{Never} & \textbf{-8.45} & \textbf{0.71} & \textbf{-11.82} & \textbf{0.00 } \\
\textbf{  2} & \textbf{Facebook - LinkedIn} & \textbf{Never} & \textbf{-5.83} & \textbf{0.70} & \textbf{-8.28} & \textbf{0.00 } \\
\textbf{  3} & \textbf{Facebook - X} & \textbf{Never} & \textbf{-7.02} & \textbf{0.62} & \textbf{-11.35} & \textbf{0.00 } \\
\textbf{  4} & \textbf{Instagram - LinkedIn} & \textbf{Never} & \textbf{2.62} & \textbf{0.81} & \textbf{3.22} & \textbf{0.01 } \\
  5 & Instagram - X & Never & 1.43 & 0.74 & 1.94 & 0.21 \\ 
  6 & LinkedIn - X & Never & -1.18 & 0.73 & -1.62 & 0.36 \\ 
\textbf{  7} & \textbf{Facebook - Instagram} & \textbf{Less than 4 times/year} & \textbf{-4.66} & \textbf{0.73} & \textbf{-6.37} & \textbf{0.00 } \\
\hline
\end{tabular}
\captionof{table}{Full reporting of the comparative statistics of Figure 2B in the main text: 
    Pairwise Comparisons of Connection Value Percentile for different Platforms within different bins for the frequency in which ego meets alter. 
    \textbf{Highlighted contrasts} are significant at the 1\% level. DF = 70743.}
\end{center}

\subsubsection{Figure 2C: Connection Value Percentile by Platform \& Whether Ego is Connected with Alter on Other Platform.}

      \begin{center}
  \renewcommand{\arraystretch}{0.63}  
\begin{tabular}{@{\extracolsep{5pt}} ccccccc} 
\\[-3.8ex]\hline 
\hline \\[-1.8ex] 
 & contrast & Platform & estimate(\%) & SE & t.ratio & p.value \\ 
\hline \\[-1.8ex] 
\textbf{1} & \textbf{No - Yes} & \textbf{Facebook} & \textbf{-6.917} & \textbf{0.722} & \textbf{-9.583} & \textbf{0 } \\
\textbf{2} & \textbf{No - Yes} & \textbf{Instagram} & \textbf{-2.928} & \textbf{0.620} & \textbf{-4.722} & \textbf{0.00000 } \\
\textbf{3} & \textbf{No - Yes} & \textbf{LinkedIn} & \textbf{-3.515} & \textbf{0.605} & \textbf{-5.813} & \textbf{0 } \\
\textbf{4} & \textbf{No - Yes} & \textbf{X} & \textbf{-2.287} & \textbf{0.454} & \textbf{-5.038} & \textbf{0.00000 } \\
\hline \\[-3.8ex] 
\end{tabular} 
  \captionof{table}{Full reporting of the comparative statistics of Figure 2C in the main text: Pairwise Comparisons of 
          Connection Value Percentile between yes and no  response to the question whether ego is connected with alter in other platform for the different platforms. \textbf{Highlighted
contrasts} are significant at the 1\% level. DF = 70759.} \label{complementarity_table}
\end{center} 

\begin{center}
  \renewcommand{\arraystretch}{0.63} 
\footnotesize
\begin{tabular}{rllrrrr}
  \hline
 & contrast & Are you con. with Alter on other Platform? & estimate(\%) & SE & t.ratio & p.value \\ 
  \hline
\textbf{1} & \textbf{Facebook- Instagram} & \textbf{No} & \textbf{-7.94} & \textbf{0.87} & \textbf{-9.11} & \textbf{0.00 } \\
\textbf{  2} & \textbf{Facebook- LinkedIn} & \textbf{No} & \textbf{-5.50} & \textbf{0.85} & \textbf{-6.44} & \textbf{0.00 } \\
\textbf{  3} & \textbf{Facebook- X} & \textbf{No} & \textbf{-5.88} & \textbf{0.80} & \textbf{-7.34} & \textbf{0.00 } \\
\textbf{  4} & \textbf{Instagram - LinkedIn} & \textbf{No} & \textbf{2.45} & \textbf{0.71} & \textbf{3.46} & \textbf{0.00 } \\
\textbf{  5} & \textbf{Instagram - X} & \textbf{No} & \textbf{2.06} & \textbf{0.64} & \textbf{3.21} & \textbf{0.01 } \\
  6 & LinkedIn - X & No & -0.38 & 0.62 & -0.62 & 0.93 \\ 
\textbf{  7} & \textbf{Facebook- Instagram} & \textbf{Yes} & \textbf{-3.95} & \textbf{0.38} & \textbf{-10.38} & \textbf{0.00 } \\
\textbf{  8} & \textbf{Facebook- LinkedIn} & \textbf{Yes} & \textbf{-2.09} & \textbf{0.40} & \textbf{-5.27} & \textbf{0.00 } \\
\textbf{  9} & \textbf{Facebook- X} & \textbf{Yes} & \textbf{-1.25} & \textbf{0.29} & \textbf{-4.26} & \textbf{0.00 } \\
\textbf{  10} & \textbf{Instagram - LinkedIn} & \textbf{Yes} & \textbf{1.86} & \textbf{0.50} & \textbf{3.73} & \textbf{0.00 } \\
\textbf{  11} & \textbf{Instagram - X} & \textbf{Yes} & \textbf{2.70} & \textbf{0.42} & \textbf{6.43} & \textbf{0.00 } \\
  12 & LinkedIn - X & Yes & 0.84 & 0.44 & 1.94 & 0.21 \\ 
   \hline
\end{tabular}
\captionof{table}{Full reporting of the comparative statistics of Figure 2C in the main text: Pairwise Comparisons of Connection Value Percentile for the different Platforms for whether ego is connected with alter in other platform. \textbf{Highlighted
contrasts} are significant at the 1\% level. DF = 70759.} 
\end{center}

\subsubsection{Figure 2D: Connection Value Percentile by Platform \& Whether Ego is Connected with Alter on Each other Platform}

 \begin{center}
  \renewcommand{\arraystretch}{0.63}
 \footnotesize
\begin{tabular}{rllrrrr}
  \hline
 & contrast & Are you connected with Alter also in Facebook? & estimate (\%) & SE & t-stat & p-value \\ 
  \hline
1 & Instagram - LinkedIn& No & 2.09 & 0.79 & 2.64 & 0.02 \\ 
  2 & Instagram - X & No & 1.96 & 0.71 & 2.77 & 0.02 \\ 
  3 & LinkedIn- X & No & -0.13 & 0.70 & -0.19 & 0.98 \\ 
\textbf{  4} & \textbf{Instagram - LinkedIn} & \textbf{Yes} & \textbf{1.90} & \textbf{0.51} & \textbf{3.76} & \textbf{0.00 } \\
\textbf{  5} & \textbf{Instagram - X} & \textbf{Yes} & \textbf{2.73} & \textbf{0.43} & \textbf{6.40} & \textbf{0.00 } \\
  6 & LinkedIn- X & Yes & 0.83 & 0.44 & 1.87 & 0.15 \\ 
   \hline
\end{tabular}
\captionof{table}{Full reporting of the comparative statistics of Figure 2D (top left) in the main text: Pairwise Comparisons of 
           Connection Value Percentile between different platforms for the case that Ego and Alter are connected also in Facebook or not. \textbf{Highlighted
 contrasts} are significant at the 1\% level. DF = 36045.} 
\end{center}

 \begin{center}
  \renewcommand{\arraystretch}{0.63}
\begin{tabular}{@{\extracolsep{5pt}} ccccccc} 
\\[-1.8ex]\hline 
\hline \\[-1.8ex] 
 & contrast & Platform & estimate (\%) & SE & t-stat & p-value \\ 
\hline \\[-1.8ex] 
\textbf{1} & \textbf{No - Yes} & \textbf{Instagram} & \textbf{-2.610} & \textbf{0.666} & \textbf{-3.919} & \textbf{0.0001 } \\
\textbf{2} & \textbf{No - Yes} & \textbf{LinkedIn} & \textbf{-2.796} & \textbf{0.663} & \textbf{-4.217} & \textbf{0.00002 } \\
\textbf{3} & \textbf{No - Yes} & \textbf{X} & \textbf{-1.838} & \textbf{0.489} & \textbf{-3.761} & \textbf{0.0002 } \\
\hline \\[-1.8ex] 
\end{tabular} 
  \captionof{table}{Full reporting of the comparative statistics of Figure 2D (top left) in the main text: Pairwise Comparisons of 
          Connection Value Percentile between whether Alter is also connected with Ego in Facebook or not for the different platforms. \textbf{Highlighted
contrasts} are significant at the 1\% level. DF = 36045} 
\end{center}

  \begin{center}
  \renewcommand{\arraystretch}{0.63}
 \footnotesize
\begin{tabular}{rllrrrr}
  \hline
 & contrast & Are you connected with Alter also on Instagram? & estimate (\%) & SE & t-stat & p-value \\ 
  \hline
\textbf{1} & \textbf{Facebook - LinkedIn} & \textbf{No} & \textbf{-2.02} & \textbf{0.47} & \textbf{-4.28} & \textbf{0.00} \\
  2 & Facebook - X & No & -0.91 & 0.45 & -2.01 & 0.11 \\ 
  3 & LinkedIn - X & No & 1.11 & 0.56 & 1.98 & 0.12 \\ 
  4 & Facebook - LinkedIn & Yes & -0.61 & 0.47 & -1.31 & 0.39 \\ 
  5 & Facebook - X & Yes & -0.08 & 0.33 & -0.26 & 0.96 \\ 
  6 & LinkedIn - X & Yes & 0.53 & 0.48 & 1.10 & 0.51 \\ 
   \hline
\end{tabular}
\captionof{table}{Full reporting of the comparative statistics of Figure 2D (top right) in the main text: Pairwise Comparisons of 
           Connection Value Percentile between different platforms for the case that Ego and Alter are connected also on Instagram or not. \textbf{Highlighted
 contrasts} are significant at the 1\% level. DF = 60799.} 
\end{center}

 \begin{center}
  \renewcommand{\arraystretch}{0.63}
\begin{tabular}{@{\extracolsep{5pt}} ccccccc} 
\\[-1.8ex]\hline 
\hline \\[-1.8ex] 
 & contrast & Platform & estimate (\%) & SE & t-stat & p-value \\ 
\hline \\[-1.8ex] 
\textbf{1} & \textbf{No - Yes} & \textbf{Facebook} & \textbf{-3.518} & \textbf{0.321} & \textbf{-10.944} & \textbf{0 } \\
\textbf{2} & \textbf{No - Yes} & \textbf{LinkedIn} & \textbf{-2.104} & \textbf{0.580} & \textbf{-3.625} & \textbf{0.0003 } \\
\textbf{3} & \textbf{No - Yes} & \textbf{X} & \textbf{-2.694} & \textbf{0.455} & \textbf{-5.919} & \textbf{0  } \\
\hline \\[-1.8ex] 
\end{tabular} 
  \captionof{table}{Full reporting of the comparative statistics of Figure 2D (top right) in the main text: Pairwise Comparisons of 
          Connection Value Percentile between whether Alter is also connected with Ego on Instagram or not for the different platforms. \textbf{Highlighted
contrasts} are significant at the 1\% level. DF = 60799.} 
\end{center}

   \begin{center}
  \renewcommand{\arraystretch}{0.63}
  \footnotesize
\begin{tabular}{rllrrrr}
  \hline
 & contrast & Are you connected with Alter also in LinkedIn? & estimate & SE & t-value & p-value \\ 
  \hline
\textbf{1} & \textbf{Facebook - Instagram} & \textbf{No} & \textbf{-0.04} & \textbf{0.00} & \textbf{-10.30} & \textbf{0.00 } \\
\textbf{  2} & \textbf{Facebook - X} & \textbf{No} & \textbf{-0.01} & \textbf{0.00} & \textbf{-3.88} & \textbf{0.00 } \\
\textbf{  3} & \textbf{Instagram - X} & \textbf{No} & \textbf{0.03} & \textbf{0.00} & \textbf{6.07} & \textbf{0.00 } \\
\textbf{  4} & \textbf{Facebook - Instagram} & \textbf{Yes} & \textbf{-0.02} & \textbf{0.01} & \textbf{-2.92} & \textbf{0.01 } \\
  5 & Facebook - X & Yes & -0.00 & 0.00 & -0.57 & 0.83 \\ 
  6 & Instagram - X & Yes & 0.02 & 0.01 & 2.55 & 0.03 \\ 
   \hline
\end{tabular}
\captionof{table}{Full reporting of the comparative statistics of Figure 2D (bottom left) in the main text: Pairwise Comparisons of 
           Connection Value Percentile between different platforms for the case that Ego and Alter are also connected on LinkedIn or not. \textbf{Highlighted
 contrasts} are significant at the 1\% level. DF = 61069} 
\end{center}

   \begin{center}
  \renewcommand{\arraystretch}{0.63}
\begin{tabular}{@{\extracolsep{5pt}} ccccccc} 
\\[-1.8ex]\hline 
\hline \\[-1.8ex] 
 & contrast & Platform & estimate (\%) & SE & t-value & p-value \\ 
\hline \\[-1.8ex] 
1 & No - Yes & Facebook & $$-$0.008$ & $0.004$ & $$-$2.164$ & $0.030$ \\ 
2 & No - Yes & Instagram & $0.010$ & $0.007$ & $1.395$ & $0.163$ \\ 
3 & No - Yes & X & $0.003$ & $0.004$ & $0.628$ & $0.530$ \\ 
\hline \\[-1.8ex] 
\end{tabular} 
 \captionof{table}{Full reporting of the comparative statistics of Figure 2D (bottom left) in the main text: Pairwise Comparisons of 
          Connection Value Percentile between whether Alter is also connected with Ego on LinkedIn or not for the different platforms. \textbf{Highlighted
contrasts} are significant at the 1\% level. DF = 61069.} 
\end{center}

  \begin{center}
  \renewcommand{\arraystretch}{0.63}
 \footnotesize
\begin{tabular}{rllrrrr}
  \hline
 & contrast & Are you connected with Alter also on X? & estimate (\%) & SE & t-stat & p-value \\ 
  \hline
\textbf{1} & \textbf{Facebook - Instagram} & \textbf{No} & \textbf{-3.05} & \textbf{0.42} & \textbf{-7.25} & \textbf{0.00 } \\
\textbf{  2} & \textbf{Facebook - LinkedIn} & \textbf{No} & \textbf{-1.52} & \textbf{0.41} & \textbf{-3.73} & \textbf{0.00 } \\
\textbf{  3} & \textbf{Instagram - LinkedIn} & \textbf{No} & \textbf{1.53} & \textbf{0.51} & \textbf{2.97} & \textbf{0.01 } \\
\textbf{  4} & \textbf{Facebook - Instagram} & \textbf{Yes} & \textbf{-3.77} & \textbf{0.53} & \textbf{-7.16} & \textbf{0.00 } \\
  5 & Facebook - LinkedIn & Yes & -0.62 & 0.57 & -1.08 & 0.53 \\ 
\textbf{  6} & \textbf{Instagram - LinkedIn} & \textbf{Yes} & \textbf{3.15} & \textbf{0.68} & \textbf{4.67} & \textbf{0.00 } \\
   \hline
\end{tabular}
\captionof{table}{Full reporting of the comparative statistics of Figure 2D (bottom right) in the main text: Pairwise Comparisons of 
           Connection Value Percentile between different platforms for the case that Ego and Alter are also connected on X or not. \textbf{Highlighted
 contrasts} are significant at the 1\% level. DF = 51402.} 
\end{center}

  \begin{center}
  \renewcommand{\arraystretch}{0.63}
\begin{tabular}{@{\extracolsep{5pt}} ccccccc} 
\\[-1.8ex]\hline 
\hline \\[-1.8ex] 
 & contrast & Platform & estimate (\%) & SE & t-stat & p-value \\ 
\hline \\[-1.8ex] 
\textbf{1} & \textbf{No - Yes} & \textbf{Facebook} & \textbf{-1.495} & \textbf{0.336} & \textbf{-4.448} & \textbf{0.00001 } \\
\textbf{2} & \textbf{No - Yes} & \textbf{Instagram} & \textbf{-2.217} & \textbf{0.584} & \textbf{-3.799} & \textbf{0.0001 } \\
3 & No - Yes & LinkedIn & -0.591 & 0.615 & -0.961 & 0.336 \\ 
\hline \\[-1.8ex] 
\end{tabular} 
  \captionof{table}{Full reporting of the comparative statistics of Figure 2D (bottom right) in the main text: Pairwise Comparisons of 
          Connection Value Percentile between whether Alter is also connected with Ego on X or not for the different platforms. \textbf{Highlighted
contrasts} are significant at the 1\% level. DF = 51402.} 
\end{center}

\subsubsection{Figure 3A: Connection Value Percentile by Platform \& Ego and Alter Gender}

  \begin{center}
  \renewcommand{\arraystretch}{0.63}
  \scriptsize
\begin{tabular}{@{\extracolsep{5pt}} ccccccc} 
\\[-3.8ex]\hline 
\hline \\[-1.8ex] 
 & contrast & Platform & estimate (\%) & SE & t-stat & p-value \\ 
\hline \\[-1.8ex] 
1 & (Ego Male / Alter Male) - (Ego Female / Alter Female) & Facebook & -0.893 & 0.408 & -2.188 & 0.126 \\ 
\textbf{2} & \textbf{(Ego Male / Alter Male) - (Ego Female / Alter Male)} & \textbf{Facebook} & \textbf{-1.932} & \textbf{0.476} & \textbf{-4.060} & \textbf{0.0003 } \\
\textbf{3} & \textbf{(Ego Male / Alter Male) - (Ego Male / Alter Female)} & \textbf{Facebook} & \textbf{-3.439} & \textbf{0.490} & \textbf{-7.012} & \textbf{0 } \\
4 & (Ego Female / Alter Female) - (Ego Female / Alter Male) & Facebook & -1.039 & 0.466 & -2.229 & 0.115 \\
\end{tabular} 
  \captionof{table}{Full reporting of the comparative statistics of Figure 3A in the main text: Pairwise Comparisons of 
          Connection Value Percentile between different Gender mix of Ego and Alter for the different platforms. \textbf{Highlighted
contrasts} are significant at the 1\% level. DF = 67567} 
\end{center}

  \begin{center}
  \renewcommand{\arraystretch}{0.61}
  \footnotesize
\begin{tabular}{@{\extracolsep{5pt}} ccccccc} 
\\[-3.8ex]\hline 
\hline \\[-1.8ex] 
 & contrast & gender mix & estimate (\%) & SE & t-stat & p-value \\   
\hline \\[-1.8ex] 
1 & Facebook - Instagram & Ego Male / Alter Male & -2.233 & 0.726 & -3.075 & 0.011 \\ 
2 & Facebook - LinkedIn & Ego Male / Alter Male & -1.527 & 0.603 & -2.530 & 0.055 \\ 
3 & Facebook - X & Ego Male / Alter Male & -1.009 & 0.442 & -2.284 & 0.102 \\ 
4 & Instagram - LinkedIn & Ego Male / Alter Male & 0.706 & 0.846 & 0.835 & 0.838 \\
\end{tabular} 
 \captionof{table}{Full reporting of the comparative statistics of Figure 3A in the main text: Pairwise Comparisons of 
          Connection Value Percentile between different platforms for different gender contrast between Ego and Alter. \textbf{Highlighted
contrasts} are significant at the 1\% level. DF = 70739}
\end{center}

\clearpage

\subsubsection{Figure 3B: Connection Value Percentile by Platform \& Ego and Alter Race}

  \begin{center}
  \renewcommand{\arraystretch}{0.63}
  \scriptsize
\begin{tabular}{@{\extracolsep{5pt}} ccccccc} 
\\[-1.8ex]\hline 
\hline \\[-1.8ex] 
 & contrast & Platform & estimate (\%) & SE & t-stat & p-value \\ 
\hline \\[-1.8ex] 
\textbf{1} & \textbf{Ego} \& \textbf{Alter both Whites - (Ego White - Alter Non-White)} & \textbf{Facebook} & \textbf{1.991} & \textbf{0.494} & \textbf{4.029} & \textbf{0.0003 } \\
2 & Ego \& Alter both Whites - (Ego Non-White - Alter White) & Facebook & 1.692 & 0.712 & 2.377 & 0.082 \\ 
3 & Ego \& Alter both Whites - (Both Ego and Alter Non-Whites) & Facebook & -0.263 & 0.410 & -0.642 & 0.918 \\ 
4 & (Ego White - Alter Non-White) - (Ego Non-White - Alter White) & Facebook & -0.300 & 0.813 & -0.368 & 0.983 \\
\end{tabular} |
  \captionof{table}{Full reporting of the comparative statistics of Figure 3B in the main text: Pairwise Comparisons of 
          Connection Value Percentile between different Race mix for Ego and Alter for the different platforms. \textbf{Highlighted
contrasts} are significant at the 1\% level. DF = 66762.} 
\end{center}

  \begin{center}
  \renewcommand{\arraystretch}{0.55}
  \footnotesize
\begin{tabular}{@{\extracolsep{5pt}} ccccccc} 
\\[-3.8ex]\hline 
\hline \\[-1.8ex] 
 & contrast & Relative Ego and Alter Race & estimate (\%) & SE & t-stat & p-value \\ 
\hline \\[-1.8ex] 
\textbf{1} & \textbf{Facebook - Instagram} & \textbf{Ego} \& \textbf{Alter both Whites} & \textbf{-3.519} & \textbf{0.469} & \textbf{-7.500} & \textbf{0 } \\
2 & Facebook - LinkedIn & Ego \& Alter both Whites & -0.739 & 0.454 & -1.626 & 0.364 \\ 
3 & Facebook - X & Ego \& Alter both Whites & -0.661 & 0.368 & -1.798 & 0.274 \\ 
\textbf{4} & \textbf{Instagram - LinkedIn} & \textbf{Ego} \& \textbf{Alter both Whites} & \textbf{2.781} & \textbf{0.580} & \textbf{4.792} & \textbf{0.00001 } \\
\end{tabular} 
  \captionof{table}{Full reporting of the comparative statistics of Figure 3B in the main text: Pairwise Comparisons of 
           Connection Value Percentile between different platforms for different Race mix of Ego and Alter. \textbf{Highlighted
 contrasts} are significant at the 1\% level. DF = 66762.} 
\end{center}

\subsubsection{Figure 3C: Connection Value Percentile by Platform \& Relative Age of Ego and Alter.}

  \begin{center}
  \renewcommand{\arraystretch}{0.63}
  \footnotesize
\begin{tabular}{@{\extracolsep{5pt}} ccccccc} 
\\[-3.8ex]\hline 
\hline \\[-1.8ex] 
 & contrast & Platform & estimate (\%) & SE & t-stat & p-value \\ 
\hline \\[-1.8ex] 
1 & ego younger than alter - ego same age as alter & Facebook & 0.272 & 0.398 & 0.683 & 0.773 \\ 
2 & ego younger than alter - ego older than alter & Facebook & -0.022 & 0.422 & -0.053 & 0.998 \\ 
3 & ego same age as alter - ego older than alter & Facebook & -0.294 & 0.384 & -0.766 & 0.724 \\ 
\textbf{4} & \textbf{ego younger than alter - ego same age as alter} & \textbf{Instagram} & \textbf{-2.623} & \textbf{0.712} & \textbf{-3.685} & \textbf{0.001 } \\
\textbf{5} & \textbf{ego younger than alter - ego older than alter} & \textbf{Instagram} & \textbf{-3.124} & \textbf{0.821} & \textbf{-3.804} & \textbf{0.0004 } \\
6 & ego same age as alter - ego older than alter & Instagram & -0.501 & 0.722 & -0.694 & 0.767 \\ 
7 & ego younger than alter - ego same age as alter & LinkedIn & 1.501 & 0.692 & 2.168 & 0.077 \\ 
8 & ego younger than alter - ego older than alter & LinkedIn & 0.779 & 0.804 & 0.970 & 0.596 \\ 
9 & ego same age as alter - ego older than alter & LinkedIn & -0.721 & 0.789 & -0.914 & 0.632 \\ 
10 & ego younger than alter - ego same age as alter & X & -0.671 & 0.566 & -1.186 & 0.462 \\ 
11 & ego younger than alter - ego older than alter & X & -0.103 & 0.603 & -0.171 & 0.984 \\ 
12 & ego same age as alter - ego older than alter & X & 0.568 & 0.479 & 1.187 & 0.461 \\ 
\hline \\[-1.8ex] 
\end{tabular} 
  \captionof{table}{Full reporting of the comparative statistics of Figure 3C in the main text: Pairwise Comparisons of 
          Connection Value Percentile between different Age mix of Ego and Alter for the different platforms. \textbf{Highlighted
contrasts} are significant at the 1\% level. DF = 67965.} 
\end{center}

  \begin{center}
  \renewcommand{\arraystretch}{0.63}
\begin{tabular}{@{\extracolsep{5pt}} ccccccc} 
\\[-3.8ex]\hline 
\hline \\[-1.8ex] 
 & contrast & Ego Age & estimate (\%) & SE & t-stat & p-value \\ 
\hline \\[-1.8ex] 
1 & Facebook - Instagram & younger than alter & -1.231 & 0.651 & -1.890 & 0.232 \\ 
\textbf{2} & \textbf{Facebook - LinkedIn} & \textbf{younger than alter} & \textbf{-2.091} & \textbf{0.588} & \textbf{-3.559} & \textbf{0.002 } \\
3 & Facebook - X & younger than alter & -0.410 & 0.567 & -0.723 & 0.888 \\ 
4 & Instagram - LinkedIn & younger than alter & -0.860 & 0.762 & -1.129 & 0.671 \\
\end{tabular} 
  \captionof{table}{Full reporting of the comparative statistics of Figure 3C in the main text: Pairwise Comparisons of 
          Connection Value Percentile between different platforms for different relative age between Ego and Alter. \textbf{Highlighted
contrasts} are significant at the 1\% level. DF = 67965.} 
\end{center}

\subsubsection{Figure 3D: Connection Value Percentile by Platform \& Absolute Age of Alter.}

  \begin{center}
  \renewcommand{\arraystretch}{0.63}
  \scriptsize
\begin{tabular}{@{\extracolsep{5pt}} ccccccc} 
\\[-3.8ex]\hline 
\hline \\[-1.8ex] 
 & contrast & Platform & estimate (\%) & SE & t-stat & p-value \\ 
\hline \\[-1.8ex] 
1 & Under 18 years old - (18-24 years old) & Facebook& 0.035 & 0.993 & 0.035 & 1 \\ 
2 & Under 18 years old - (25-34 years old) & Facebook& 0.723 & 0.964 & 0.750 & 0.976 \\ 
3 & Under 18 years old - (35-44 years old) & Facebook& 0.672 & 0.981 & 0.685 & 0.984 \\ 
4 & Under 18 years old - (45-54 years old) & Facebook& 1.381 & 1.033 & 1.337 & 0.764 \\
\end{tabular} 
  \captionof{table}{Full reporting of the comparative statistics of Figure 3D in the main text: Pairwise Comparisons of 
          Connection Value Percentile between different Age bins of Alter for the different platforms. \textbf{Highlighted
contrasts} are significant at the 1\% level. DF = 67953.} 
\end{center}

  \begin{center}
  \renewcommand{\arraystretch}{0.63}
\begin{tabular}{rllrrrr}
  \hline
 & contrast & Alter Age & estimate (\%) & SE & t-stat & p-value \\ 
  \hline
\textbf{1} & \textbf{Facebook - Instagram} & \textbf{Under 18 years old} & \textbf{-6.91} & \textbf{1.59} & \textbf{-4.34} & \textbf{0.00 } \\
  2 & Facebook - LinkedIn & Under 18 years old & -1.11 & 2.58 & -0.43 & 0.97 \\ 
  3 & Facebook - X & Under 18 years old & -3.54 & 1.56 & -2.27 & 0.11 \\ 
  4 & Instagram - LinkedIn & Under 18 years old & 5.80 & 2.73 & 2.12 & 0.15 \\ 
  5 & Instagram - X & Under 18 years old & 3.37 & 1.81 & 1.86 & 0.24 \\ 
  6 & LinkedIn - X & Under 18 years old & -2.43 & 2.72 & -0.90 & 0.81 \\ 
\textbf{  7} & \textbf{Facebook - Instagram} & \textbf{18-24 years old} & \textbf{-3.65} & \textbf{0.66} & \textbf{-5.49} & \textbf{0.00 } \\
\hline
\end{tabular}
\captionof{table}{Full reporting of the comparative statistics of Figure 3D in the main text: Pairwise Comparisons of 
          Connection Value Percentile between different platforms for different Age bin of Alter. \textbf{Highlighted
contrasts} are significant at the 1\% level. DF = 67953.} 
\end{center}

\subsubsection{Figure 4A: Total Platform Valuation by Platform Feature Use}

    \begin{center}
  \renewcommand{\arraystretch}{0.63}
  \scriptsize
\begin{tabular}{lrr}
\toprule
Contrast & estimate (\$) & p-value\\
\midrule
Messenger - Posting\_content\_on\_your\_news\_feed & 5.361 & 0.326\\
Messenger - Engaging\_with\_content\_from\_your\_friends\_on\_your\_news\_feed & 9.135 & 0.090\\
Messenger - Facebook\_Live & 10.334 & 0.058\\
Messenger - Joining\_groups & 13.468 & 0.011\\
\textbf{Messenger - Playing\_games} & \textbf{15.781} & \textbf{0.002}\\
\textbf{Messenger - Organize\_attend\_events} & \textbf{16.241} & \textbf{0.004}\\
\textbf{Messenger - Marketplace} & \textbf{19.520} & \textbf{0.000}\\
\bottomrule
\end{tabular}
  \captionof{table}{Full reporting of the comparative statistics of Figure 4A ({\bf Facebook}) in the main text: Pairwise Comparisons of the increase in TPV associated with feature use. \textbf{Highlighted contrasts} are significant at the 1\% level.} 
\end{center}

    \begin{center}
  \renewcommand{\arraystretch}{0.63}
\begin{tabular}{lrr}
\toprule
Comparison & Estimate & p-value\\
\midrule
Post\_Instagram\_reels vs Post\_photos\_as\_a\_post & 6.293 & 0.237\\
Post\_Instagram\_reels vs Direct\_messages & 7.632 & 0.158\\
Post\_Instagram\_reels vs Like\_and\_comment\_on\_stories & 7.950 & 0.158\\
Post\_Instagram\_reels vs Post\_Instagram\_stories & 9.596 & 0.080\\
Post\_Instagram\_reels vs Like\_and\_comment\_on\_reels & 9.815 & 0.077\\
\addlinespace
Post\_Instagram\_reels vs Instagram\_live & 12.999 & 0.016\\
\textbf{Post\_Instagram\_reels vs Like\_and\_comment\_on\_posts} & \textbf{17.487} & \textbf{0.001}\\
\bottomrule
\end{tabular}
\end{center}
  \captionof{table}{Full reporting of the comparative statistics of Figure 4A ({\bf Instagram}) in the main text: Pairwise Comparisons of the increase in TPV associated with feature use. \textbf{Highlighted contrasts} are significant at the 1\% level.}

   \begin{center}
  \renewcommand{\arraystretch}{0.63}
\begin{tabular}{lrr}
\toprule
Comparison & Estimate & p-value\\
\midrule
Spaces vs Bookmark\_tweets & 2.449 & 0.908\\
Spaces vs X\_Blue\_Premium & 3.018 & 0.908\\
Spaces vs Like\_reply\_retweet\_tweets & 4.670 & 0.832\\
Spaces vs Direct\_messages & 4.748 & 0.816\\
Spaces vs Post\_tweets & 4.990 & 0.816\\
\addlinespace
Spaces vs X\_lists & 9.482 & 0.816\\
Bookmark\_tweets vs X\_Blue\_Premium & 0.568 & 0.985\\
Bookmark\_tweets vs Like\_reply\_retweet\_tweets & 2.220 & 0.908\\
Bookmark\_tweets vs Direct\_messages & 2.299 & 0.908\\
Bookmark\_tweets vs Post\_tweets & 2.540 & 0.908\\
\addlinespace
Bookmark\_tweets vs X\_lists & 7.032 & 0.816\\
X\_Blue\_Premium vs Like\_reply\_retweet\_tweets & 1.652 & 0.938\\
X\_Blue\_Premium vs Direct\_messages & 1.731 & 0.938\\
X\_Blue\_Premium vs Post\_tweets & 1.972 & 0.938\\
X\_Blue\_Premium vs X\_lists & 6.464 & 0.816\\
\addlinespace
Like\_reply\_retweet\_tweets vs Direct\_messages & 0.078 & 0.985\\
Like\_reply\_retweet\_tweets vs Post\_tweets & 0.320 & 0.985\\
Like\_reply\_retweet\_tweets vs X\_lists & 4.812 & 0.816\\
Direct\_messages vs Post\_tweets & 0.241 & 0.985\\
Direct\_messages vs X\_lists & 4.733 & 0.816\\
\addlinespace
Post\_tweets vs X\_lists & 4.492 & 0.816\\
\bottomrule
\end{tabular}
  \captionof{table}{Full reporting of the comparative statistics of Figure 4A ({\bf X}) in the main text: Pairwise Comparisons of the increase in TPV associated with feature use. \textbf{Highlighted contrasts} are significant at the 1\% level.} 
\end{center}

   \begin{center}
  \renewcommand{\arraystretch}{0.63}
\begin{tabular}{lrr}
\toprule
Comparison & Estimate & p-value\\
\midrule
LinkedIn\_groups vs Keep\_profile\_up\_to\_date & 1.402 & 0.719\\
LinkedIn\_groups vs like\_comment\_and\_share\_content & 6.756 & 0.148\\
LinkedIn\_groups vs Search\_and\_apply\_for\_jobs & 8.402 & 0.148\\
Keep\_profile\_up\_to\_date vs like\_comment\_and\_share\_content & 5.353 & 0.254\\
Keep\_profile\_up\_to\_date vs Search\_and\_apply\_for\_jobs & 7.000 & 0.148\\
like\_comment\_and\_share\_content vs Search\_and\_apply\_for\_jobs & 1.646 & 0.719\\
\bottomrule
\end{tabular}
  \captionof{table}{Full reporting of the comparative statistics of Figure 4A ({\bf LinkedIn}) in the main text: Pairwise Comparisons of the increase in TPV associated with feature use. \textbf{Highlighted contrasts} are significant at the 1\% level.} 
\end{center}

\clearpage

\subsubsection{Figure 4B: Total Platform Valuation by Platform Feature Use and Income}

   \begin{center}
  \renewcommand{\arraystretch}{0.63}
  \small
\begin{tabular}{@{\extracolsep{5pt}} ccccccc} 
\\[-3.8ex]\hline 
\hline \\[-1.8ex] 
  & contrast & income level & estimate (\$) & SE & t-stat & p-value \\
\hline \\[-1.8ex] \\
\textbf{1} & \textbf{Messenger(FALSE) - Messenger(TRUE)} & \textbf{High Income} & \textbf{-14.719} & \textbf{4.086} & \textbf{-3.602} & \textbf{0.0003} \\
\textbf{2} & \textbf{Messenger(FALSE) - Messenger(TRUE)} & \textbf{Low Income} & \textbf{-26.949} & \textbf{5.229} & \textbf{-5.154} & \textbf{0.00000} \\
\hline \\[-3.8ex] 
\end{tabular} 
  \captionof{table}{Full reporting of the comparative statistics of \textit{Figure 4B (top left)} in the main text: Pairwise Comparisons of TPV for differential use of Facebook Messenger for different Income levels. \textbf{Highlighted
contrasts} are significant at the 1\% level. DF $=$ 2466.} 
\end{center}

   \begin{center}
  \renewcommand{\arraystretch}{0.63}
%
 
  \captionof{table}{Full reporting of the comparative statistics of \textit{Figure 4B (top left)} in the main text: Pairwise Comparisons of TPV for different Income level for Messenger Users and Non Users . \textbf{Highlighted
contrasts} are significant at the 1\% level. DF = 2466.}
\end{center}

   \begin{center}
  \renewcommand{\arraystretch}{0.63}
  \small
%
 
  \captionof{table}{Full reporting of the comparative statistics of \textit{Figure 4B (top right)} in the main text: Pairwise Comparisons of TPV for differential use of Reels and by different Income level. \textbf{Highlighted
contrasts} are significant at the 1\% level. DF = 2484.} 
\end{center}

   \begin{center}
  \renewcommand{\arraystretch}{0.63}
%
 
  \captionof{table}{Full reporting of the comparative statistics of \textit{Figure 4B (top right)} in the main text: Pair-
wise Comparisons of TPV for different Income level for Users that either post Instagram Reels or not. \textbf{Highlighted
contrasts} are significant at the 1\% level. DF = 2484.} 
\end{center}

   \begin{center}
  \renewcommand{\arraystretch}{0.63}
%
 
  \captionof{table}{Full reporting of the comparative statistics of \textit{Figure 4B (bottom right)} in the main text: Pairwise Comparisons of TPV for differential use of X Lists and by different Income level. \textbf{Highlighted
contrasts} are significant at the 1\% level. DF = 2233.} 
\end{center}

   \begin{center}
  \renewcommand{\arraystretch}{0.63}
%
 
  \captionof{table}{Full reporting of the comparative statistics of \textit{Figure 4B (bottom right)} in the main text: Pairwise Comparisons of TPV for different Income level for users and non-users of X Lists. \textbf{Highlighted
contrasts} are significant at the 1\% level. DF = 2233.} 
\end{center}

   \begin{center}
  \renewcommand{\arraystretch}{0.63}
  \small
%
 
  \captionof{table}{Full reporting of the comparative statistics of Figure 4B (bottom left) in the main text: Pair-
wise Comparisons of TPV for differential use of LinkedIn groups for different Income levels. \textbf{Highlighted
contrasts} are significant at the 1\% level. DF = 2196.} 
\end{center}

   \begin{center}
  \renewcommand{\arraystretch}{0.63}
%
 
  \captionof{table}{Full reporting of the comparative statistics of \textit{Figure 4B (bottom left)} in the main text: Pairwise Comparisons of TPV for different Income level for users and non-users of LinkedIn groups. \textbf{Highlighted
contrasts} are significant at the 1\% level. DF = 2196.} 
\end{center}

\centering
\begin{talltblr}[         
entry=none,label=none,
note{}={* p \num{< 0.1}, ** p \num{< 0.05}, *** p \num{< 0.01}},
note{ }={Odds ratios reported. Strata by ego \$\textbackslash{}times\$ ranking stage.},
]                     
{                     
colspec={Q[]Q[]},
column{2}={}{halign=c,},
column{1}={}{halign=l,},
hline{48}={1,2}{solid, black, 0.05em},
}                     
\toprule
& (1) \\ \midrule 
How you know: I only know them online &  \\[-1.8ex] 
How you know: Family member or family friend & \num{1.304}*** \\[-1.8ex]
How you know: None of these & \num{1.027} \\[-1.8ex]
How you know: School (K-12) & \num{1.014} \\[-1.8ex]
How you know: Work & \num{1.028} \\[-1.8ex]
How you know: Shared interest in real life & \num{1.071}** \\[-1.8ex]
How you know: College or University & \num{1.068}** \\[-1.8ex]
Meeting frequency: Never &  \\[-1.8ex]
Meeting frequency: I live with this person & \num{2.738}*** \\[-1.8ex]
Meeting frequency: Less than 4 times/year & \num{1.326}*** \\[-1.8ex]
Meeting frequency: About once a month & \num{1.563}*** \\[-1.8ex]
Meeting frequency: About once a week & \num{1.651}*** \\[-1.8ex]
Meeting frequency: Multiple times per week & \num{1.855}*** \\[-1.8ex]
Connected on other platform: Yes & \num{1.263}*** \\[-1.8ex]
Alter gender: Male &  \\[-1.8ex]
Alter gender: Female & \num{1.066}*** \\[-1.8ex]
Alter age: Over 65 years old &  \\[-1.8ex]
Alter age: Under 18 years old & \num{1.121}** \\[-1.8ex]
Alter age: 18-24 years old & \num{1.108}** \\[-1.8ex]
Alter age: 25-34 years old & \num{1.100}** \\[-1.8ex]
Alter age: 35-44 years old & \num{1.078}** \\[-1.8ex]
Alter age: 45-54 years old & \num{1.015} \\[-1.8ex]
Alter age: 55-64 years old & \num{1.058} \\[-1.8ex]
Alter race: Asian / Pacific Islander &  \\[-1.8ex]
Alter race: White & \num{1.042} \\[-1.8ex]
Alter race: Black or African American & \num{1.018} \\[-1.8ex]
Alter race: Hispanic or Latino & \num{1.037} \\[-1.8ex]
Alter race: Native American or American Indian & \num{1.022} \\[-1.8ex]
Num.Obs. & \num{230669} \\
\bottomrule
\end{talltblr}

\centering
\begin{talltblr}[         
entry=none,label=none,
note{}={* p \num{< 0.1}, ** p \num{< 0.05}, *** p \num{< 0.01}},
note{ }={Odds ratios reported. Strata by ego \$\textbackslash{}times\$ ranking stage.},
]                     
{                     
colspec={Q[]Q[]},
column{2}={}{halign=c,},
column{1}={}{halign=l,},
hline{38}={1,2}{solid, black, 0.05em},
}                     
\toprule
& (1) \\ \midrule 
How you know: I only know them online &  \\[-1.8ex] 
How you know: Family member or family friend & \num{1.299}*** \\[-1.8ex] 
How you know: None of these & \num{1.029} \\[-1.8ex] 
How you know: School (K-12) & \num{1.013} \\[-1.8ex] 
How you know: Work & \num{1.030} \\[-1.8ex] 
How you know: Shared interest in real life & \num{1.070}** \\[-1.8ex] 
How you know: College or University & \num{1.068}** \\[-1.8ex] 
Meeting frequency: Never &  \\[-1.8ex]
Meeting frequency: I live with this person & \num{2.722}*** \\[-1.8ex] 
fq\_5\_friendnum\_Less than 4 times/year & \num{1.325}*** \\[-1.8ex] 
Meeting frequency: About once a month & \num{1.560}*** \\[-1.8ex] 
Meeting frequency: About once a week & \num{1.647}*** \\[-1.8ex] 
Meeting frequency: Multiple times per week & \num{1.851}*** \\[-1.8ex] 
Connected on other platform: No &  \\[-1.8ex] 
Connected on other platform: Yes & \num{1.262}*** \\[-1.8ex]
Gender Mix: Ego Male - Alter Male &  \\[-1.8ex] 
Gender Mix: Ego Female - Alter Female & \num{1.042}** \\[-1.8ex] 
Gender Mix: Ego Male - Alter Female & \num{1.092}*** \\[-1.8ex] 
Age Mix: younger than alter &  \\[-1.8ex] 
Age Mix: same age as alter & \num{1.054}*** \\[-1.8ex] 
Age Mix: older than alter & \num{1.044}** \\[-1.8ex] 
Race Mix: Ego Non-White - Alter White & \\[-1.8ex]
Race Mix: Ego \& Alter both Whites & \num{1.052}** \\[-1.8ex] 
Race Mix: Both Ego and Alter Non-Whites & \num{1.057}* \\[-1.8ex] 
Num.Obs. & \num{230669} \\
\bottomrule
\end{talltblr}

\centering
\begin{talltblr}[         
entry=none,label=none,
note{}={* p \num{< 0.1}, ** p \num{< 0.05}, *** p \num{< 0.01}},
note{ }={Odds ratios reported. Strata by ego \$\textbackslash{}times\$ ranking stage.},
]                     
{                     
colspec={Q[]Q[]},
column{2}={}{halign=c,},
column{1}={}{halign=l,},
hline{38}={1,2}{solid, black, 0.05em},
}                     
\toprule
& (1) \\ \midrule 
How you know: I only know them online &  \\[-1.8ex] 
How you know: None of these & \num{1.029} \\[-1.8ex] 
How you know: School (K-12) & \num{1.013} \\[-1.8ex] 
How you know: Work & \num{1.029} \\[-1.8ex] 
How you know: Shared interest in real life & \num{1.069}** \\[-1.8ex] 
How you know: College or University & \num{1.068}** \\[-1.8ex] 
How you know: Family member or family friend & \num{1.298}*** \\[-1.8ex] 
Meeting frequency: Never &  \\[-1.8ex]
Meeting frequency: Less than 4 times/year & \num{1.325}*** \\[-1.8ex] 
Meeting frequency: About once a month & \num{1.560}*** \\[-1.8ex] 
Meeting frequency: About once a week & \num{1.647}*** \\[-1.8ex] 
Meeting frequency: Multiple times per week & \num{1.850}*** \\[-1.8ex] 
Meeting frequency: I live with this person & \num{2.719}*** \\[-1.8ex] 
Connected on other platform: No &  \\[-1.8ex] 
Connected on other platform: Yes & \num{1.262}*** \\[-1.8ex] 
Gender: Alter Female & \num{1.067}*** \\[-1.8ex] 
Gender: Ego and Alter Different Gender & \num{1.024}** \\[-1.8ex] 
Age Mix: younger than alter &  \\[-1.8ex] 
Age Mix: same age as alter & \num{1.054}*** \\[-1.8ex] 
Age Mix: ego older than alter & \num{1.044}** \\[-1.8ex] 
Race: alter\_nonwhite & \num{1.011} \\[-1.8ex] 
Race: same\_race & \num{1.061}*** \\[-1.8ex] 
Num.Obs. & \num{230669} \\
\bottomrule
\end{talltblr}

\clearpage
\newpage

\bibliography{SuppMat.bib}
\bibliographystyle{naturemag.bst}
\pagebreak
\huge{Wave 1 Instrument}
\includepdf[pages=-]{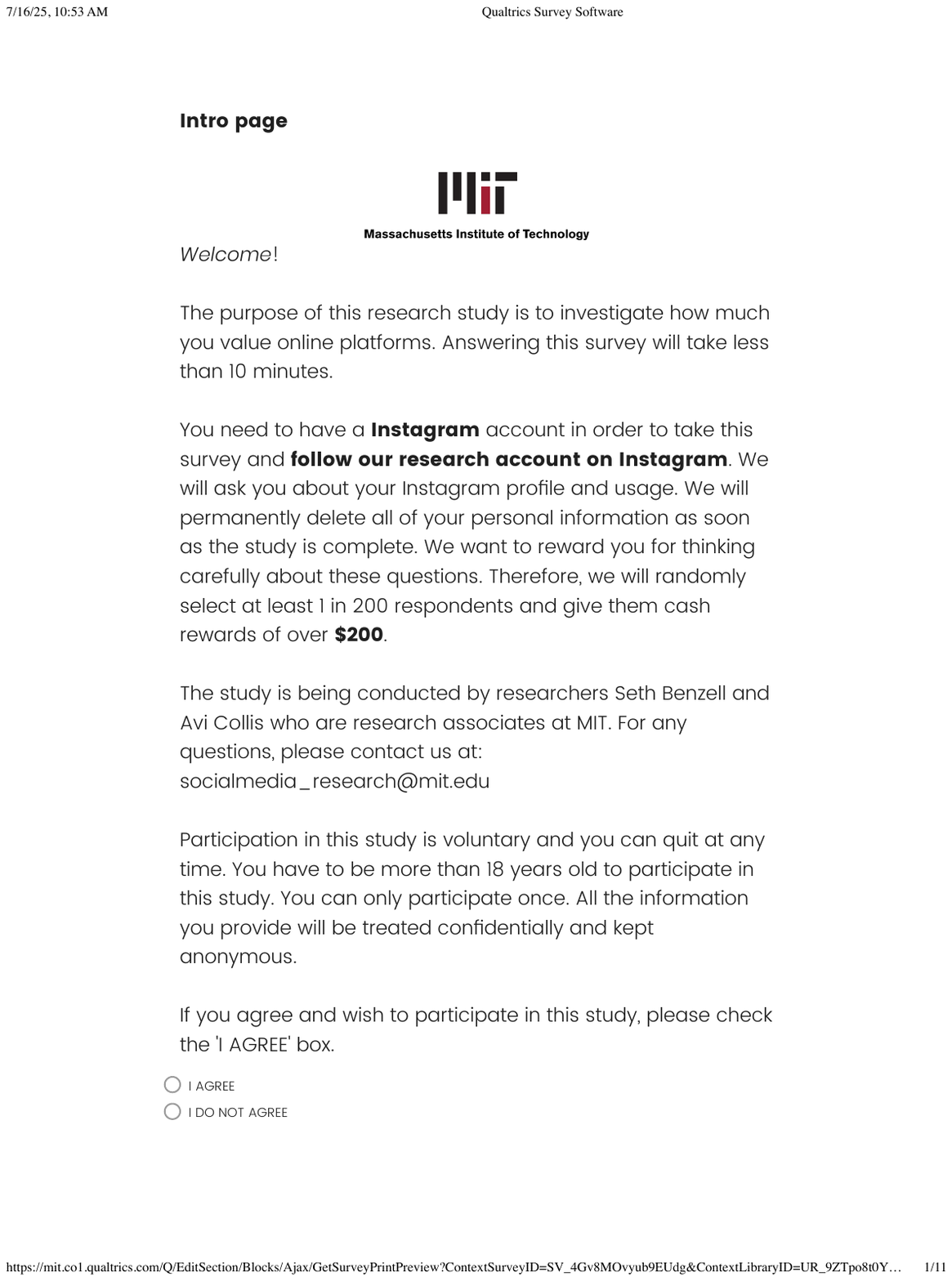}
\huge{Wave 2 Instrument}
\includepdf[pages=-]{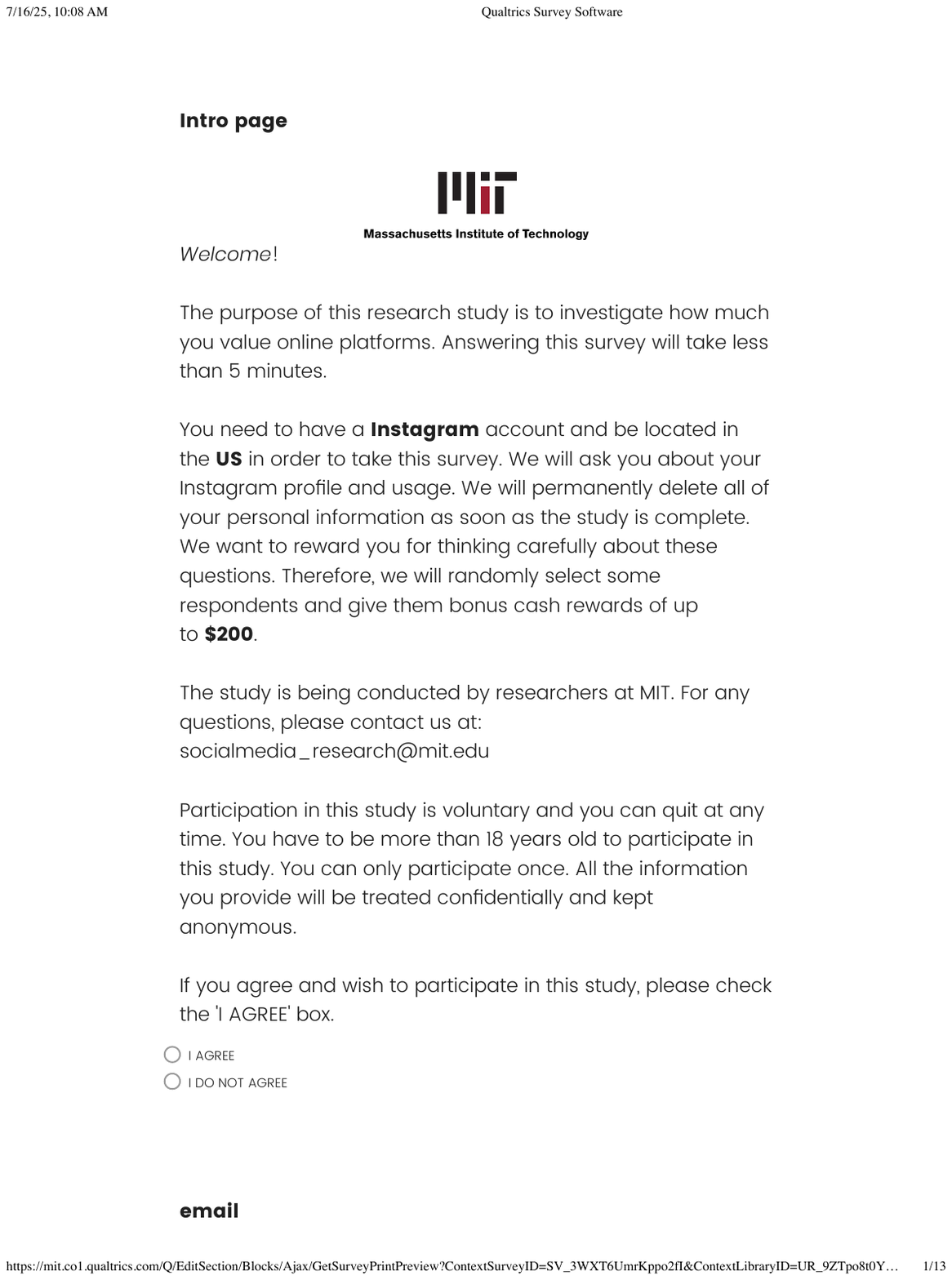}